\documentclass[12pt]{article}
\usepackage{graphicx}

\newcommand{\cd}{\makebox[0.08cm]{$\cdot$}}
\newcommand{\MSbar} {\hbox{$\overline{\hbox{\tiny MS}}$}}
\newcommand{\barMS} {\overline{\rm MS}}
\newcommand{\ket}[1]{\,\left|\,{#1}\right\rangle}
\renewcommand{\bar}[1]{\overline{#1}}
\newcommand{\VEV}[1]{\left\langle{#1}\right\rangle}
\newcommand{\ie}{{\em i.e.}}
\newcommand{\etal}{{\em et al.}}
\newcommand{\eg}{{\em e.g.}}
\newcommand{\gsim} {\buildrel > \over {_\sim}}
\newcommand{\M}{\mathcal{M}}

\begin {document}
\begin{flushright}
{\small
SLAC--PUB--10811\\
October 2004\\}
\end{flushright}

\vfill
\begin{center}
{{\bf\LARGE Testing Quantum Chromodynamics with\\[1ex] Antiprotons}
\footnote{Work supported by the Department of Energy under contract
number DE--AC02--76SF00515.}}

\bigskip
{\it Stanley J. Brodsky \\
Stanford Linear Accelerator Center \\
Stanford University, Stanford, California 94309 \\
E-mail:  sjbth@slac.stanford.edu}
\medskip
\end{center}

\vfill

\begin{center}
{\it Presented at the  \\
International School of Physics ``Enrico Fermi''\\
Course CLVIII -- ``Hadron Physics'' \\
Varenna, Italy\\
June 22--July 2, 2004 }\\
\end{center}

\vfill \newpage

\vfill

\begin{center}
{\bf Abstract}
\end{center}

The antiproton storage ring HESR to be constructed at GSI will open
up a new range of perturbative and nonperturbative tests of QCD in
exclusive and inclusive reactions. I discuss 21 tests of QCD using
antiproton beams which can illuminate novel features of QCD. The
proposed experiments include the formation of exotic hadrons,
measurements of timelike generalized parton distributions, the
production of charm at threshold, transversity measurements in
Drell-Yan reactions, and searches for single-spin asymmetries.  The
interactions of antiprotons in nuclear targets will allow tests of
exotic nuclear phenomena such as color transparency, hidden color,
reduced nuclear amplitudes, and the non-universality of nuclear
antishadowing. The central tool used in these lectures are
light-front Fock state wavefunctions which encode the bound-state
properties of hadrons in terms of their quark and gluon degrees of
freedom at the amplitude level.  The freedom to choose the
light-like quantization four-vector provides an explicitly covariant
formulation of light-front quantization and can be used to determine
the analytic structure of light-front wave functions. QCD becomes
scale free and conformally symmetric in the analytic limit of zero
quark mass and zero $\beta$ function.  This ``conformal
correspondence principle" determines the form of the expansion
polynomials for distribution amplitudes and the behavior of
non-perturbative wavefunctions which control hard exclusive
processes at leading twist.  The conformal template also can be used
to derive commensurate scale relations which connect observables in
QCD without scale or scheme ambiguity.  The AdS/CFT correspondence
of large $N_C$ supergravity theory in higher-dimensional  anti-de
Sitter space with supersymmetric QCD in 4-dimensional space-time has
important implications for hadron phenomenology in the conformal
limit, including the nonperturbative derivation of counting rules
for exclusive processes and the behavior of structure functions at
large $x_{bj}.$ String/gauge duality also predicts the QCD power-law
fall-off of light-front Fock-state hadronic wavefunctions with
arbitrary orbital angular momentum at high momentum transfer. I also
review recent work which shows that the diffractive component of
deep inelastic scattering, single spin asymmetries, as well as
nuclear shadowing and antishadowing, cannot be computed from the
LFWFs of hadrons in isolation.

\newpage

\section{Introduction}

Quantum Chromodynamics is a remarkable theory. Not only does it
provide a consistent description of the strong and nuclear
interactions in terms of quark and gluon degrees of freedom at
short distances, but its non-Abelian Yang Mills gauge theory
structure also provides the foundation for the electroweak
interactions and the eventual unification of the electrodynamic,
weak, and hadronic forces at very short distances. The theory has
extraordinary properties, such as color
confinement~\cite{Greensite:2003bk}, asymptotic
freedom~\cite{Gross:1973id,Politzer:1973fx}, a complex
vacuum structure, and predicts an array of new forms of hadronic
matter such as gluonium and hybrid states~\cite{Klempt:2000ud}.
The phase structure of QCD~\cite{Rajagopal:2000wf} implies
the formation of a quark-gluon plasma in high energy heavy ion
collisions~\cite{Rischke:2003mt} as well insight into the
evolution of the early universe~\cite{Schwarz:2003du}.

The asymptotic freedom property of QCD explains why the strong
interactions become weak at short distances, thus allowing hard
processes to be interpreted directly in terms of the perturbative
interactions of quark and gluon quanta. This in turn leads  to
factorization theorems~\cite{Collins:1989gx,Bodwin:1984hc} for
both inclusive and exclusive processes~\cite{Brodsky:1989pv} which
separate the hard scattering subprocesses which control the
reaction from the nonperturbative physics of the interacting
hadrons.

More recently, a remarkable duality has been established between
supergravity string theory in 10 dimensions and conformal
supersymmetric extensions of
QCD~\cite{Maldacena:1997re,Polchinski:2001tt,Brower:2002er,Andreev:2002aw}.
The AdS/CFT correspondence is now leading to a new understanding
of QCD at strong coupling and the implications of its
nearly-conformal structure.

The  Lagrangian density of QCD~\cite{Fritzsch:1973pi} has a deceptively
simple form:
\begin{equation}
{\cal L}=   {\overline \psi}(i\gamma_\mu D^\mu-m)\psi
-{1\over 4} G^2_{\mu \nu}
\end{equation}
where the covariant derivative is $i
D_\mu = i \partial_\mu - g A_\mu$ and where the gluon field
strength is $G_{\mu \nu} = {i\over g}[D_\mu, D_\nu]$. The
structure of the QCD Lagrangian is dictated by two principles: (i)
local $SU(N_C)$ color gauge invariance -- the theory is invariant
when a quark field is rotated in color space and
transformed in phase by an arbitrary unitary matrix $\psi(x) \to
U(x) \psi(x)$ locally at any point $x^\mu$ in space and time;  and (ii)
renormalizability, which requires the appearance of dimension four
interactions. In principle, the only parameters of QCD are the
quark masses and the QCD coupling determined from a single
observable at a single scale.

Solving QCD is extremely challenging because of the non-Abelian
three-point and four-point gluonic couplings contained in its Lagrangian.
Exact solutions are known  for $QCD(1+1)$ at $N_C \to \infty$  by
't Hooft~\cite{'tHooft:1974hx}. The one-space one-time theory can
be solved numerically to any precision at finite $N_C$ for any
coupling strength and number of quark flavors using discretized
light-cone quantization
(DLCQ)~\cite{Pauli:1985pv,Hornbostel:1988fb,burkardt89,Brodsky:1997de}.
One can use DLCQ to calculate the entire spectrum of virtually any 1+1
theory, its discrete bound states as well as the scattering continuum.
The main emphasis of the DLCQ method applied to QCD is the determination
of the wavefunctions of the hadrons from first principles.

Currently the most important computational tool for making
predictions in strong-coupling QCD(3+1) is lattice gauge
theory~\cite{Wilson:1976zj} which has made enormous progress in
recent years particularly in computing mass spectra and decay
constants. Lattice gauge theory can only provide limited dynamical
information because of the difficulty of continuing predictions
from Euclidean to Minkowski space. At present, results are limited
to large quark and pion masses such that the $\rho$ meson is
stable~\cite{DeGrand:2003xu}. The DLCQ solutions for 1+1 quantum
field theories could provide a powerful test of lattice methods.

Other important nonperturbative QCD methods are Dyson-Schwinger
techniques~\cite{Maris:2003vk} and the transverse
lattice~\cite{Dalley:2004rq} which combines DLCQ methods for the
one space and one time theory with lattice methods in transverse
space. The Dyson-Schwinger methods account well for running quark
mass effects, and in principle can give important hadronic
wavefunction information. The transverse lattice method has
recently provided the first computation of the generalized parton
distributions of the pion~\cite{Dalley:2004rq}.

{\it Light Front Wavefunctions}~\cite{Brodsky:1997de}: The concept
of a wave function of a hadron as a composite of relativistic
quarks and gluons is naturally formulated in terms of the
light-front Fock expansion at fixed light-front
time~\cite{Dirac:1949cp}, $\tau=x \cdot \omega$.  The four-vector
$\omega$, with $\omega^2 = 0$, determines the orientation of the
light-front plane; the freedom to choose $\omega$ provides an
explicitly covariant formulation of light-front
quantization~\cite{cdkm}. The light-front wave functions (LFWFs)
$\psi_n(x_i,k_{\perp_i},\lambda_i)$, with $x_i={k_i \cdot
\omega\over P\cdot \omega}$, $\sum^n_{i=1} x_i=1, $
$\sum^n_{i=1}k_{\perp_i}=0_\perp$, are the coefficient functions
for $n$ partons in the Fock expansion, providing a general
frame-independent representation of the hadron state. The
$\lambda_i$ are the eigenvalues of the spin projections $S^z$ in
the $\hat z$ direction.  Angular momentum conservation for each
Fock state implies
\begin{equation}
J^z= \sum_i^{n} S^z_i + \sum_i^{n-1} L^z_i
\end{equation}
where $L_z$ is one of the
$n-1$ relative orbital angular momentum.  Relativity and quantum
theory require that the number of Fock states cannot be
bounded.  However, the probability of massive Fock states with
invariant mass $\cal M$ falls-off at least as fast as $1/{\cal
M}^2.$

The LFWFs are boost invariant; i.e., independent of Lorentz frame.
In principle, they are solutions of the LF Heisenberg equation
where $H_{LF}$ is computed from the theory quantized at fixed
$\tau.$  As I will review in these lectures, given the LFWFs, one
can calculate a myriad of dynamical processes. The LFWFs of
hadrons are thus centerposts of the theory~\cite{Brodsky:2004sb}.

The Light-Front Fock state expansion provides a number of new perspectives
for QCD:

{\it Intrinsic Glue and Sea:} Even though QCD was motivated by the
successes of the parton model, QCD predicts many new features
which go well beyond a three-quark bound state description of the
proton.  Since the number of Fock components cannot be limited in
relativity and quantum mechanics,  the nonperturbative
wavefunction of a proton contains gluons and sea quarks, including
heavy quarks, at any resolution scale.  Thus there is no scale
$Q_0$ in deep inelastic lepton-proton scattering where the proton
can be approximated by its valence quarks. The sea-quark
distributions $Q(x)$ and $\bar Q(x)$ are not
equal~\cite{Brodsky:1996hc}.

{\it Initial and Final-State Interactions:} Although it has been
more than 35 years since the discovery of Bjorken
scaling~\cite{Bjorken:1968dy} in
electroproduction~\cite{Bloom:1969kc}, there are still many issues
in deep-inelastic lepton scattering and Drell-Yan reactions which
are only now being understood from a fundamental basis in QCD. In
contrast to the parton model, final-state interactions in deep
inelastic  scattering and initial state interactions in hard
inclusive reactions cannot be neglected -- leading to $T-$odd
single spin
asymmetries~\cite{Brodsky:2002cx,Belitsky:2002sm,Collins:2002kn}
and diffractive
contributions~\cite{Brodsky:2002ue,Brodsky:2004hi}. This in turn
implies that  the structure functions measured in deep inelastic
scattering are not probability distributions computed from the
square of the LFWFs computed in isolation~\cite{Brodsky:2002ue}.

{\it Novel Nuclear Phenomena:} In the case of nuclei, QCD predicts
that nuclear wavefunctions contain ``hidden
color"~\cite{Brodsky:1983vf} components: color configurations not
dual to the usual nucleonic degrees of freedom. For example, the
scaling of the deuteron's reduced form factor suggests that the
probability of six-quark Fock states with a color configuration
orthogonal to that of the proton and neutron is of order of
$15\%.$ I will also discuss in these lectures other surprising
features of QCD which contrast with standard nuclear physics
descriptions, such as ``color transparency"~\cite{Brodsky:1988xz},
the physical origin of antishadowing~\cite{Brodsky:1989qz} and its
nonuniversal character~\cite{Brodsky:2004qa}. As I will discuss,
the antishadowing of nuclear structure functions is quark-flavor
specific; this implies that part of the anomalous
NuTeV~\cite{Zeller:2001hh} result for $\sin^2\theta_W$ could be
due to the non-universality  of nuclear antishadowing for charged
and neutral currents.

I will also several topics in which the underlying conformal
symmetry of QCD plays a crucial role:

{\it AdS/CFT Correspondence and QCD:} The AdS/CFT
correspondence~\cite{Maldacena:1997re,Polchinski:2001tt,%
Brower:2002er,Andreev:2002aw} between superstring theory in 10 dimensions and
supersymmetric Yang Mills theory in 3+1 dimensions can provide important information on
QCD phenomena without reliance on perturbation theory. As I will discuss in these
lectures, one can use this connection to establish the form of QCD wavefunctions at large
transverse momentum $k^2_\perp \to \infty$ and at $x \to 1$~\cite{Brodsky:2003px}. The
AdS/CFT correspondence has important implications for hadron phenomenology in the
conformal limit, including an all-orders demonstration of counting
rules~\cite{Brodsky:1973kr,Matveev:ra,Brodsky:1974vy} for hard exclusive
processes~\cite{Polchinski:2001tt}, as well as determining essential aspects of hadronic
light-front wavefunctions~\cite{Brodsky:2003px}.

{\it The Conformal Correspondence
Principle~}\cite{Brodsky:2004qb,Brodsky:2003dn}: The recent
investigations using the AdS/CFT correspondence has reawakened
interest in the conformal features of QCD. QCD becomes scale free
and conformally symmetric in the analytic limit of zero quark mass
and zero $\beta$ function~\cite{Parisi:zy}.  This correspondence
principle provides a new tool, the conformal template, which is
very useful for theory analyses, such as the expansion polynomials
for distribution
amplitudes~\cite{Brodsky:1980ny,Brodsky:1984xk,Brodsky:1985ve,Braun:2003rp},
the non-perturbative wavefunctions which control exclusive
processes at leading twist~\cite{Lepage:1979zb,Brodsky:2000dr}.
The conformal template also can be used to derive commensurate
relations~\cite{Brodsky:1994eh,Rathsman:2001xe} which connect
observables in QCD without scale or scheme ambiguity.

The classical Lagrangian of QCD for massless quarks is conformally
symmetric.  Since it has no intrinsic mass scale, the classical
theory is invariant under the $SO(4,2)$ translations, boosts, and
rotations of the Poincare  group, plus the dilatations and other
transformations of the conformal group. Scale invariance and
therefore conformal symmetry is destroyed in the quantum theory by
the renormalization procedure which introduces a renormalization
scale as well as by quark masses. Conformal symmetry is thus
broken in physical QCD; nevertheless, we can still recover the
underlying features of the conformally invariant theory by
evaluating any expression in QCD in the analytic limit of zero
quark mass and zero $\beta$ function:
\begin{equation}
\lim_{m_q \to 0, \beta \to 0} \mathcal{O}_{QCD} = \mathcal{ O}_{\rm
conformal\ QCD} \ .
\end{equation} This conformal correspondence limit is analogous
to Bohr's correspondence principle where one recovers predictions
of classical theory from quantum theory in the limit of zero
Planck constant.  The contributions to an expression in QCD from
its nonzero $\beta$-function can be systematically
identified~\cite{Brodsky:2000cr,Rathsman:2001xe,Grunberg:2001bz}
order-by-order in perturbation theory using the Banks-Zaks
procedure~\cite{Banks:1981nn}.

There are other important consequences of near-conformal behavior:
the conformal approximation with zero $\beta$ function can be used
as template for QCD analyses~\cite{Brodsky:1985ve,Brodsky:1984xk}
such as the form of the expansion polynomials for distribution
amplitudes~\cite{Braun:2003rp,Braun:1999te}. The ``conformal
correspondence principle" also dictates the form of the expansion
basis for hadronic distribution amplitudes.

{\it Commensurate Scale Relations:} The near-conformal behavior of
QCD is the basis for commensurate scale
relations~\cite{Brodsky:1994eh} which relate observables to each
other without renormalization scale or scheme
ambiguities~\cite{Brodsky:2000cr}. An important example is the
generalized Crewther relation~\cite{Brodsky:1995tb}.  In this
method the effective charges of observables are related to each
other in conformal gauge theory; the effects of the nonzero QCD
$\beta-$ function are then taken into account using the BLM
method~\cite{Brodsky:1982gc} to set the scales of the respective
couplings. The magnitude of the  effective
charge~\cite{Brodsky:1997dh} defined from the ratio of elastic pion and
photon-to-pion transition form factors
$\alpha^{\rm exclusive}_s(Q^2) = {F_\pi(Q^2)/ 4\pi Q^2 F^2_{\gamma
\pi^0}(Q^2)}$  is connected to other effective
charges and observables by commensurate scale relations.  Its magnitude,
$\alpha^{\rm exclusive}_s(Q^2) \sim 0.8$ at small $Q^2,$  is sufficiently
large as to explain the observed magnitude of exclusive amplitudes such
as the pion form factor using the asymptotic distribution amplitude. An
analytic effective charge such as the pinch scheme~\cite{Cornwall:1981zr}
provides a method to unify the electroweak and strong couplings and
forces.

{\it Fixed Point Behavior:} Although the QCD coupling decreases
logarithmically at high virtuality  due to asymptotic freedom,
theoretical and phenomenological evidence is now accumulating that
QCD couplings based on a physical observable, such as hadronic
$\tau$ decay, becomes constant at small virtuality.  It thus
develops an infrared fixed
point~\cite{Brodsky:2002nb,Baldicchi:2002qm,Furui:2004bq,
Badalian:2004ig,Ackerstaff:1998yj}. This is in contradiction to
the usual assumption of singular growth in the infrared. The
near-constant behavior of effective couplings also suggests that
QCD can be approximated as a conformal theory even at relatively
small momentum transfer.

{\it The Abelian Correspondence Principle:} One can consider QCD
predictions as functions of analytic variables of the number of
colors $N_C$ and flavors $N_F$. At $N_C \to \infty$ at fixed $N_C
\alpha_s,$ calculations in QCD greatly simplify since only planar
diagrams enter. However, the $N_C \to 0$ limit is also very
interesting. Remarkably, one can show at all orders of
perturbation theory~\cite{Brodsky:1997jk} that PQCD predictions
reduce to those of an Abelian theory similar to QED at $N_C \to 0$
with $C_F \alpha_s$ and $N_F\over T_F C_F$ held fixed, where
$C_F={N^2_C-1\over 2 N_C}$ and $T_F=1/2.$ The resulting theory
corresponds to the group ${1/U(1)}$ which means that
light-by-light diagrams acquire a particular topological factor.
The $N_C \to 0$ limit provides an important check on QCD analyses;
QCD formulae and phenomena must match their Abelian analog. The
renormalization scale is effectively fixed by this requirement.
Commensurate scale relations obey the Abelian Correspondence
principle, giving the correct Abelian relations between
observables in the limit $N_C \to 0.$

\section{Twenty-One Key Antiproton Experiments}

Experiment is critical for testing QCD and unravelling its novel
features. The advent of the new antiproton storage ring HESR to be
constructed at GSI and the new {\it PANDA} detector open up a new
range of perturbative and nonperturbative tests of QCD in
exclusive and inclusive reactions. These include the formation of
exotic hadrons and novel tests involving timelike generalized
parton distributions, the effects of charm at threshold,
transversity, and single-spin asymmetries.  The interactions of
antiprotons in nuclear targets allows tests of exotic nuclear
phenomena such as color transparency, hidden color, reduced
nuclear amplitudes, and the non-universality of nuclear
antishadowing. I will also discuss the physics of the heavy-quark
sea and the role of conformal symmetry in hard exclusive
processes. Most of these key experiments can be performed with
stored antiprotons of moderate energy $E^{\bar p}_{\rm lab} < 15
~{\rm GeV}$ interacting in an internal target.

\begin{enumerate}

\item {\it Total Annihilation.} The antiproton and proton can
annihilate into a multi-hadron inclusive state, a system
potentially rich in gluonic matter. Specific predictions for the
inclusive distributions can be made in soliton-anti-soliton
models~\cite{Ellis:2001xc}. Skyrmion-anti-Skyrmion annihilation
provides a fairly accurate description of low-energy
baryon-antibaryon annihilation. A statistical approach may also be
useful~\cite{Bjorken:1969wi}. The production of charmonium states
is particularly interesting in view of the anomalously large
signal observed at Belle~\cite{Abe:2002rb} for $e^+ e^- \to J/\psi
\eta_c.$  The process $p \bar p \to {c \bar c} X$ where $X$ is a
glueball state could provide an important tagged source of gluonic
excitations~\cite{Brodsky:2003hv}.

\item {\it Exotic Resonance Formation in $\bar p p$ Reactions.} The strongest hadronic
interactions and thus the strongest opportunity to form resonances occurs when
constituents have the same 4-velocity or rapidity.  In the case of $\bar p p$
interactions, one can have total annihilation of the incident quarks and antiquarks into
open or hidden charmed hadrons; one can have ``diquark" anti-diquark $qq + \bar q \bar q
$ annihilation into charm quarks. These then can coalesce with the remaining valence
quarks to produce charmed  hadrons or a  $\ket{q \bar q c \bar c}$ ``quartoquark"
resonance.  If a $q$ and $\bar q$ annihilate to charm quarks, the remaining quarks can
produce``hectoquark" state $\ket{q q \bar q \bar q c \bar c}.$ One can have $c \bar c$
production where all of the incident quarks and antiquarks appear in the final state. In
this last case, one can produce ``octoquark" resonances $\ket{uud \bar u \bar u \bar d c
\bar c}.$  Thus there can be several kinematic regimes where novel charmed hadrons can
naturally appear.  The octoquark is the analog of $J=1, L=1, S=1$ $\ket{uud uud c \bar
c}$ state postulated to cause the large $A_{NN}$ at the charm threshold in transversely
polarized $p p$ collisions~\cite{Brodsky:1987xw}.

\item {\it Tests of Dimensional Counting Rules and Conformal
Scaling for Hard  Exclusive
Processes}~\cite{Brodsky:1973kr,Brodsky:1974vy,Matveev:1973ra,Lepage:1980fj}.
The counting rule for $\bar p p$ annihilation into two hadrons,
photons, or leptons is
\begin{equation}
{d\sigma\over dt}(\bar p p \to A  B) =
{|F_{AB}(t/s)|^2 \over s^{ 4 +n_A + n_B}} .
\end{equation}
where $n_I $ is the minimum number of
Fock-state particles in each final-state hadron; in the case of
leptons or photons, $n_I=1.$  For example, $n= 8$ for
${d\sigma\over dt}(\bar p p \to K^+ K^-) $ and $n= 7$ for timelike
photoproduction $ {d\sigma\over dt}(\bar p p \to \pi^0 \gamma) $
at fixed $t/s$ or $\theta_{\rm CM}.$ In the case of a
multiparticle final state where all particles are produced at
distinct $\theta^I_{\rm CM}$,
\begin{equation}
\Delta \sigma(\bar p p \to A, B, C, ...) \sim s^{-5 - \sum_{I=A,B,C,
...}{(n_I-1)}} .
\end{equation}
For example, $\Delta \sigma(\bar p p \to K^+ K^- \pi^0) \sim s^{-8}.$ These rules can be
derived from PQCD; they also follow from conformal symmetry and the LFWFs derived from
AdS/CFT. The power-law predictions are modified by logarithmic corrections from the
non-zero QCD $\beta$ function and the evolution of the hadronic distribution amplitudes
of each hadron. One also finds that Regge trajectories must become flat, approaching
negative integers at large negative $t$~\cite{Blankenbecler:1973kt}. This has recently
been demonstrated within the context of AdS/CFT by Andreev and
Siegel~\cite{Andreev:2004sy}.

\item {\it Hadron helicity Conservation in Exclusive
Processes}~\cite{Brodsky:1981kj}. The helicity and the angular
dependence of large-momentum-transfer exclusive processes such as
$\bar p p \to A B$ can be used to test gluon spin and other
basic elements of perturbative QCD. These processes isolate QCD
hard-scattering subprocesses in situations where the helicities of
all the interacting quarks are controlled. The predictions can be
summarized in terms of a general spin selection rule which states
that the total hadron helicity is conserved:
\begin{equation}
\sum_{initial} \lambda_H = \sum_{final} \lambda_H ,
\end{equation}
up to corrections falling as an inverse
power in the momentum transfer.

\item {\it Tests of Quark Interchange Dominance in Exclusive
Processes.} The angular distributions $F_{AB}(t/s)$ appearing in
the fixed-angle scaling laws are sensitive to the scattering
mechanism as well as the shapes of the hadron distribution
amplitudes. In the limit of large $N_C$, the dominant scattering
amplitude derives from quark
interchange~\cite{Gunion:1973ex,Blankenbecler:1973kt,Brodsky:2003px}.
For example, the dominant scattering mechanism for $K^+ p \to K^+
p$ derives from the exchange of the common $u$ quark.

The  quark interchange amplitude~\cite{Gunion:1973ex} for $ A B
\to C D$ can be written as a convolution of the four light-front
wavefunctions appearing in the process
\begin{equation}
\int d^2k_\perp dx \psi_{A} \times \psi_B \times \psi_{C} \times \psi_{D}~
[M^2_A+M^2_B - {\cal M}^2]
\end{equation}
 where {\cal M} is the invariant mass of the
constituents. The complete expression is given in
Ref.~\cite{Gunion:1973ex}. In the case of $K^+ p \to K^+ p$, the
interchange amplitude scales as $1/ u t^2$ and thus
$d\sigma/dt(K^+ p \to K^+ p) \simeq { 1/ s^2 u^2 t^4}.$ This
agrees with the observed scaling and angular dependence of the
fixed- CM angle data. If the $u$-quark exchange mechanism is
dominant, then one can predict the amplitude for $\bar p p \to K^+
K^-$  via $s \leftrightarrow t$ crossing. Thus the crossed
amplitude $\bar p p \to K^+ K^-$ must scale as $1/u t^2$ and the
cross section is $d\sigma/dt(\bar p p \to K^+ K^-) \simeq { 1/ u^2
s^6}.$

\item {\it Anomalous Regge Behavior.} At fixed $t$ and large $s
>>> -t$, one can use the Regge expansion
\begin{equation}
M(\bar p p \to A B) = \sum_R \beta_R(t) s^\alpha_R(t) \zeta_R
\end{equation}
 where
$\alpha_R(t)$ parameterizes the Regge trajectory for spacelike $t$
and $\zeta_R(t)$ is the signature factor which determines the
phase of the amplitude. Remarkably, perturbative QCD and conformal
scaling require that these trajectories approach negative integers
at large $-t$ ($-1$ for meson exchange, $-2$ for baryon exchange
in the $t$ channel), rather than the conventional linear
trajectories normally used in Regge
theory~\cite{Blankenbecler:1973kt}. The power behavior of
$\beta_R(t)$ is also determined.

\item {\it  Timelike Compton
Scattering}~\cite{Brodsky:1981rp,Brodsky:2001hv}: $\bar p p \to
\gamma \gamma$. The scaling, normalization, and angular
distribution of this fundamental process has been computed at
lowest order in the PQCD factorization
framework~\cite{Brooks:2000nb}. Conformal symmetry predicts the
scaling $s^6 {d\sigma\over dt}(\bar p p \to \gamma \gamma) =
F(\theta_{cm})$ at large $s$ and $t.$ The ratio of the timelike
Compton amplitude to the timelike proton form factor is important
since uncertainties from the baryon coupling $F_p$ to three quarks
(the same coupling which controls proton decay!) and the QCD
coupling at timelike virtuality cancel out. This normalization
thus can expose the importance of higher order corrections in
$\alpha_s.$ The angular distributions of timelike Compton
scattering including spin correlations are highly sensitive to the
shape of the proton distribution amplitude $\phi_p(x_i,Q),$ the
basic three quark wavefunction.

\item {\it Timelike Deeply Virtual Compton Scattering $\bar p p
\to \gamma^* \gamma $ (DVCS). } At large photon virtuality, this
amplitude is computable from the convolution of the $\bar q q \to
\gamma^* \gamma$ amplitude with the timelike generalized parton
distributions. Phase information can be obtained from the
interference of the DVCS amplitude $\bar p p \to \ell^+ \ell^-
\gamma$ with the bremsstrahlung amplitude $\bar p \bar p \to
\gamma^* \to \ell^+ \ell^- \gamma$ derived from the timelike
proton form factor which causes an $\ell^+ \ell^-$ asymmetry.  The
handbag contribution to the DVCS amplitude can be computed from
the overlap of proton light-front
wavefunctions~\cite{Brodsky:2000xy,Diehl:2000xz} including
contributions from Fock states with the same parton number $n =
n^\prime$, and states differing by the presence of an extra $q
\bar q$: $n= n^\prime +2.$

\item {\it The $J=0$ Fixed pole}: One of the most distinctive
features of QCD is the presence of a $J=0$ fixed Regge pole
contribution to the Compton amplitude reflecting the fact that the
two photons can act quasi-locally on the same
quark~\cite{Brodsky:1972vv}. This contribution can be observed in
timelike DVCS: $\bar p p \to \gamma \gamma^*$ from its distinctive
kinematic properties: the amplitude from the $J=0$ term is
independent of $t$ at fixed $s$, independent of photon virtuality
at fixed $s.$

\item {\it Time-like Proton Form Factors.}  Leading-twist PQCD
predictions for hard exclusive amplitudes~\cite{Lepage:1980fj}
are written  in a factorized form as the product of hadron
distribution amplitudes $\phi_I(x_i,Q)$ for each hadron $I$
convoluted with  the hard scattering amplitude $T_H$ obtained by
replacing each hadron with collinear on-shell quarks with
light-front momentum fractions $x_i = k^+_i/P^+.$ The hadron
distribution amplitudes are obtained by integrating the $n-$parton
valence light-front wavefunctions:
\begin{equation}
\phi(x_i,Q) = \int^Q
\Pi^{n-1}_{i=1} d^2 k_{\perp i} ~ \psi_{\rm val}(x_i,k_\perp).
\end{equation}
Thus the distribution amplitudes are $L_z=0$ projections of the LF
wavefunction, and the sum of the spin projections of the valence
quarks must equal the $J_z$ of the parent hadron. Higher orbital
angular momentum components lead to power-law suppressed exclusive
amplitudes~\cite{Lepage:1980fj,Ji:2003fw}. Since quark masses can
be neglected at leading twist in $T_H$, one has quark helicity
conservation, and thus, finally, hadron-helicity conservation: the
sum of initial hadron helicities equals the sum of final
helicities. In particular, since the hadron-helicity violating
Pauli form factor is computed from states with $\Delta L_z = \pm
1,$  PQCD predicts $F_2(Q^2)/F_1(Q^2) \sim 1/Q^2 $ [modulo
logarithms].  A detailed analysis shows that the asymptotic
fall-off takes the form $F_2(Q^2)/F_1(Q^2) \sim \log^2
Q^2/Q^2$~\cite{Belitsky:2002kj}. One can also construct other
models~\cite{Brodsky:2003pw} incorporating the leading-twist
perturbative QCD prediction which are consistent with the JLab
polarization transfer data~\cite{Jones:1999rz} for the ratio of
proton Pauli and Dirac form factors.  This analysis can also be
extended to study the spin structure of scattering amplitudes at
large transverse momentum and other processes which are dependent
on the scaling and orbital angular momentum structure of
light-front wavefunctions. Recently, Afanasev, Carlson, Chen,
Vanderhaeghen, and I~\cite{Chen:2004tw} have shown that the
interfering two-photon exchange contribution to elastic
electron-proton scattering, including inelastic intermediate
states, can account for the discrepancy between Rosenbluth and
Jefferson Lab spin transfer polarization data~\cite{Jones:1999rz}.

A crucial prediction of models for proton form factors is the
relative phase of the timelike form factors, since this can be
measured from the proton single spin symmetries in $e^+ e^- \to p
\bar p$ or $p \bar p \to \ell \bar \ell$~\cite{Brodsky:2003gs}.
The Zemach radius of the proton is known to a precision of better
than 2\% from the comparison of hydrogen and muonium hyperfine
splittings; this constraint needs to be incorporated into any
analysis~\cite{Brodsky:2004ck}.

The annihilation process $\bar p p \to \ell^+ \ell^-$ thus
provides a primary test of proton structure. Its angular
distribution allows a direct separation of the $G_E(s)$ and
$G_M(s)$ timelike form factors. Carl Carlson, John Hiller, Dae
Sung Hwang and I~\cite{Brodsky:2003gs} have shown that
measurements of the proton's polarization strongly discriminate
between the analytic forms of models which fit the proton form
factors in the spacelike region. In particular, the single-spin
asymmetry normal to the scattering plane measures the relative
phase difference between the timelike $G_E$ and $G_M$ form
factors. The dependence on proton polarization in the timelike
region is expected to be large in most models, of the order of
several tens of percent.  The continuation of the spacelike form
factors to the timelike don=main $t = s > 4 M^2_p$ is very
sensitive to the analytic form of the form factors; in particular
it is very sensitive to the form of the PQCD predictions including
the corrections to conformal scaling. The forward-backward $\ell^+
\ell^-$ asymmetry measures the interference of one-photon and
two-photon contributions to $\bar p p \to \ell^+ \ell^-.$

\item {\it Tests of Color Transparency. } The small transverse
size fluctuations of a hadron wavefunction with a small color
dipole moment will have minimal interactions in a
nucleus~\cite{Bertsch:1981py,Brodsky:1988xz}. Color transparency
can be tested in quasi-elastic antiproton-nucleus reactions such
as $\bar p A \to  \pi^+ \pi^- (A-1)$ where the proton annihilates
in the nucleus leaving a recoiling nucleus with one less proton.
According to color transparency,  at large ${\cal M}_{\pi^+ \pi^-}$ the
small size wavefunction fluctuations of the incident $\bar p$ and
outgoing pions which enter the hard scattering exclusive $\bar p p
\to \pi^+ \pi^-$ amplitude will not be absorbed in the nucleus so
that the ideal rate is proportional to the number $Z$ of protons
in the nucleus. In contrast, the standard nuclear physics
prediction is the number $Z^{1/3}$ of protons on the periphery of
the nucleus.

\item {\it Intrinsic Charm}~\cite{Brodsky:1980pb}.  The
probability for Fock states of a light hadron such as the proton
to have an extra heavy quark pair decreases as $1/m^2_Q$ in
non-Abelian gauge theory~\cite{Franz:2000ee,Brodsky:1984nx}.  The
relevant matrix element is the cube of the QCD field strength
$G^3_{\mu \nu}.$ This is in contrast to abelian gauge theory where
the relevant operator is $F^4_{\mu \nu}$ and the probability of
intrinsic heavy leptons in QED bound state is suppressed as
$1/m^4_\ell.$  The intrinsic Fock state probability is maximized
at minimal off shellness. The maximum probability occurs at $x_i =
{ m^i_\perp /\sum^n_{j = 1} m^j_\perp}$; {\em i.e.}, when the
constituents have equal rapidity.   Thus the heaviest constituents
have the highest momentum fractions and highest $x$. Intrinsic
charm thus predicts that the charm structure function has support
at large $x_{bj}$  in excess of DGLAP
extrapolations~\cite{Brodsky:1980pb}; this is in agreement with
the EMC measurements~\cite{Harris:1995jx}.

Intrinsic charm allows charm production to occur close to it's
kinematic threshold~\cite{Brodsky:2000zc}. Charm and bottom
production near threshold is sensitive to the multi-quark,
gluonic, and hidden-color correlations of hadronic and nuclear
wavefunctions in QCD since all of the target's constituents must
act coherently within the small interaction volume of the heavy
quark production subprocess. Although such multi-parton subprocess
cross sections are suppressed by powers of $1/m^2_Q$, they have
less phase-space suppression and can dominate the contributions of
the leading-twist single-gluon subprocesses in the threshold
regime. In fact, an anomalous signal was observed at CESR in
$J/\psi$ photoproduction near threshold~\cite{Gittelman:1975ix}.
Similarly, intrinsic charm predicts anomalously large rates for
open and hidden charm in $\bar p p $ collisions such as $\bar p p
\to \Lambda_C X$ and $\bar p p \to J/\psi X$ even at relatively
small antiproton energies.  The rate for threshold channels will
be significantly enhanced in nuclear targets.

\item {\it Anomalous Deuteron Reactions and Hidden Color.} In
general, the six-quark wavefunction of a deuteron is a mixture of
five different color-singlet states~\cite{Brodsky:1983vf}. The
dominant color configuration at large distances corresponds to the
usual proton-neutron bound state where transverse momenta are  of
order ${\vec k}^2 \sim 2 M_d \epsilon_{BE}.$ However, at small
impact space separation, all five Fock color-singlet components
eventually acquire equal weight, {\em i.e.}, the deuteron
wavefunction evolves to 80\% hidden color.  At high $Q^2$ the
deuteron form factor is sensitive to wavefunction configurations
where all six quarks overlap within an impact separation $b_{\perp
i} < {\cal O} (1/Q).$ The normalization of the deuteron form
factor observed at large $Q^2$~\cite{Arnold:1975dd}, as well as
the presence of two mass scales in the scaling behavior of the
reduced deuteron form factor~\cite{Brodsky:1976rz} $f_d(Q^2)=
F_d(Q^2)/F^2(Q^2/4)$, suggests sizable hidden-color contributions
such as $\ket{(uud)_{8_C} (ddu)_{8_C}}$ with probability  of order
$15\%$ in the deuteron wavefunction~\cite{Farrar:1991qi}. See
Fig.~\ref{reduced}. Perturbative QCD and conformal symmetry can
also be directly applied to exclusive antiproton-deuteron
reactions, such as the fixed angle scaling $s^{12} d\sigma
/dt(\bar p d \to \pi^- p) $ corresponding to 14 participating
elementary fields.   Such hard-scattering nuclear reactions are
sensitive to the minimal six-quark hidden-color Fock states of the
deuteron.

\begin{figure}[htb]
\centering
\includegraphics[width=4.3in]   
{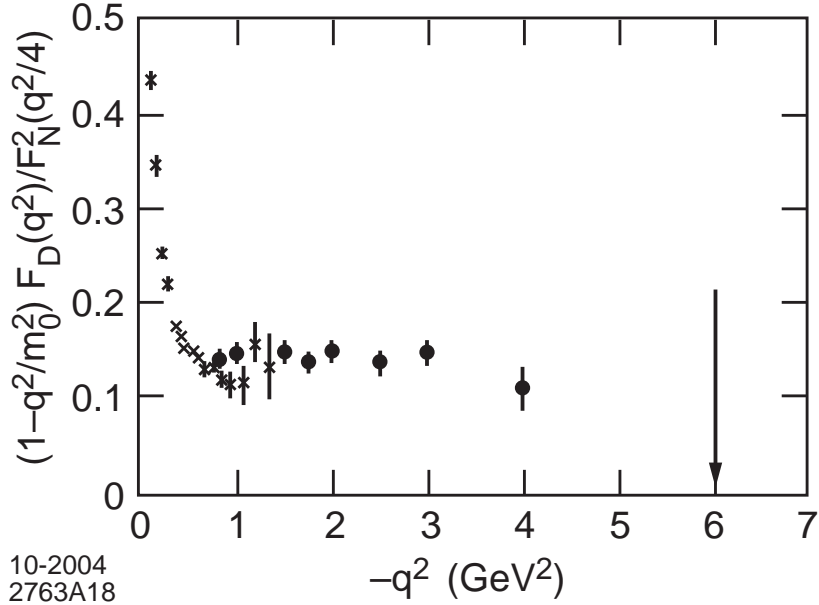} \caption{Reduced deuteron form factor testing the
scaling predicted by perturbative QCD and conformal scaling.  The
data show two regimes: a fast-falling behavior at small $Q^2$
characteristic of normal nuclear binding, and a hard scattering
regime with monopole fall-off controlled by the scale $m^2_0 =
0.28~{\rm GeV^2}.$ The latter contribution is attributable to
non-nucleonic hidden-color components of the deuteron's six-quark
Fock state. From Ref.~\cite{Brodsky:1976rz}.} \label{reduced}
\end{figure}

\item {\it  Transversity and the Drell-Yan
reaction}~\cite{Miyama:1999yp}.  The production of massive pairs
in anti-proton--proton reactions $p \bar p \to \ell^+ \ell^- X$ is
the ideal prototype of a hard inclusive reaction.  In particular,
it provides the most direct test of the correlation between
transversely-polarized quarks in a transversely polarized
proton---via the $A_{TT}$ correlation.

\item {\it  Single-Spin asymmetries in Drell-Yan
Processes}~\cite{Collins:2002kn,Brodsky:2002rv}. Initial-state
interactions between the annihilating antiquark of the $\bar p$
and the spectators of the target proton will produce a $T-$odd
``Sivers effect" single-spin asymmetry proportional to the
correlation $\vec S_p \cdot \vec q \times  \vec{ \bar p}.$ Here
$S_p$ is the spin-vector of the target proton. The asymmetry is
predicted to be equal but opposite in sign to the corresponding
single-spin asymmetry $\vec S_p \cdot \vec q \times \vec p^\prime
$ in semi-inclusive $\gamma^* p \to  p^\prime X$ deep inelastic
scattering. The same initial-state interactions which produce
single-spin asymmetries also produce a dramatic $\cos  2 \phi$
correlation of the lepton-pair plane and the $\vec q \to \vec{
\bar p}$ plane~\cite{Boer:2002ju}.

\item {\it  Diffractive Drell-Yan reactions.}  The reaction $\bar
p p \to \gamma^* p^\prime X,$ where the target nucleon remains
intact and has a rapidity gap with the other final state hadrons,
provides a novel look at the hard pomeron and its origin. The
behavior of the Drell-Yan cross section at large $x_F$ probes the
$x \to 1$ behavior of the antiproton structure function. The power
behavior is predicted by perturbative QCD counting
rules~\cite{Lepage:1980fj,Brodsky:1994kg} and conformal arguments.
In this regime DGLAP evolution is quenched~\cite{Lepage:1980fj}
since the annihilating $\bar q$ is far off shell: $k^2 \sim
-{k^2_\perp}/(1-x).$

\item {\it Higher-Twist Processes.}  Leading-twist perturbative
QCD predicts a classic $1+ \cos^\theta$ distribution for the
lepton angular distribution in the lepton pair rest frame.  The
data for pion-induced reactions shows significant deviations
especially at large $x_F$ for the lepton pair where higher-twist
subprocesses such as $\pi q \to \gamma^* q$ are
favored~\cite{Berger:1979du} and the pion itself enters the hard
subprocess. The net result is a $\sin^2 \theta\over Q^2$
contribution which dominates the cross section at $x_F \to 1$. In
the case of antiproton beams, higher-twist processes such as
$(\bar q \bar q) q \to \gamma^* \bar q$ arise from diquark
correlations can be studied in detail in $\bar p p \to \gamma^*
X.$

\item {\it Nuclear Antishadowing.} The Drell-Yan reaction $p \bar
A \to \ell^+ \ell^- X$ in nuclear targets provides an important
measure of nuclear antishadowing. Recent work has shown that
antishadowing is non-universal; it depends on the flavor of the
individual quark components of the nuclear wavefunction. The
shadowing and antishadowing of nuclear structure functions in the
Gribov-Glauber picture is due respectively to the destructive and
constructive interference of amplitudes arising from the
multiple-scattering of quarks in the nucleus. The effective
quark-nucleon scattering amplitude includes Pomeron and Odderon
contributions from multi-gluon exchange as well as Reggeon
quark-exchange contributions~\cite{Brodsky:1989qz}.  The coherence
of these multiscattering nuclear processes leads to shadowing and
antishadowing of the electromagnetic nuclear structure functions
in agreement with measurements. Recently, Ivan Schmidt, Jian-Jun
Yang, and I~\cite{Brodsky:2004qa} have shown that this picture
leads to substantially different antishadowing for charged and
neutral current reactions, thus affecting the extraction of the
weak-mixing angle $\sin^2\theta_W$. We find that part of the
anomalous NuTeV result for $\sin^2\theta_W$ could be due to the
non-universality  of nuclear antishadowing for charged and neutral
currents.  Detailed measurements of the nuclear dependence of
individual quark structure functions, including measurements of
the Drell-Yan process in $\bar p A$ reactions, are thus needed to
establish the distinctive phenomenology of shadowing and
antishadowing and to make the NuTeV results definitive.

\item {\it Heavy Quark Asymmetries.} The binding of the strange
quarks in the nucleon produces an asymmetry between the strange
and antistrange distributions. This can be tested in the
asymmetric production of charmed-strange hadrons. $\bar p p \to
D_s X$ versus $\bar p p \to \bar D_s X$ or exclusive $\bar p \to
D_s \bar D_s$ because of the coalescence of the $s$ or $\bar s$
arising from the $\ket{\bar u \bar d \bar d \bar s s}$ Fock state
of the projectile antiproton.

\item {Search for the Odderon.} The interference of odderon and
pomeron contributions leads to baryon-antibaryon
asymmetries~\cite{Brodsky:1999mz}.  For example, the asymmetry in
the fractional energy of charm versus anticharm or strange versus
antistrange jets produced in high energy diffractive
photoproduction is sensitive to the interference of the Odderon
$(C = -)$ and Pomeron $(C = +)$ exchange amplitudes in QCD.

\item { The Exclusive-Inclusive Connection.}  The Drell-Yan
process provides and interesting arena for testing duality in QCD.
In the exclusive limit of small $M_X,$ $\bar p p \to \gamma^* X$
will approach the double-resonant regime: $\bar p p \to \gamma^*
M^*$ where the massive system $M^*$ could be a baryon-antibaryon
system, a meson pair or even a single meson.

\end{enumerate}

The following sections in these lectures expand on the physics
issues underlying the above tests.

\section{QCD on the Light Front }

One of the central problems in particle physics is determining the
structure of hadrons such as the proton and neutron in terms of
their fundamental QCD quark and gluon degrees of freedom.  The
bound-state structure of hadrons plays a critical role in
virtually every area of particle physics phenomenology.  For
example, in the case of the spacelike and timelike nucleon form
factors, pion electroproduction $ep \rightarrow e^\prime \pi^+n$,
and timelike Compton scattering $\bar p p \to \gamma \gamma,$ the
cross sections depend not only on the nature of the quark
currents, but also on the coupling of the quarks to the initial
and final hadronic states.  Exclusive decay amplitudes such as $B
\rightarrow K^*\gamma$, processes which are studied at $B$
factories, depend not only on the underlying weak transitions
between the quark flavors, but also the wavefunctions which
describe how the $B$ and $K^*$ mesons are assembled in terms of
their fundamental quark and gluon constituents.  Unlike the
leading-twist structure functions measured in deep inelastic
scattering, such exclusive channels are sensitive to the structure
of the hadrons at the amplitude level and to the coherence between
the contributions of the various quark currents and multi-parton
amplitudes.

\begin{figure}[htb]
\centering
\includegraphics[width=4.3in]
{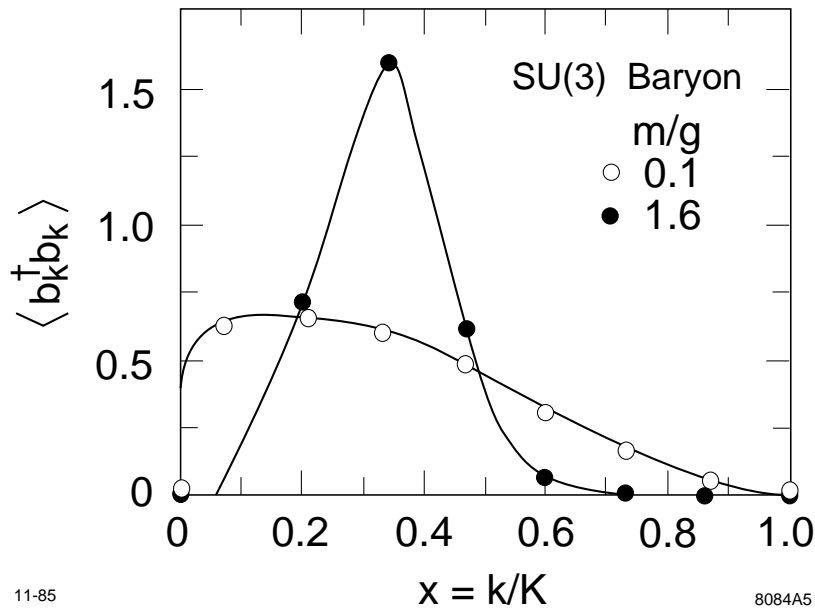} \caption{Valence contribution to the baryon
structure function in QCD$_{1+1}$, as a function of the light-cone
longitudinal momentum fraction.  The gauge group is SU(3), $m$ is
the quark mass, and $g$ is the gauge coupling. (From Ref.
~\cite{Hornbostel:1988fb}.)} \label{fig1}
\end{figure}

Light-front Fock state wavefunctions $\psi_{n/H}(x_i,\vec k_{\perp
i},\lambda_i)$ encode the bound-state quark and gluon properties
of hadrons, including their spin and flavor correlations, in the
form of universal process- and frame- independent amplitudes.
Because the generators of certain Lorentz boosts are kinematical,
knowing the LFWFs in one frame allows one to obtain it in any
other frame.  LFWFs underlie virtually all areas of QCD
phenomenology. The hadronic distribution amplitudes which control
hard exclusive processes are computed from the valence Fock state
LFWFs. Matrix elements of space-like local operators for the
coupling of photons, gravitons, and the moments of deep inelastic
structure functions all can be expressed as overlaps of
light-front wavefunctions with the same number of Fock
constituents.   Similarly, the exclusive decays of heavy hadrons
such as the $B$ meson are computed from overlaps of LFWFs. The
unintegrated parton distributions and generalized parton
distributions measured in deeply virtual Compton scattering can be
constructed from LFWFs.  Hadronization phenomena such as the
coalescence mechanism for leading heavy hadron production are
computed from LFWF overlaps.  Diffractive jet production provides
another phenomenological window into the structure of LFWFs.
However, some leading-twist phenomena such as the diffractive
component of deep inelastic scattering, single spin asymmetries,
nuclear shadowing and antishadowing cannot be computed from the
LFWFs of hadrons in isolation.

Formally, the light-front expansion is constructed by quantizing
QCD at fixed light-cone time~\cite{Dirac:1949cp} $\tau = t + z/c$
and forming the invariant light-front Hamiltonian: $ H^{QCD}_{LF}
= P^+ P^- - {\vec P_\perp}^2$ where $P^\pm = P^0 \pm
P^z$~\cite{Brodsky:1997de}.  The momentum generators $P^+$ and
$\vec P_\perp$ are kinematical; {\em i.e.}, they are independent
of the interactions.  The generator $P^- = i {d\over d\tau}$
generates light-cone time translations, and the eigen-spectrum of
the Lorentz scalar $ H^{QCD}_{LF}$ gives the mass spectrum of the
color-singlet hadron states in QCD together with their respective
light-front wavefunctions.  For example, the proton state
satisfies $ H^{QCD}_{LF} \ket{\psi_p} = M^2_p \ket{\psi_p}$.  The
expansion of the proton eigensolution $\ket{\psi_p}$ on the
color-singlet $B = 1$, $Q = 1$ eigenstates $\{\ket{n} \}$ of the
free Hamiltonian $ H^{QCD}_{LF}(g = 0)$ gives the light-front Fock
expansion:
\begin{eqnarray}
&&\ket{\psi_p(P^+, {\vec P_\perp} )} = \sum_{n}\ \prod_{i=1}^{n}
{{\rm d}x_i\, {\rm d}^2 {\vec k_{\perp i}} \over \sqrt{x_i}\,
16\pi^3} \, 16\pi^3 \nonumber
\\&&\times \ \delta\left(1-\sum_{i=1}^{n} x_i\right)\,
\delta^{(2)}\left(\sum_{i=1}^{n} {\vec k_{\perp i}}\right)
\label{a318}\nonumber
\\
&& \rule{0pt}{4.5ex} \times \psi_{n/H}(x_i,{\vec k_{\perp i}},
\lambda_i) \ket{ n;\, x_i P^+, x_i {\vec P_\perp} + {\vec k_{\perp
i}}, \lambda_i}. \nonumber
\end{eqnarray}
The light-cone momentum fractions $x_i = k^+_i/P^+$ and ${\vec
k_{\perp i}}$ represent the relative momentum coordinates of the
QCD constituents.  The physical transverse momenta are ${\vec
p_{\perp i}} = x_i {\vec P_\perp} + {\vec k_{\perp i}}.$ The
$\lambda_i$ label the light-cone spin projections $S^z$ of the
quarks and gluons along the quantization direction $z$.  The
physical gluon polarization vectors $\epsilon^\mu(k,\ \lambda =
\pm 1)$ are specified in light-cone gauge by the conditions $k
\cdot \epsilon = 0,\ \eta \cdot \epsilon = \epsilon^+ = 0.$
Light-front quantization in the doubly-transverse light-cone
gauge~\cite{Tomboulis:jn,Srivastava:2000cf}  has a number of
advantages, including explicit unitarity, a physical Fock
expansion, exact representations of current matrix elements, and
the decoupling properties needed to prove factorization theorems
in high momentum transfer inclusive and exclusive reactions.

\begin{figure}[htb]
\centering
\includegraphics[width=4.3in]   
{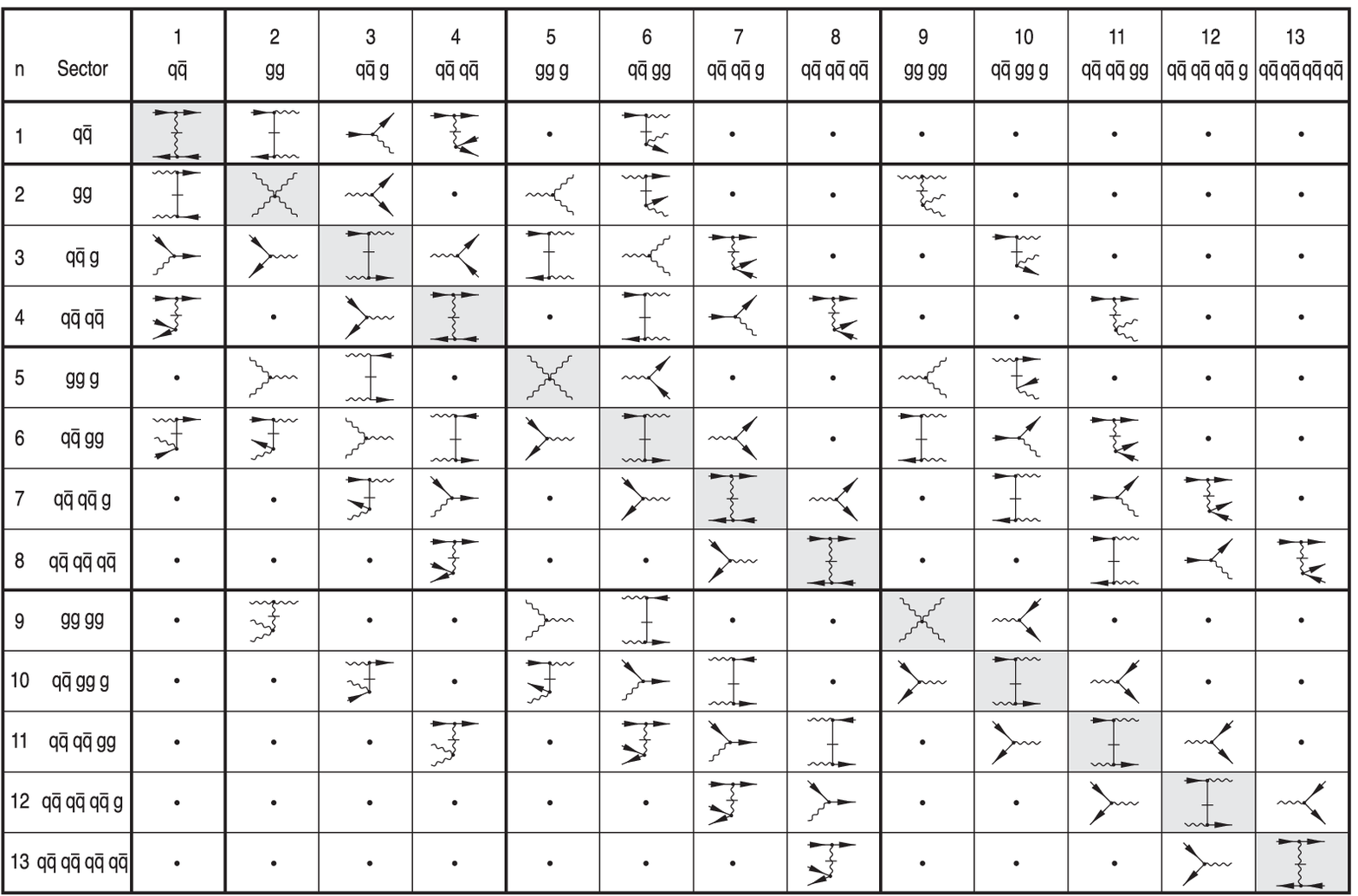} \caption{Graphical Representation of QCD light-front
Hamiltonian interactions in the Light-Front Fock Space. The figure
illustrates the matrix elements between Fock states from the
three-point and four-point interactions of the theory plus the
contributions from instantaneous gluon and quark exchange.  From
Ref.~\cite{Brodsky:1997de}.} \label{LFHam}
\end{figure}

The matrix elements of the light-front Hamiltonian are illustrated
in Fig.~\ref{LFHam}. Light-front four-point instantaneous gluon
and quark interactions appear when one eliminates the dependent
quark and gluon fields using the QCD equation of motion in
light-cone gauge $A^+=0.$  This is the analog of the Coulomb
interactions which appear in a gauge theory Hamiltonian when
quantized in radiation gauge.

The solutions of $ H^{QCD}_{LF} \ket{\psi_p} = M^2_p \ket{\psi_p}$
are independent of $P^+$ and ${\vec P_\perp}$;  thus given the
eigensolution Fock projections $ \langle n; x_i, {\vec k_{\perp
i}}, \lambda_i |\psi_p \rangle  = \psi_n(x_i, {\vec k_{\perp i}},
\lambda_i) ,$ the wavefunction of the proton is determined in any
frame~\cite{Lepage:1980fj}.  In contrast, in equal-time
quantization,  a Lorentz boost always mixes dynamically with the
interactions, so that computing a wavefunction in a new frame
requires solving a nonperturbative problem as complicated as the
Hamiltonian eigenvalue problem itself. The LFWFs
$\psi_{n/H}(x_i,\vec k_{\perp i},\lambda_i)$ are properties of the
hadron itself; they are thus universal and process independent.

One can also define the light-front Fock expansion using a
covariant generalization of light-front time: $\tau=x \cd \omega$.
The four-vector $\omega$, with $\omega^2 = 0$, determines the
orientation of the light-front plane.  The freedom to choose
$\omega$ provides an explicitly covariant formulation of
light-front quantization~\cite{cdkm}: all observables such as
matrix elements of local current operators, form factors, and
cross sections are light-front invariants and thus must be
independent of $\omega_\mu.$ In recent work, Dae Sung Hwang, John
Hiller, Volodya Karmonov,  and I~\cite{Brodsky:2003pw} have
studied the analytic structure of LFWFs using the explicitly
Lorentz-invariant formulation of the front form.  Eigensolutions
of the Bethe-Salpeter equation have specific angular momentum as
specified by the Pauli-Lubanski vector.  The corresponding LFWF
for an $n$-particle Fock state evaluated at equal light-front time
$\tau = \omega\cdot x$ can be obtained by integrating the
Bethe-Salpeter solutions over the corresponding relative
light-front energies.  The resulting LFWFs $\psi^I_n(x_i, k_{\perp
i})$ are functions of the light-cone momentum fractions $x_i=
{k_i\cdot \omega / p \cdot \omega}$ and the invariant mass squared
of the constituents $M_0^2= (\sum^n_{i=1} k_i^\mu)^2 =\sum_{i
=1}^n \big [\frac{k^2_\perp + m^2}{x}\big]_i,$ each multiplying
spin-vector and polarization tensor invariants which can involve
$\omega^\mu.$  They are eigenstates of the Karmanov--Smirnov
kinematic angular momentum operator~\cite{Karmanov:1991fv}. Thus
LFWFs satisfy all Lorentz symmetries of the front form, including
boost invariance, and they are proper eigenstates of angular
momentum.

In principle, one can solve for the LFWFs directly from the
fundamental theory using methods such as discretized light-front
quantization (DLCQ), the transverse lattice, lattice gauge theory
moments, or Bethe--Salpeter techniques.  DLCQ has been remarkably
successful in determining the entire spectrum and corresponding
LFWFs in 1+1 field theories, including supersymmetric examples.
Reviews of nonperturbative light-front methods may be found in
references~\cite{Brodsky:1997de,cdkm,Dalley:ug,Brodsky:2003gk}.
One can also project the known solutions of the Bethe--Salpeter
equation to equal light-front time, thus producing hadronic
light-front Fock wave functions.  A potentially important method
is to construct the $q\bar q$ Green's function using light-front
Hamiltonian theory, with DLCQ boundary conditions and
Lippmann-Schwinger resummation.  The zeros of the resulting
resolvent projected on states of specific angular momentum $J_z$
can then generate the meson spectrum and their light-front Fock
wavefunctions.  The DLCQ properties and boundary conditions allow
a truncation of the Fock space while retaining the kinematic boost
and Lorentz invariance of light-front quantization.  For a recent
review of light-front methods and references, see
Ref.~\cite{Brodsky:2003gk}.

Even without explicit solutions, much is known about the explicit
form and structure of LFWFs.  They can  be matched to
nonrelativistic Schrodinger wavefunctions at soft scales.  LFWFs
at large $k_\perp$ and $x_i \to 1$ are constrained at high momenta
by arguments based on conformal symmetry, the operator product
expansion, or perturbative QCD.  The pattern of higher Fock states
with extra gluons is given by ladder relations.  The structure of
Fock states with nonzero orbital angular momentum is also
constrained.

\section{Light-Front Wavefunctions and QCD Phenomenology}

Given the light-front wavefunctions, one can compute the
unintegrated parton distributions in $x$ and $k_\perp$ which
underlie  generalized parton distributions for nonzero skewness.
As shown by Diehl, Hwang, and myself~\cite{Brodsky:2000xy},  one
can give a complete representation of virtual Compton scattering
$\gamma^* p \to \gamma p$ at large initial photon virtuality $Q^2$
and small momentum transfer squared $t$ in terms of the light-cone
wavefunctions of the target proton.  One can then verify the
identities between the skewed parton distributions $H(x,\zeta,t)$
and $E(x,\zeta,t)$ which appear in deeply virtual Compton
scattering and the corresponding integrands of the Dirac and Pauli
form factors $F_1(t)$ and $F_2(t)$ and the gravitational form
factors $A_{q}(t)$ and $B_{q}(t)$ for each quark and anti-quark
constituent.  We have illustrated the general formalism for the
case of deeply virtual Compton scattering on the quantum
fluctuations of a fermion in quantum electrodynamics at one loop.

The integrals of the unintegrated parton distributions over
transverse momentum at  zero skewness provide the helicity and
transversity distributions measurable in polarized deep inelastic
experiments~\cite{Lepage:1980fj}.  For example, the polarized
quark distributions at resolution $\Lambda$ correspond to
\begin{eqnarray}
&& q_{\lambda_q/\Lambda_p}(x, \Lambda) =\nonumber \\
&&\times\  \sum_{n,q_a} \int\prod^n_{j=1} dx_j d^2 k_{\perp
j}\sum_{\lambda_i} \vert \psi^{(\Lambda)}_{n/H}(x_i,\vec k_{\perp
i},\lambda_i)\vert^2
\nonumber \\
&& \times\ \delta\left(1- \sum^n_i x_i\right) \delta^{(2)}
\left(\sum^n_i \vec k_{\perp i}\right) \delta(x - x_q)\nonumber \\
&& \times\  \delta_{\lambda_a, \lambda_q} \Theta(\Lambda^2 - {\cal
M}^2_n)\ , \nonumber
\end{eqnarray}
where the sum is over all quarks $q_a$ which match the quantum
numbers, light-cone momentum fraction $x,$ and helicity of the struck
quark.

As shown by Raufeisen and myself~\cite{Raufeisen:2004dg}, one can
construct a ``light-front density matrix" from the complete set of
light-front wavefunctions which is a Lorentz scalar. One can also
define a light-front partition function $Z_{LF}$ as an outer
product of light-front wavefunctions. The deeply virtual Compton
amplitude and generalized parton distributions can then be
computed as the trace $Tr[Z_{LF} {\cal O}],$ where $\cal O$ is the
appropriate local operator~\cite{Raufeisen:2004dg}.  This
partition function formalism can be extended to multi-hadronic
systems and systems in statistical equilibrium to provide a
Lorentz-invariant description of relativistic
thermodynamics~\cite{Raufeisen:2004dg}. This form can be used at
finite temperature to give a boost invariant formulation of
thermodynamics.  At zero temperature the light-front density
matrix is directly connected to the Green's function for quark
propagation in the hadron as well as deeply virtual Compton
scattering.  In addition, moments of transversity distributions
and off-diagonal helicity convolutions are defined from the
density matrix of the light-cone wavefunctions. The light-front
wavefunctions also specify the multi-quark and gluon correlations
of the hadron.  For example,  the distribution of spectator
particles in the final state which could be measured in the proton
fragmentation region in deep inelastic scattering at an
electron-proton collider or in the Drell-Yan process $\bar p p \to
\ell^+ \ell^- X$ which can be studied in antiproton collisions at
GSI are in principle encoded in the light-front wavefunctions.

Matrix elements of local operators such as spacelike proton form
factors can be computed simply from the overlap integrals of light
front wave functions in analogy to nonrelativistic Schr\"odinger
theory. Thus given the $\psi^{(\Lambda)}_{n/H},$ one can construct
any spacelike electromagnetic, electroweak, or gravitational form
factor or local operator product matrix element of a composite or
elementary system from the diagonal overlap of the
LFWFs~\cite{Brodsky:1980zm}. Exclusive semi-leptonic $B$-decay
amplitudes involving timelike currents such as $B\rightarrow A
\ell \bar{\nu}$ can also be evaluated exactly in the light-front
formalism~\cite{Brodsky:1998hn}.  In this case, the timelike decay
matrix elements require the computation of both the diagonal
matrix element $n \rightarrow n$ where parton number is conserved
and the off-diagonal $n+1\rightarrow n-1$ convolution such that
the current operator annihilates a $q{\bar{q'}}$ pair in the
initial $B$ wavefunction.  This term is a consequence of the fact
that the time-like decay $q^2 = (p_\ell + p_{\bar{\nu}} )^2 > 0$
requires a positive light-cone momentum fraction $q^+ > 0$.
Conversely for space-like currents, one can choose $q^+=0$, as in
the Drell-Yan-West representation of the space-like
electromagnetic form factors.

One can also compute the generalized parton distributions which
appear the deeply virtual Compton amplitude (DVCS) in the handbag
approximation from overlap of light-front
wavefunctions~\cite{Brodsky:2000xy,Diehl:2000xz}.  An interesting
aspect of DVCS is the prediction from QCD of a $J=0$ fixed Regge
pole contribution to the real part of the Compton amplitude which
has constant energy $s^0 F(t)$ dependence at any momentum transfer
$t$ or photon virtuality~\cite{Brodsky:1971zh,Brodsky:1973hm}. This
unique contribution is due to the quasi-local coupling of two photons to
the quark current coming from the quark $Z$-graph in time-ordered
perturbation theory or, equivalently, the instantaneous quark propagator
arising in light-front quantization.

\begin{figure}
\begin{center}
\includegraphics{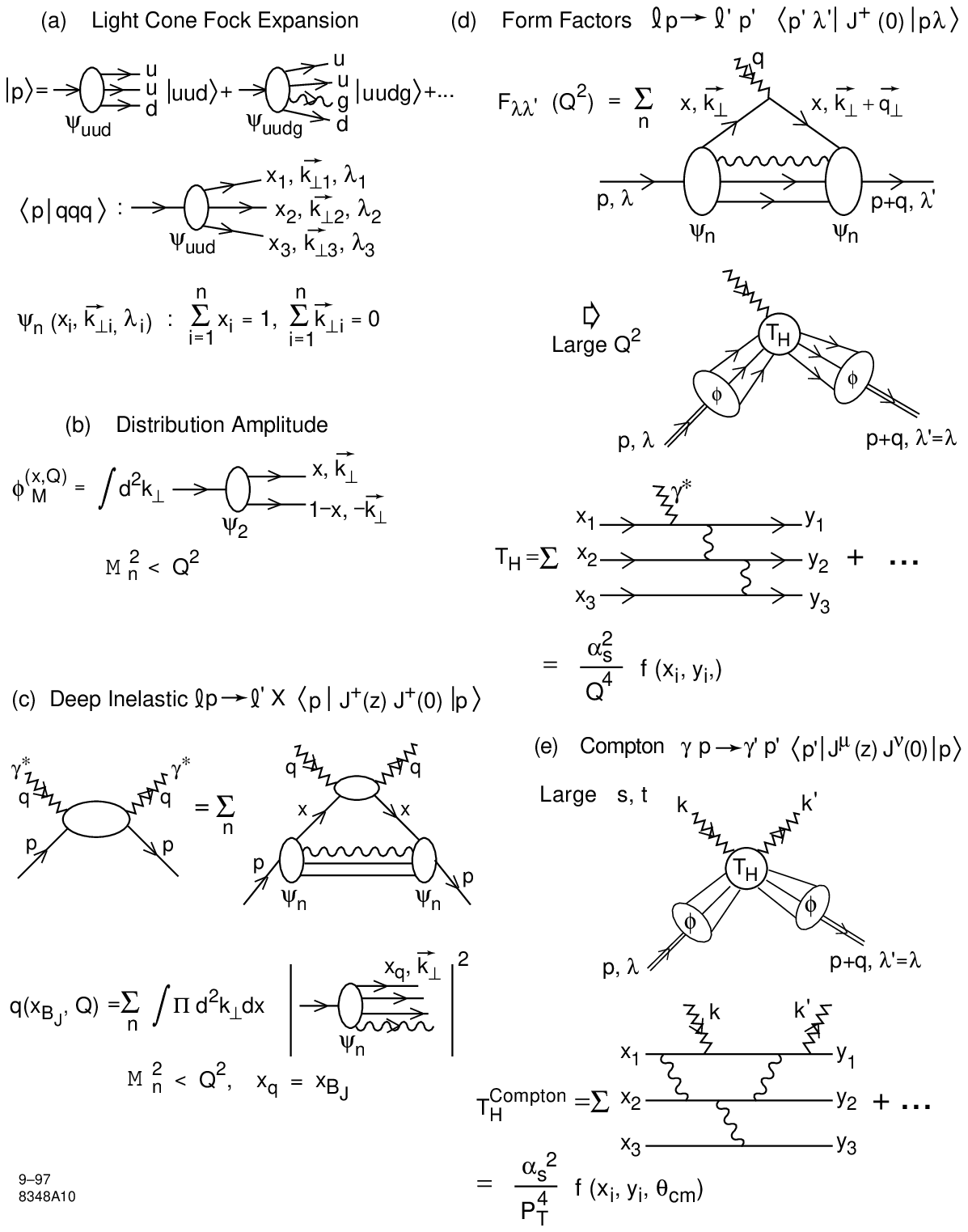}
\end{center}
\caption[*]{ Representation of QCD hadronic processes in the
light-cone Fock expansion. (a) The valence $uud$  and higher Fock $uudg$
contributions to the light-cone Fock expansion for the proton. (b)
The distribution amplitude $\phi(x,Q)$ of a meson expressed as an
integral over its valence light-cone wavefunction restricted to $q
\bar q$ invariant mass less than $Q$.  (c) Representation of deep
inelastic scattering and the quark distributions $q(x,Q)$  as
probabilistic measures of the light-cone Fock wavefunctions. The
sum is over the Fock states with invariant mass less than $Q$. (d)
Exact representation of spacelike form factors of the proton in
the light-cone Fock basis.  The sum is over all Fock components.
At large momentum transfer the leading-twist contribution
factorizes as the product of the hard scattering amplitude $T_H$
for the scattering of the valence quarks collinear with the
initial to final direction convoluted with the proton distribution
amplitude.  (e) Leading-twist factorization of the Compton
amplitude at large momentum transfer. \label{Fig:repc1}}
\end{figure}

\begin{figure}[htbp]
\includegraphics{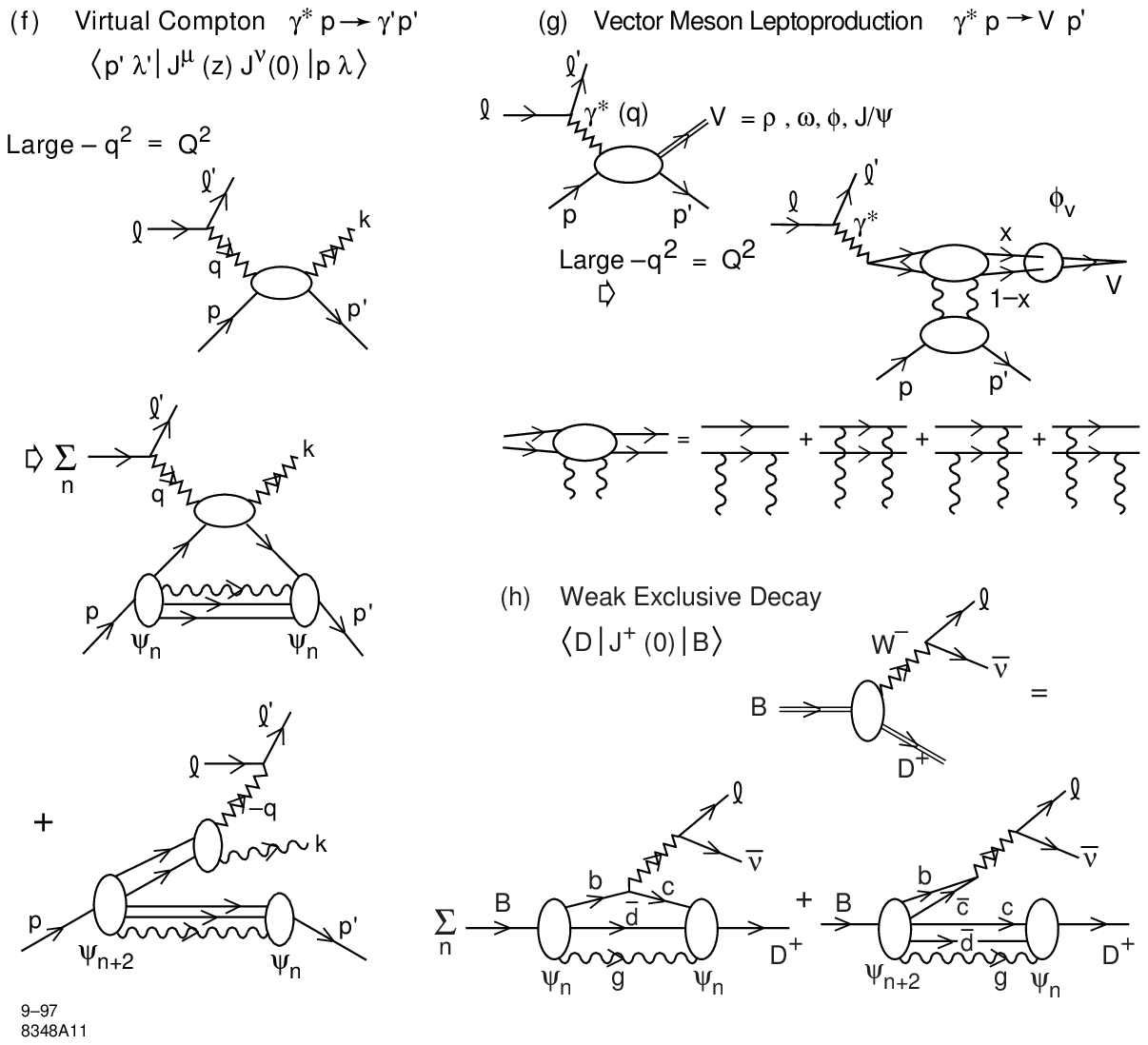}
\caption{ (f) Representation of deeply virtual Compton scattering
in the light-cone Fock expansion at leading twist. Both diagonal
$n  \to n$ and off-diagonal $n+2 \to n$ contributions are
required.  (g) Diffractive vector meson production at large photon
virtuality $Q^2$ and longitudinal polarization.  The high energy
behavior involves two gluons in the $t$ channel coupling to the
compact color dipole structure of the upper vertex.  The
bound-state structure of the vector meson enters through its
distribution amplitude.  (h) Exact representation of the weak
semileptonic decays of heavy hadrons in the light-cone Fock
expansion.   Both diagonal $n  \to n$ and off-diagonal pair
annihilation $n+2 \to n$ contributions are required.
\label{Fig:repc2}}
\end{figure}

The relationship of QCD processes to the hadron LFWFs is illustrated in
Figs.~\ref{Fig:repc1} and \ref{Fig:repc2}.

\vspace{.5cm}

\vspace{.5cm}

Other applications include two-photon exclusive reactions, and
diffractive dissociation into jets.  The universal light-front
wave functions and distribution amplitudes control hard exclusive
processes such as form factors, deeply virtual Compton scattering,
high momentum transfer photoproduction, and two-photon processes.

The light-front Fock representation thus provides an exact
formulation of current matrix elements of local and bi-local
operators.  In contrast, in equal-time Hamiltonian theory, one
must evaluate connected time-ordered diagrams where the gauge
particle or graviton couples to particles associated with vacuum
fluctuations.  Thus even if one knows the equal-time wavefunction
for the initial and final hadron, one cannot determine the current
matrix elements.  In the case of the covariant Bethe-Salpeter
formalism, the evaluation of the matrix element of the current
requires the calculation of an infinite number of irreducible
diagram contributions. One can also prove that the anomalous
gravitomagnetic moment $B(0)$ vanishes for any composite
system~\cite{Brodsky:2000ii}. This property follows directly from
the Lorentz boost properties of the light-front Fock
representation and holds separately for each Fock state component.

One of the central issues in the analysis of fundamental hadron
structure is the presence of non-zero orbital angular momentum in
the bound-state wave functions. The evidence for a ``spin crisis"
in the Ellis-Jaffe sum rule signals a significant orbital
contribution in the proton wave
function~\cite{Jaffe:1989jz,Ji:2002qa}.  The Pauli form factor of
nucleons is computed from the overlap of LFWFs differing by one
unit of orbital angular momentum $\Delta L_z= \pm 1$.  Thus the
fact that the anomalous moment of the proton is non-zero requires
nonzero orbital angular momentum in the proton
wavefunction~\cite{Brodsky:1980zm}.  In the light-front method,
orbital angular momentum is treated explicitly; it includes the
orbital contributions induced by relativistic effects, such as the
spin-orbit effects normally associated with the conventional Dirac
spinors.

\section{Perturbative QCD and Exclusive Processes}

There are a number of fundamental tests of QCD based on exclusive
reactions. There has been considerable progress analyzing
exclusive and diffractive reactions at large momentum transfer
from first principles in QCD. Rigorous statements can be made on
the basis of asymptotic freedom and factorization theorems which
separate the underlying hard quark and gluon subprocess amplitude
from the nonperturbative physics of the hadronic wavefunctions.
The leading-power contribution to exclusive hadronic amplitudes
such as quarkonium decay, heavy hadron decay,  and scattering
amplitudes where hadrons are scattered with large momentum
transfer can often be factorized as a convolution of distribution
amplitudes $\phi_H(x_i,\Lambda)$ and hard-scattering quark/gluon
scattering amplitudes $T_H$ integrated over the light-cone
momentum fractions of the valence quarks~\cite{Lepage:1980fj}:
\begin{eqnarray}
\M_{\rm Hadron} &=&\int
 \prod \phi_H^{(\Lambda)} (x_i,\lambda_i)\, T_H^{(\Lambda)} dx_i\ .
\label{eq:e}
\end{eqnarray}
Here $T_H^{(\Lambda)}$ is the underlying quark-gluon subprocess
scattering amplitude in which each incident and final hadron is
replaced by valence quarks with collinear momenta $k^+_i =x_i
p^+_H$, $\vec k_{\perp i} = x_i \vec p_{\perp H }.$ The invariant
mass of all intermediate states in $T_H$ is evaluated above the
separation scale $\M^2_n > \Lambda^2$. The essential part of the
hadronic wavefunction is the distribution
amplitude~\cite{Lepage:1980fj}, defined as the integral over
transverse momenta of the valence (lowest particle number) Fock
wavefunction; \eg\ for the pion
\begin{equation}
\phi_\pi (x_i,Q) \equiv \int d^2k_\perp\, \psi^{(Q)}_{q\bar q/\pi} (x_i,
\vec k_{\perp
i},\lambda) \label{eq:f}
\end{equation}
where the separation scale $\Lambda$ can be taken to be order of
the characteristic momentum transfer $Q$ in the process. It should
be emphasized that the hard scattering amplitude $T_H$ is
evaluated in the QCD perturbative domain where the propagator
virtualities are above the separation scale.

The leading power fall-off of the hard scattering amplitude as
given by dimensional counting rules follows from the nominal
scaling of the hard-scattering amplitude: $T_H \sim 1/Q^{n-4}$,
where $n$ is the total number of fields (quarks, leptons, or gauge
fields) participating in the hard
scattering~\cite{Brodsky:1974vy,Matveev:1973ra}. Thus the reaction
is dominated by subprocesses and Fock states involving the minimum
number of interacting fields.  In the case of $2 \to 2$ scattering
processes, this implies
\begin{equation}
{d\sigma\over dt}(A B \to C D) ={F_{A B \to C D}(t/s)/ s^{n-2}}
\end{equation}
where $n = N_A + N_B + N_C +N_D$ and $n_H$ is the minimum number of
constituents of $H$.

In the case of form factors, the dominant helicity conserving
amplitude has the nominal power-law falloff  $F_H(t) \sim (1/t)^{n_H-1},$
The complete predictions from PQCD modify the nominal scaling by
logarithms from the running coupling and the evolution of the distribution
amplitudes. In some cases, such as large angle $pp \to p p $
scattering, there can be ``pinch"
contributions~\cite{Landshoff:ew} when the scattering can occur
from a sequence of independent near-on shell quark-quark
scattering amplitudes at the same CM angle.  After inclusion of
Sudakov suppression form factors, these contributions also have a
scaling behavior close to that predicted by constituent counting.

The constituent counting
rules were originally derived
in 1973~\cite{Brodsky:1974vy,Matveev:1973ra} before the development of
QCD in anticipation that the underlying theory of hadron physics would
be renormalizable and close to a conformal theory.  The factorized
structure of hard exclusive amplitudes in terms of a convolution of
valence hadron wavefunctions times a hard-scattering quark scattering
amplitude was also proposed~\cite{Brodsky:1974vy}.  Upon the discovery of
the asymptotic freedom in QCD, there was a systematical development of
the theory of hard exclusive reactions, including factorization theorems,
counting rules, and evolution equations for the hadronic distribution
amplitudes~\cite{Brodsky:1979qm,Lepage:1979za,Lepage:1979zb,Efremov:1980rn}.

In a remarkable recent development, Polchinski and
Strassler have shown how one can map
features of gravitational theories in higher dimensions $ (AdS_5)$
to phenomenological properties of physical QCD in ordinary 3+1
space-time~\cite{Polchinski:2001tt}. The AdS/CFT correspondence
connecting superstring theory to superconformal gauge theory has
important implications for hadron phenomenology in the conformal limit,
including an all-orders demonstration of counting rules for hard exclusive
processes as well as determining essential aspects of hadronic
light-front wavefunctions.

The distribution amplitudes which control leading-twist exclusive
amplitudes at high momentum transfer can be related to the
gauge-invariant Bethe-Salpeter wavefunction at equal light-cone
time $\tau = x^+$.  The logarithmic evolution of the hadron
distribution amplitudes $\phi_H (x_i,Q)$ with respect to the
resolution scale $Q$ can be derived from the
perturbatively-computable tail of the valence light-cone
wavefunction in the high transverse momentum regime. The DGLAP
evolution of quark and gluon distributions can also be derived in
an analogous way by computing the variation of the Fock expansion
with respect to the separation scale. Other key features of the
perturbative QCD analyses are: (a) evolution equations for
distribution amplitudes which incorporate the operator product
expansion, renormalization group invariance, and conformal
symmetry~\cite{Lepage:1980fj,Brodsky:1980ny,Muller:1994cn,%
Ball:1998ff,Braun:1999te}; (b) hadron helicity conservation which
follows from the underlying chiral structure of
QCD~\cite{Brodsky:1981kj}; (c) color transparency, which
eliminates corrections to hard exclusive amplitudes from initial
and final state interactions at leading power and reflects the
underlying gauge theoretic basis for the strong
interactions~\cite{Brodsky:1988xz} and (d) hidden color degrees of
freedom in nuclear wavefunctions, which reflect the color
structure of hadron and nuclear
wavefunctions~\cite{Brodsky:1983vf}. There have also been recent
advances eliminating renormalization scale ambiguities in
hard-scattering amplitudes via commensurate scale
relations~\cite{Brodsky:1994eh} which connect the couplings
entering exclusive amplitudes to the $\alpha_V$ coupling which
controls the QCD heavy quark potential.

Exclusive processes such as $\bar p p \to \bar p p,$ $\bar p p \to
K^+ K^-$ and $\bar p p \to \gamma \gamma$ provide a unique window
for viewing QCD processes and hadron dynamics at the amplitude
level~\cite{Brodsky:1981rp,Brodsky:2000dr}.  New tests of theory
and comprehensive measurements of hard exclusive amplitudes can
also be carried out for electroproduction at Jefferson Laboratory
and in two-photon collisions at CLEO, Belle, and
BaBar~\cite{Brodsky:2001hv}.    Hadronic exclusive processes are
closely related to exclusive hadronic $B$ decays, processes which
are essential for determining the CKM phases and the physics of
$CP$ violation. The universal light-front wavefunctions which
control hard exclusive processes such as form factors, deeply
virtual Compton scattering, high momentum transfer
photoproduction, and two-photon processes, are also required for
computing exclusive heavy hadron
decays~\cite{Beneke:2000ry,Keum:2000wi,Szczepaniak:1990dt,Brodsky:2001jw},
such as $B \to K \pi$, $B \to \ell \nu \pi$, and $B \to K  p \bar
p$~\cite{Chua:2002wn}. The same physics issues, including color
transparency, hadron helicity rules, and the question of dominance
of leading-twist perturbative QCD mechanisms enter in both realms
of physics.

\section{The Pion Form Factor}

The pion spacelike form factor provides an important illustration
of the perturbative QCD formalism. The proof of factorization
begins with the exact Drell-Yan-West
representation~\cite{Drell:1970km,West:1970av,Brodsky:1980zm} of
the current in terms of the light-cone Fock wavefunctions (see
Section 7.) The integration over the momenta of the constituents
of each wavefunction can be divided into two domains
$\mathcal{M}^2_n < \Lambda^2 $ and $\mathcal{M}^2_n > \Lambda^2, $
where $\mathcal{M}^2_n$ is the invariant mass of the n-particle
state. $\Lambda$ plays the role of a separation scale.  In
practice, it can be taken to be of order of the momentum transfer.

Consider the contribution of the two-particle Fock state.  The
argument of the final state pion wavefunction is $k_\perp + (1-x)
q_\perp$.  First take $k_\perp$ small.  At high momentum transfer
where
\begin{equation}
\mathcal{M}^2  \sim {(1-x)^2 q^2_\perp \over x(1-x)} = {Q^2 (1-x)\over x} >
\Lambda^2,
\end{equation}
one can iterate the equation of motion for the valence light-front
wavefunction using the one gluon exchange kernel.  Including all
of the hard scattering domains, one can organize the result into
the factorized form:
\begin{equation}
F_\pi(Q^2) = \int^1_0 dx \int^1_0 dy \phi_{\pi}(y,\Lambda)
T_H(x,y,Q^2) \phi_{\pi}(x,\Lambda) ,\end{equation} where $T_H$ is
the hard-scattering amplitude $\gamma^* (q \bar q) \to (q \bar q)$
for the production of the valence quarks nearly collinear with
each meson, and $\phi_M(x,\Lambda)$ is the distribution amplitude
for finding the valence $q$ and $\bar q$ with light-cone fractions
of the meson's momentum, integrated over invariant mass up to
$\Lambda.$ The process independent distribution amplitudes contain
the soft physics intrinsic to the nonperturbative structure of the
hadrons.  Note that $T_H$ is non-zero only if ${(1-x)Q^2\over x}
> \Lambda^2 $ and ${(1-y)Q^2\over y} > \Lambda^2 .$ In this
hard-scattering domain, the transverse momenta in the formula for
$T_H$ can be ignored at leading power, so that the structure of
the process has the form of hard scattering on collinear quark and
gluon constituents: $T_H(x,y,Q^2) = {16 \pi C_F
\alpha_s(Q^{*2})\over (1-x) (1-y) Q^2}\left(1 +
\mathcal{O}(\alpha_s)\right)$
and thus~\cite{Brodsky:1979qm,Lepage:1979za,Lepage:1979zb,%
Lepage:1980fj,Efremov:1980rn,Chernyak:1977fk,%
Chernyak:1980dk,Farrar:1979aw,Duncan:1980hi}
\begin{equation}
F_{\pi}(Q^2) = {16 \pi C_F \alpha_s(Q^{*2}) \over Q^2} \int^{\widehat x} _0 dx
{\phi_\pi(x,\Lambda)\over (1-x)} \int^{\widehat y}_0 dy
{\phi_\pi(y,\Lambda)\over (1-y)}
,
\end{equation} to leading order in $\alpha_s(Q^{*2})$ and leading
power in $1/Q.$ Here $C_F = 4/3 $ and $Q^*$ can be taken as the
BLM scale~\cite{Brodsky:1998dh}. The endpoint regions of
integration $1-x < { \Lambda^2\over Q^2 } = 1-\widehat x $ and
$1-y < { \Lambda^2\over Q^2 }= 1- \widehat y$ are to be explicitly
excluded in the leading-twist formula. However, since the
integrals over $x$ and $y$ are convergent, one can formally extend
the integration range to $0< x < 1$ and $0< y < 1$ with an error
of higher twist. This is only done for convenience; the actual
domain only encompasses the off-shell regime.  The contribution
from the endpoint regions of integration, $x \sim 1$ and $y \sim
1,$ are power-law and Sudakov suppressed and thus contribute
corrections at higher order in
$1/Q$~\cite{Lepage:1979za,Lepage:1979zb,Lepage:1980fj}. The
contributions from non-valence Fock states and corrections from
fixed transverse momentum entering the hard subprocess amplitude
are higher twist, {\em i.e.}, power-law suppressed. Loop
corrections involving hard momenta give next-to-leading-order
(NLO) corrections in $\alpha_s$.

It is sometimes assumed that higher twist terms in the LC wave
function, such as those with $L_z \ne 0$, have flat distributions
at the $ x \to 0,1$ endpoints. This is difficult to justify since
it would correspond to bound state wavefunctions which fall-off in
transverse momentum but have no fall-off at large $k_z.$ After
evolution to $Q^2 \to \infty$, higher twist distributions can
evolve eventually to constant behavior at $x =0,1;$ however, the
wavefunctions are in practice only being probed at moderate
scales.  In fact, if the higher twist terms are evaluated in the
soft domain, then there is no evolution at all. A recent analysis
by Beneke~\cite{Beneke:2002bs} indicates that the $1/Q^4$
contribution to the pion form factor is only logarithmically
enhanced even if the twist-3 term is flat at the endpoints.  It is
also possible that contributions from the twist three  $q \bar q
g$ light-front wavefunctions may well cancel even this
enhancement.

Thus perturbative QCD can unambiguously predict the leading-twist
behavior of exclusive amplitudes.  These contributions only
involve the truncated integration domain of $x$ and $k_\perp$
momenta where the  quark and gluon propagators and couplings are
perturbative; by definition the soft regime is excluded.  The
central question is then whether the PQCD leading-twist prediction
can account for the observed leading power-law fall-off of the
form factors and other exclusive processes.  Assuming the pion
distribution amplitude is close to its asymptotic form,  one can
predict the normalization of exclusive amplitudes such as the
spacelike pion form factor $Q^2 F_\pi(Q^2)$.  Next-to-leading
order predictions are available which incorporate higher order
corrections to the pion distribution amplitude as well as the hard
scattering
amplitude~\cite{Muller:1994cn,Muller:1994fv,Melic:1998qr,Szczepaniak:1998sa}.
The natural renormalization scheme for the QCD coupling in hard
exclusive processes is $\alpha_V(Q)$, the effective charge defined
from the scattering of two infinitely-heavy quark test charges.
Assuming $\alpha_V(Q^*) \simeq 0.4$ at the BLM scale $Q^*$, the
QCD LO prediction appears to be smaller by approximately a factor
of 2 compared to the presently available data extracted from pion
electroproduction experiments~\cite{Brodsky:1998dh}. However, the
extrapolation from spacelike $t$ to the pion pole in
electroproduction may be unreliable in the same sense that
lattice gauge theory extrapolations to $m^2_\pi \to 0$ are known
to be nonanalytic.  Thus it is not clear that there is an actual
discrepancy between perturbative QCD and experiment. It would be
interesting to develop predictions for the transition form factor
$F_{q \bar q \to \pi}(t,q^2)$ which is in effect what is measured
in electroproduction.

It is interesting to compare the calculation of a meson form
factor in QCD with the calculation of the form factor of a bound
state in QED. The analog to a soft wavefunction is the
Schr\"odinger-Coulomb solution $\psi_{1s}(\vec k) \propto (1 +
{{\vec p}^2/(\alpha m_{\rm red})^2})^{-2}$, and the full
wavefunction, which incorporates transversely polarized photon
exchange, differs by a factor $(1 + {\vec p}^2/m^2_{\rm red})$.
Thus the leading-twist dominance of form factors in QED occurs at
relativistic scales $Q^2 > {m^2_{\rm red}}$~\cite{Brodsky:2000dr}.

\section{Perturbative QCD Calculation of Baryon Form Factors}

\begin{figure}[htb]
\centering
\includegraphics[width=4.3in]
{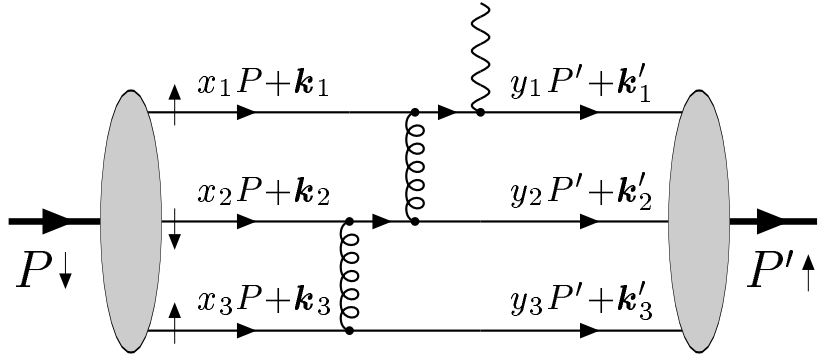} \caption{A typical leading QCD diagram contributing to
the nucleon form factors. From Ref.~\cite{Belitsky:2002kj} .}
\label{Typical}
\end{figure}

The baryon form factor at large momentum transfer provides another
important example of the application of perturbative QCD to
exclusive processes.  Away from possible special points in the
$x_i$ integrations (which are suppressed by Sudakov form factors)
baryon form factors can be written to leading order in $1/Q^2$ as
a convolution of the connected hard-scattering amplitude $T_H$
with the  baryon distribution amplitudes. An example of
a perturbative QCD contribution to $F_2$ is illustrated in
Fig.~\ref{Typical} The $Q^2$-evolution of the baryon distribution
amplitude can be derived from the operator product expansion of
three quark fields or from the gluon exchange kernel. Taking into
account the evolution of the baryon distribution amplitude, the
nucleon magnetic form factors at large $Q^2$, has the
form~\cite{Lepage:1980fj,Lepage:1979zb,Brodsky:1981kj}
\begin{equation}
G_M(Q^2)\rightarrow{\alpha^2_s(Q^2)\over Q^4}\sum_{n,m} b_{nm} \left({\rm
log}{Q^2\over
\Lambda^2}\right)^{\gamma^B_n+\gamma^B_n}
\left[1+\mathcal{O}\left(\alpha_s(Q^2),{m^2\over Q^2}\right)\right]\quad .
\end{equation}
where the $\gamma^B_n$ are computable anomalous
dimensions~\cite{Peskin:1979mn} of the baryon three-quark wave
function at short distance, and the $b_{mn}$ are determined from
the value of the distribution amplitude $\phi_B(x,Q^2_0)$ at a
given point $Q_0^2$ and the normalization of $T_H$.
Asymptotically, the dominant term has the minimum anomalous
dimension. The contribution from the endpoint regions of
integration, $x \sim 1$ and $y \sim 1,$ at finite $k_\perp$ is
Sudakov
suppressed~\cite{Lepage:1979za,Lepage:1979zb,Lepage:1980fj};
however, the endpoint region may play a significant role in
phenomenology.

\begin{figure}[htb]
\centering
\includegraphics[width=4.3in,height=3.7in]
{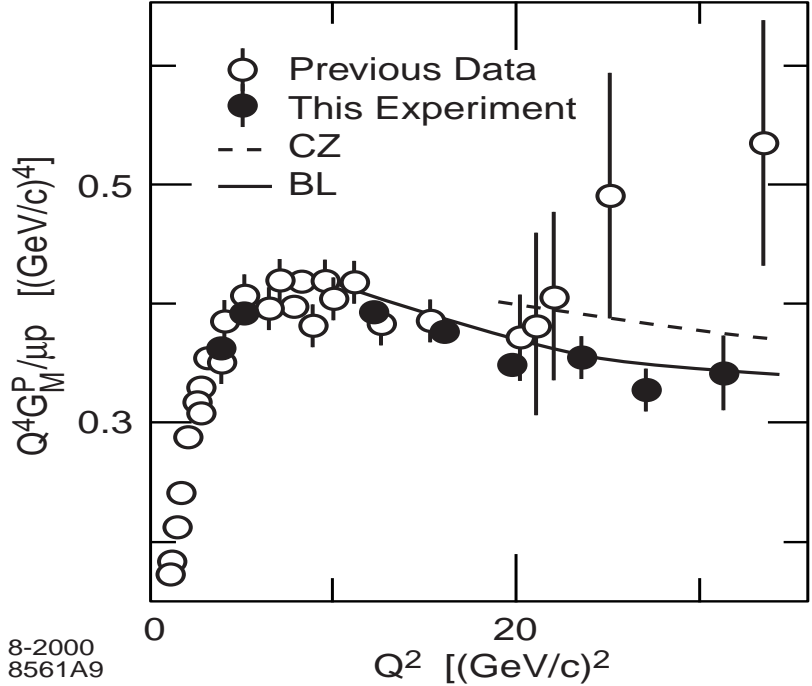} \caption[ *]{Predictions for the normalization and
sign of the proton form factor at high $Q^2$ using perturbative
QCD factorization and QCD sum rule predictions for the proton
distribution amplitude (From Ji {\em et al.}~\cite{jsl}) The curve
labelled BL has arbitrary normalization and incorporates the
fall-off of two powers of the running coupling.  The dotted line
is the QCD sum rule prediction of given by Chernyak and
Zhitnitsky~\cite{Chernyak:1984bm,Chernyak:1989nv}.  The results
are similar for the model distribution amplitudes of King and
Sachrajda~\cite{ks}, and Gari and Stefanis~\cite{gs}.}
\label{figzrpic}
\end{figure}

The proton form factor appears to scale at $Q^2 > 5 \ {\rm GeV}^2$
according to the PQCD predictions.  See Fig. \ref{figzrpic}.
Nucleon form factors are approximately described
phenomenologically by the well-known dipole form $ G_M(Q^2) \simeq
{1 / (1+Q^2/0.71\,{~\rm GeV}^2)^2}$ which behaves asymptotically
as $G_M(Q^2) \simeq (1 /Q^4)( 1- 1.42 {~\rm GeV}^2/ Q^2 +
\cdots)\,.$ This suggests that the corrections to leading twist in
the proton form factor and similar exclusive processes involving
protons become important in the range $Q^2 < 1.4\ {\rm GeV}^2$.

The shape of the distribution amplitude controls the normalization
of the leading-twist prediction for the proton form factor. If one
assumes that the proton distribution amplitude has the asymptotic
form: $\phi_N = C x_1 x_2 x_3$, then the convolution with the
leading order form for $T_H$ gives zero! If one takes a
non-relativistic form peaked at $x_i = 1/3$, the sign is negative,
requiring a crossing point zero in the form factor at some finite
$Q^2$.  The broad asymmetric distribution amplitude advocated by
Chernyak and Zhitnitsky~\cite{Chernyak:1984bm,Chernyak:1989nv}
gives a more satisfactory result. If one assumes a constant value
of $\alpha_s = 0.3$, and $f_N=5.3 \times 10^{-3}$GeV$^2$, the
leading order prediction is below the data by a factor of $\approx
3.$ However, since the form factor is proportional to $\alpha^2_s
f^2_N$, one can obtain agreement with experiment by a simple
renormalization of the parameters.  For example, if one uses the
central value of Ioffe's determination $f_N=8 \times 10^{-3}
$GeV$^2$, then good agreement is obtained~\cite{Stefanis:1999wy}.
The normalization of the proton distribution amplitude is also
important for estimating the proton decay
rate~\cite{Brodsky:1983st}. The most recent lattice
results~\cite{Kuramashi:2000hw} suggest a significantly larger
normalization for the required proton matrix elements, 3 to 5
times larger than earlier phenomenological estimates.  One can
also use PQCD to predict ratios of various baryon and isobar form
factors assuming isospin or $SU(3)$-flavor symmetry for the basic
wave function structure.  Results for the neutral weak and charged
weak form factors assuming standard $SU(2)\times U(1)$ symmetry
can also be derived~\cite{Brodsky:1981sx}.

A  useful technique for obtaining the solutions to the baryon
evolution equations is to construct completely antisymmetric
representations as a polynomial orthonormal basis for the
distribution amplitude of multi-quark bound states.  In this way.
one obtains a distinctive classification of nucleon $(N)$ and Delta
$(\Delta)$ wave functions and the corresponding $Q^2$ dependence
which discriminates $N$ and $\Delta$ form factors.
Braun and collaborators have shown how one can use conformal
symmetry to classify the eigensolutions of the baryon distribution
amplitude~\cite{Braun:1999te}. They identify a new `hidden'
quantum number which distinguishes components in the $\lambda=3/2$
distribution amplitudes with different scale dependence. They are
able to find analytic solutions of the evolution equation for
$\lambda=3/2$ and $\lambda=1/2$ baryons where the  two lowest
anomalous dimensions for the $\lambda=1/2$ operators (one for each
parity) are separated from the rest of the spectrum by a finite
`mass gap'. These special states can be interpreted as baryons
with scalar diquarks. Their results may support Carlson's
solution~\cite{Carlson:1986mm}  to the puzzle that the proton to
$\Delta$ form factor falls faster~\cite{Stoler:1993yk} than other
$p \to N^*$ amplitudes if the $\Delta$ distribution amplitude has
a symmetric $x_1 x_2 x_3$ form.

In a remarkable  development, Pobylitsa {\em et
al.}~\cite{Pobylitsa:2001cz} have shown how to compute  transition
form factors linking the proton to nucleon-pion states which have
minimal invariant mass $W$. A new soft pion theorem for high
momentum transfers allows one to compute the three quark
distribution amplitudes for the near threshold pion states from a
chiral rotation. The new soft pion results are in a good agreement
with the SLAC electroproduction data  for $W^2 < 1.4~$GeV$^2$ and
$7 < Q^2 < 30.7~$GeV$^2.$ This approach can be tested in
antiproton collisions in reactions such as $\bar p p \to \bar p p
\pi^0$ where the proton and pion have small invariant mass.

\section{Timelike Proton Form Factors}

The form factors of hadrons as measured in both the spacelike and
timelike domains provide fundamental information on the structure
and internal dynamics of hadrons. Recent
measurements~\cite{perdrisat} of the electron-to-proton
polarization transfer in $\overrightarrow e^- \, p \to e^- \,
\overrightarrow p$ scattering at Jefferson Laboratory show that
the ratio of Sachs form factors~\cite{walecka} $G^p_E(q^2)/
G^p_M(q^2)$ is monotonically decreasing with increasing
$Q^2=-q^2,$ in strong contradiction with the $G_E/G_M$ scaling
determined by the traditional Rosenbluth separation method.
Recently, Afanasev, Carlson, Chen,
Vanderhaeghen, and I~\cite{Chen:2004tw} have shown that the
interfering two-photon exchange contribution to elastic
electron-proton scattering, including inelastic intermediate
states, can account for the discrepancy between Rosenbluth and
Jefferson Lab spin transfer polarization data~\cite{Jones:1999rz}. The
Rosenbluth method thus in fact is not reliable because of
its sensitivity to the
two-photon exchange
amplitudes~\cite{Guichon:2003qm,Blunden:2003sp}. The polarization
transfer method~\cite{perdrisat,acg81} is relatively insensitive
to such corrections.

The same data which indicate that $G_E$ for protons falls faster
than $G_M$ at large spacelike $Q^2$ require in turn that $F_2/F_1$
falls more slowly than $1/Q^2$. The conventional expectation from
dimensional counting rules~\cite{Brodsky:1974vy} and perturbative
QCD~\cite{Lepage:1979za} is that the Dirac form factor $F_1$
should fall with a nominal power $1/Q^4$, and the ratio of the
Pauli and Dirac form factors, $F_2/F_1$, should fall like $1/Q^2$,
at high momentum transfers. The Dirac form factor agrees with this
expectation in the range $Q^2$ from a few GeV$^2$ to the data
limit of 31 GeV$^2$. However, the Pauli/Dirac ratio is not
observed to fall with the nominal expected power, and the
experimenters themselves have noted that the data is well fit by
$F_2/F_1 \propto 1/Q$ in the momentum transfer range 2 to 5.6
GeV$^2$.

The new Jefferson Laboratory results make it critical to carefully
identify and separate the timelike $G_E$ and $G_M$ form factors by
measuring the center-of-mass angular distribution and by measuring
the polarization of the proton in $e^+ e^- \to p \bar p$ or $\bar p
p \to \ell^+ \ell^-$ reactions. The advent of high luminosity
$e^+ e^-$ colliders at Beijing, Cornell, and Frascati provide the
opportunity to make such measurements, both directly and via
radiative return.  The new GSI antiproton facility with  a polarized
target will make measurements of the single spin-dependence of $\bar p p
\to \ell^+ \ell^-$ feasible.

Although the spacelike form factors of a stable hadron are real,
the timelike form factors have a phase structure reflecting the
final-state interactions of the outgoing hadrons. In general, form
factors are analytic functions $F_i(q^2)$ with a discontinuity for
timelike momentum above the physical threshold $q^2> 4 M^2.$ The
analytic structure and phases of the form factors in the timelike
regime are thus connected by dispersion relations to the spacelike
regime~\cite{baldini,Geshkenbein74,seealso}. The analytic form and
phases of the timelike amplitudes are also sensitive to the resonances in
the unphysical region $0 < q^2 < 4M^2$ below the physical
threshold~\cite{baldini} in the $J^{PC} = 1^{--}$ channel,
including gluonium states and di-baryon structures.

At very large center-of-mass energies, perturbative QCD
factorization predicts diminished final interactions in $e^+ e^-
\to H \bar H$, since the hadrons are initially produced with small
color dipole moments. This principle of QCD color
transparency~\cite{Brodsky:xz} is also an essential
feature~\cite{Bjorken:kk} of hard exclusive $B$
decays~\cite{Beneke:2001ev,Keum:2000wi}, and thus needs to be
tested experimentally.

\begin{figure}[htb]
\centering
\includegraphics[width=4.3in]{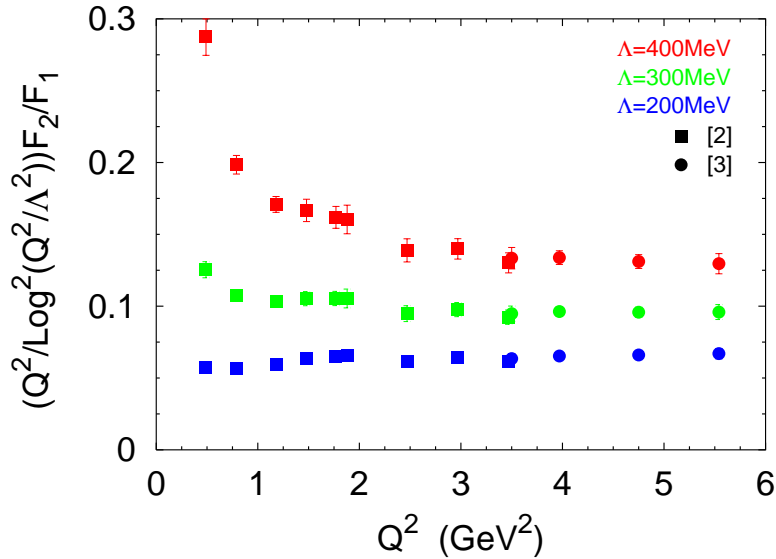}
\caption{ JLab data plotted in terms of the leading PQCD scaling.
The lower, middle, and upper data points correspond to
$\Lambda=200,300,400$, respectively. From
Ref.~\cite{Belitsky:2002kj}.} \label{ScalingF2toF1}
\end{figure}

There have been a number of explanations and theoretically
motivated fits of the $F_2/F_1$ data.  Belitsky, Ji, and
Yuan~\cite{belitsky02} have shown that factors of $\log(Q^2)$
arise from a careful QCD analysis of the form factors. The
perturbative QCD form $Q^2 F_2/ F_1 \sim \log^2{Q^2}$, which has
logarithmic factors multiplying the nominal power-law behavior,
fits the large-$Q^2$ spacelike data well. See
Fig.~\ref{ScalingF2toF1}.
Others~\cite{Ralston:2003mt,Miller:2002qb} claim to find
mechanisms that modify the traditionally expected power-law
behavior with fractional powers of $Q^2$, and they also give fits
which are in accord with the data in the experimental range.
Asymptotic behaviors of the ratio $F_2/F_1$ for general
light-front wave functions are investigated
in~\cite{Brodsky:2003pw}. Each of the model forms predicts a
specific fall-off and phase structure of the form factors from $ s
\leftrightarrow t$ crossing to the timelike domain. A fit with the
dipole polynomial or nominal dimensional counting rule behavior
would predict no phases in the timelike regime.

\section{Single-Spin Asymmetry and the Phase of Timelike Form Factors}

As noted by Dubnickova, Dubnicka, and Rekalo, and by
Rock~\cite{d}, the existence of the $T-$odd single-spin asymmetry
normal to the scattering plane in baryon pair production $e^- e^+
\rightarrow B \bar B$ requires a nonzero phase difference between
the $G_E$ and $G_M$ form factors. The phase of the ratio of form
factors $G_E/G_M$ of spin-1/2 baryons in the timelike region can
thus be determined from measurements of the polarization of one of
the produced baryons.  As discussed below, Carlson, Hiller,  Hwang  and I
have shown that measurements of the proton polarization in $e^+  e^- \to p
\bar p$ strongly discriminate between the analytic forms of  models which
have been suggested to fit the proton $G_E/G_M$ data in the spacelike
region~\cite{Brodsky:2003gs}.

The center-of-mass angular distribution provides the analog of the
Rosenbluth method for measuring the magnitudes of various helicity
amplitudes. The differential cross section for $e^- e^+
\rightarrow B \bar B$ when $B$ is a spin-1/2 baryon is given in
the center-of-mass frame by
\begin{equation}                               \label{xsctn} {d\sigma \over
d\Omega} =
{\alpha^2 \beta
\over 4 q^2}
                            \ D \ ,
\end{equation}
where $\beta = \sqrt{1-4m_B^2/q^2}$ and $D$ is given by
\begin{equation} D =  |G_M|^2 \left(1+ \cos^2\theta \right) + {1\over \tau }\,
       |G_E|^2 \sin^2\theta  \ ;
\end{equation}
where
\begin{eqnarray} G_M &=& F_1 + F_2 \ , \nonumber \\ G_E &=& F_1 + \tau F_2
\ , \end{eqnarray}
and  $\tau \equiv {q^2 / 4 m_B^2} > 1$.

The complex phases of the form
factors in the timelike region make it possible for a single
outgoing baryon to be polarized in $e^- e^+ \rightarrow B \bar B$
even without polarization in the initial state.  The corresponding
effect in the initial state produces a polarization correlation in
$\bar p p \to \ell^+ \ell^-.$ This correlation can be measured at
GSI if the target or beam baryon is polarized.

There are three polarization observables, corresponding to
polarizations in three directions, called longitudinal, sideways,
and normal but often denoted $z$, $x$, and $y$, respectively.
Longitudinal ($z$) when discussing the final state means parallel
to the direction of the outgoing or in going baryon. Sideways
($x$) means perpendicular to the direction of the baryon but in
the scattering plane. Normal ($y$) means normal to the scattering
plane, in the direction of $\vec k \times \vec p$ where $\vec k$
is the electron momentum and $\vec p$ is the baryon momentum, with
$x$, $y$, and $z$ forming a right-handed coordinate system. The
polarization ${\cal P}_y$ does not require polarization of a
lepton and is~\cite{d}
\begin{equation}                    \label{py} {\cal P}_y = {  \sin 2\theta
\, {\rm Im}
G_E^*  G_M
      \over D \sqrt{\tau} }
                    =
      { (\tau - 1 )\sin 2\theta \, {\rm Im} F_2^* F_1
      \over D  \sqrt{\tau} }
                    \ .
\end{equation}
The other two polarizations require measurement of the lepton
polarization and are discussed in detail in
Ref.~\cite{Brodsky:2003gs}. Any model which fits the spacelike
form factor data with an analytic function can be continued to the
timelike region. Spacelike form factors are usually written in
terms of $Q^2 = - q^2$. The correct relation for analytic
continuation can be obtained by examining denominators in loop
calculations in perturbation theory. The connection is $Q^2
\rightarrow q^2 e^{-i\pi}$, or
\begin{equation} \ln Q^2 = \ln (-q^2) \rightarrow \ln q^2 - i\pi  \ .
\end{equation}
If the spacelike $F_2/F_1$ is fit by a rational function of $Q^2$,
then the form factors will be  relatively real in the timelike
region also.  However,  one in general gets a complex result from
the continuation.

More sophisticated dispersion relation based continuations could
give more reliable results, if there is data also in the timelike
region to pin down the magnitudes there. So far, this is possible
for the magnetic form factor alone~\cite{baldini}, but not for both
form factors.

\begin{figure}[htb]
\begin{center}
\includegraphics[height=4in,width=0.8\textwidth]{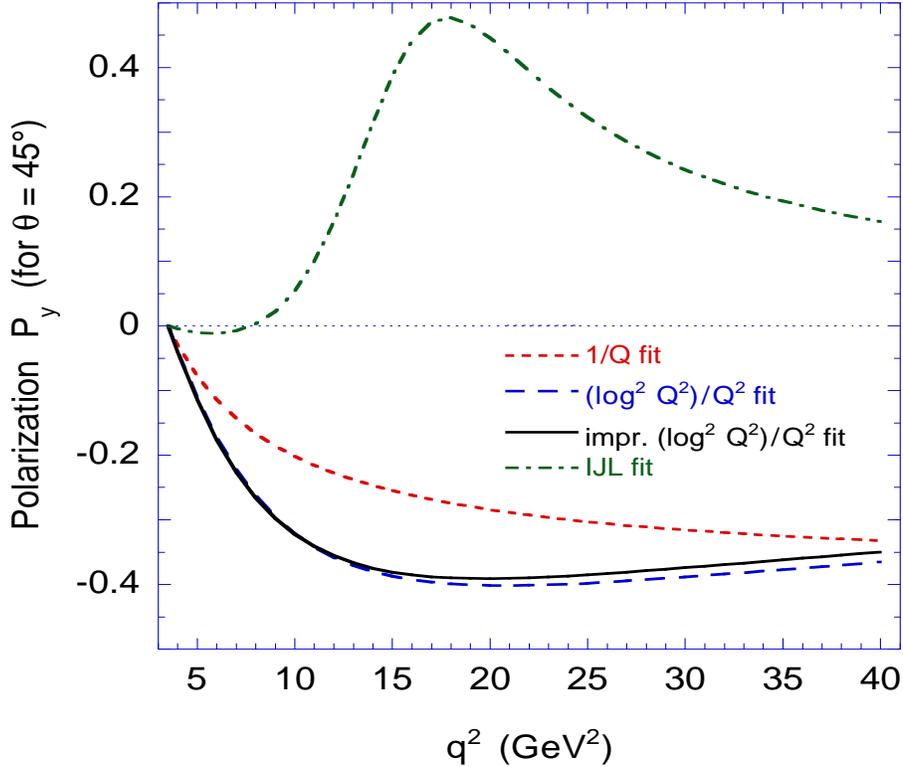}
\end{center}
\caption[*]{Predicted polarization ${\cal P}_y$ in the timelike
region for selected form factor fits described in the text.  The
plot is for $\theta = 45^\circ$. The four curves are for an
$F_2/F_1 \propto 1/Q$ fit; the $(\log^2 Q^2)/Q^2$ fit of Belitsky
{\em et al.}~\cite{Belitsky:2002kj}; an improved $(\log^2
Q^2)/Q^2$ fit; and a fit from Iachello {\em et al.}~\cite{ijl}.}
\label{figpy}
\end{figure}

The expression for polarization ${\cal P}_y$, Eq.~(\ref{py}),
leads to results shown in Fig.~\ref{figpy}.  The polarizations are
shown for four different fits  to the spacelike data as referenced
in the figure. The value of ${\cal P}_y$ should be the same for
$e^+ e^- \to p \bar p$ and $\bar p p \to \ell^+ \ell^-$ up to an
overall sign. The predicted polarizations are not small. Note that
a purely polynomial fit to the spacelike data gives zero
${\cal P}_y$. The normal polarization ${\cal P}_y$ is a
single-spin asymmetry and requires a phase difference between
$G_E$ and $G_M$. It is an example of how time-reversal-odd
observables can be nonzero if final-state or initial-state
interactions give interfering amplitudes different phases. Its
analog in the spacelike case is zero.

\section{Compton Scattering}

Compton scattering is a key test of the perturbative QCD
approach~\cite{Kronfeld:1991kp,Guichon:1998xv,Brooks:2000nb}. A
detailed recalculation of the helicity amplitudes and differential
cross section for proton Compton scattering at fixed angle has
been carried out recently by Brooks and Dixon~\cite{Brooks:2000nb}
at leading-twist and at leading order in $\alpha_s.$ They use
contour deformations to evaluate the singular integrals in the
light-cone momentum fractions arising from pinch contributions.
The shapes and scaling behavior predicted by perturbative QCD
agree well with the existing data~\cite{Shupe:vg}. In order to
reduce uncertainties associated with $\alpha_s$ and the
three-quark wave function normalization, Brooks and Dixon have
normalized the Compton cross section using the proton elastic form
factor. The theoretical predictions for the ratio of Compton
scattering to electron-proton scattering is about an order of
magnitude below existing experimental data. However, this
discrepancy of a factor of 3 in the relative normalization of the
amplitudes could be attributed to the fact that the number of
diagrams contributing to the Compton amplitude at next-to-leading
order ($\alpha_s^3$) is much larger in Compton scattering compared
to the proton form factor.

The Compton amplitude is predicted to have a real Regge
contribution~\cite{Brodsky:1971zh,Brodsky:1973hm}
\begin{equation}
M(\gamma p \to \gamma^\prime p^\prime) =
    -2 {\epsilon \cdot \epsilon^\prime}\sum_q e^2_q F_0(t)
    \end{equation}
which is constant in energy for any $t$. Thus unlike ordinary
Regge exchange in which the Regge power varies with $t$, the
$J=0$ term is independent of $t$ as well photon virtualities
$q^2_1, q^2_2$ at fixed $t.$ The spin-zero form factor $F^q_0(t)$
is the  $1/x_q$ matrix element of the proton
\begin{equation}
F^q_0(t) = \VEV{p^\prime \left\vert\, \frac{1}{x_q}\,\right\vert
p}
\end{equation}
evaluated at spacelike momentum transfer $t=(p-p^\prime)^2.$ This
$C=+$ form factor is a new object: it is relevant to the
non-spin-flip Higgs coupling to the proton.  At large $t,$ it falls
as $\sim 1/t^2.$ A similar Pauli-like contribution also enters the
spin-flip Compton amplitude. This contribution can be observed in
timelike DVCS: $\bar p p \to \gamma \gamma^*$ from its distinctive
kinematic properties: the amplitude from the $J=0$ term is
independent of $t$ at fixed $s$, independent of photon virtuality
at fixed $s.$

In the case of QED with scalar leptons, the fixed-pole contribution to the Compton
amplitude is due to the 4-point seagull interaction. In QCD it arises from the quark
$Z$-graphs; in the LF-quantized theory, it comes from the quasi-local instantaneous quark
exchange interaction for $\gamma q \to \gamma q.$ It is analogous to Thompson scattering
on the electrons of an atom:
\begin{equation}
M(\gamma A \to \gamma^\prime A^\prime) =
-2 {\sum e^2}{M_A\over m_e} \epsilon \cdot \epsilon^\prime .
\end{equation}
In
the case of hadrons in QCD $\VEV{\frac{1}{x_q}} $ plays the role of
$M_A\over m_e.$  The moment $\VEV{\frac{1}{x_q}} $ is particularly
interesting since, using the Feynman-Hellmann theorem at
$t=0$, it is precisely the expression for the change in proton
mass in the light-cone kinetic energy as one varies a given quark
mass~\cite{Weisberger:1972hk}:
\begin{equation}
{\partial M^2_p\over \partial m^2_q} = \VEV{\frac{1}{x_q}} .
\end{equation}

The $J=0$ contribution can be isolated in various ways.  Damashek
and Gilman found a signal in forward high
energy Compton scattering using analyticity and the optical
theorem~\cite{Damashek:1969xj}. They found that the phenomenological
value for the $J=0$ term in the forward Compton amplitude was
surprisingly close to the value $\sum_q e^2_q F^q_0(t) = 1$, the same
result that one has from Thompson scattering on an elementary proton. Its
role in large $t$ real Compton scattering was studied in
Ref.~\cite{Brodsky:1971zh}.

The $J=0$ term also plays an important role in recent
work~\cite{Chen:2004tw} correcting Rosenbluth from two-photon
exchange because it  has a real phase and a $\pi^2$ enhancement.

A debate has continued on whether processes such as the pion and
proton form factors and elastic Compton scattering $\gamma p \to
\gamma p$ might be dominated by higher twist mechanisms until very
large momentum
transfers~\cite{Isgur:1989iw,Radyushkin:1998rt,%
Bolz:1996sw,Diehl:1998kh,Huang:2001ej}. For example, if one assumes
that the light-cone wavefunction of the pion has the form $\psi_{\rm
soft}(x,k_\perp) = A \exp \left(-b {k_\perp^2\over x(1-x)}\right)$,
then the Feynman endpoint contribution to the overlap integral at
small $k_\perp$ and $x \simeq 1$ will dominate the form factor
compared to the hard-scattering contribution until very large $Q^2$.
However, this form for $\psi_{\rm soft}(x,k_\perp)$ does not
fall-off at all for $k_\perp = 0$ and $k_z \to - \infty$.  A soft
QCD wavefunction would be expected to be exponentially suppressed in
this regime, as in the BHL model $\psi^{\rm soft}_n(x_i, k_{\perp
i}) = A\exp \left(- b\ \sum^n_i [{{\vec k}^2_\perp + m^2\over
x}]_i\right)$~\cite{Lepage:1982gd}. The endpoint contributions are
also suppressed by a QCD Sudakov form factor~\cite{Li:1992nu},
reflecting the fact that a near-on-shell quark must radiate if it
absorbs large momentum.  If the endpoint contribution dominates
proton Compton scattering, then both photons will interact on the
same quark line in a local fashion and the amplitude is real, in
strong contrast to the QCD predictions which have a complex phase
structure.  The perturbative QCD
predictions~\cite{Kronfeld:1991kp,Guichon:1998xv,Brooks:2000nb} for
the Compton amplitude phase can be tested in virtual Compton
scattering by interference with Bethe-Heitler
processes~\cite{Brodsky:1972vv}.

The ``handbag" approximation to Compton
scattering~\cite{Diehl:1998kh,Huang:2001ej} has been applied to
$\gamma \gamma \to p \bar p$ and $\bar p p \to \gamma \gamma$
reactions at large energy~\cite{Weiss:2002ec}. In this case, one
assumes that the process occurs via the exchange of a diquark with
light-cone momentum fraction $x \sim 0,$ so that the hard
subprocess is $\bar q q \to \gamma \gamma$ where nearly on-shell quarks
annihilate with the full energy of the baryons. The critical question is
whether the proton wavefunction has significant support when the massive
diquark has zero light-front momentum fraction, since the diquark
light-cone kinetic energy and the bound state wavefunction become far-off
shell $k_F^2 \sim - (m^2+k^2_\perp)/x \to - \infty $ in this
domain.

Measurements of timelike Compton scattering in $\gamma \gamma \to
p \bar p$ and $\bar p p \to \gamma \gamma$ at large $s$ and $t$
are thus critical for settling these issues.

\section{Hadron Helicity Conservation}

Hadron helicity conservation (HHC) is a QCD selection rule
concerning the behavior of helicity amplitudes at high momentum
transfer, such as fixed CM scattering. Since the convolution of
$T_H$ with the light-cone wavefunctions projects out states with
$L_z=0$, the leading hadron amplitudes conserve hadron
helicity~\cite{Brodsky:1981kj,Chernyak:1999cj}.  Thus the dominant
amplitudes are those in which the sum of hadron helicities in the
initial state equals the sum of hadron helicities in the final
state; other helicity amplitudes are relatively suppressed by an
inverse power in the momentum transfer.

In the case of electron-proton scattering, hadron helicity
conservation states that the proton helicity-conserving form
factor ( which is proportional to $G_M$) dominates over the proton
helicity-flip amplitude  (proportional to $G_E/\sqrt \tau $) at
large momentum transfer.  Here $\tau = Q^2/4M^2, Q^2 = -q^2.$ Thus
HHC predicts ${G_E(Q^2) / \sqrt \tau G_M(Q^2)} \to 0 $ at large
$Q^2.$  The new data from Jefferson Laboratory~\cite{Jones:uu}
which shows a decrease in the ratio ${G_E(Q^2)/  G_M(Q^2)} $ is
not itself in disagreement with the HHC prediction.

The leading-twist QCD motivated form $Q^4 G_M(Q^2) \simeq {{\rm
const} / Q^4 \ln Q^2\Lambda^2}$ provides a good guide to both the
time-like and spacelike proton form factor data at $Q^2
> 5$ GeV$^2$~\cite{Ambrogiani:1999bh}.   The Jefferson
Laboratory data~\cite{Jones:uu} appears to suggest $Q
F_2(Q^2)/F_1(Q^2) \simeq {\rm const},$  for the ratio of the
proton's  Pauli and Dirac form factors in contrast to the fall-off
$Q^2 F_2(Q^2)/F_1(Q^2) \simeq {\rm const}$ (modulo logarithms)
expected from PQCD. It should however be noted that a PQCD-motivated
fit is not precluded.  For example, the form~\cite{Brodsky:2003gs}
\begin{equation}
{F_2(Q^2)\over F_1(Q^2)} = {\mu_A \over 1 + (Q^2/c) \ln^b(1+ Q^2/a)}
\end{equation}
 with
$\mu_A = 1.79,$ $a = 4 m^2_\pi = 0.073~$GeV$^2,$ $ b = -0.5922,$ $
c = 0.9599~$GeV$^2$ which is consistent with leading-twist hadron
helicity conservation also fits the data well. More recently,
Belitsky, Ji and Yuan~\cite{Belitsky:2002kj} have demonstrated
that the perturbative QCD prediction has the asymptotic form $ Q^2
{F_2(Q^2)\over F_1(Q^2)} \sim \log^2 Q^2$ and also fits the data
well.

The study of time-like hadronic form factors using $\bar p p \to
\ell^+ \ell^-$ annihilation and  $e^+ e^-$ colliding beams can
provide very sensitive tests of HHC, since the virtual photon
always has spin $\pm 1$ along the lepton axis at high energies in
the CM system. Angular momentum conservation implies that the
virtual photon can ``decay" with one of only two possible angular
distributions in the center of momentum frame: $(1+ \cos^2\theta)$
for $\vert\lambda_A - \lambda_B \vert = 1$ and $\sin^2 \theta$ for
$\vert \lambda_A - \lambda_B \vert = 0$ where $\lambda_A$ and
$\lambda_B$ are the helicities of the outgoing hadrons.  Hadronic
helicity conservation, as required by QCD, greatly restricts the
possibilities.  It implies that $\lambda_A + \lambda_B = 0$.
Consequently, angular momentum conservation requires $\vert
\lambda_A\vert = \vert \lambda_B \vert = l/2$ for baryons, and
$\vert \lambda_A\vert = \vert \lambda_B \vert = 0$ for mesons;
thus the angular distributions for any sets of hadron pairs are
now completely determined at leading twist: $ {d \sigma \over d
\cos\theta}(e^+ e^- = B \bar B) \propto 1 + \cos^2 \theta $ and $
{d \sigma \over d \cos \theta} (e^+ e^- = M \bar M) \propto \sin^2
\theta  .$ Verifying these angular distributions for vector mesons
and other higher spin mesons and baryons would verify the vector
nature of the gluon in QCD and the validity of PQCD applications
to exclusive reactions.

It is usually assumed that a heavy quarkonium state such as the
$J/\psi$ always decays to light hadrons via the annihilation of
its heavy quark constituents to gluons.  However, as Karliner and
I~\cite{Brodsky:1997fj} have shown, the transition $J/\psi \to
\rho \pi$ can also occur by the rearrangement of the $c \bar c$
from the $J/\psi$ into the $\ket{ q \bar q c \bar c}$ intrinsic
charm Fock state of the $\rho$ or $\pi$.  On the other hand, the
overlap rearrangement integral in the decay $\psi^\prime \to \rho
\pi$ will be suppressed since the intrinsic charm Fock state
radial wavefunction of the light hadrons will evidently not have
nodes in its radial wavefunction.  This observation provides a
natural explanation of the long-standing puzzle why the $J/\psi$
decays prominently to two-body pseudoscalar-vector final states in
conflict with HHC, whereas the $\psi^\prime$ does not. If the
intrinsic charm explanation is correct, then this mechanism will
complicate the analysis of virtually all heavy hadron decays such
as $J/\psi \to p \bar p.$ In addition, the existence of intrinsic
charm Fock states, even at a few percent level, provides new,
competitive decay mechanisms for $B$ decays which are nominally
CKM-suppressed~\cite{Brodsky:2001yt}. For example, the weak decays
of the B-meson to two-body exclusive states consisting of strange
plus light hadrons, such as $B\to\pi K,$ are expected to be
dominated by penguin contributions since the tree-level $b\to s
u\bar u$ decay is CKM suppressed. However, higher Fock states in
the B wave function containing charm quark pairs can mediate the
decay via a CKM-favored $b\to s c\bar c$ tree-level transition.
The presence of intrinsic charm in the $b$ meson can be checked by
the observation of final states containing three charmed quarks,
such as $B \to J/\psi D \pi$~\cite{Chang:2001yf}.

\section{Other Hard Exclusive Processes}

There are a large number of measured exclusive reactions in which
the empirical power law fall-off predicted by dimensional counting
and PQCD appears to be accurate over a large range of momentum
transfer.
\begin{figure}[htb]
\centering
\includegraphics[width=4.3in]{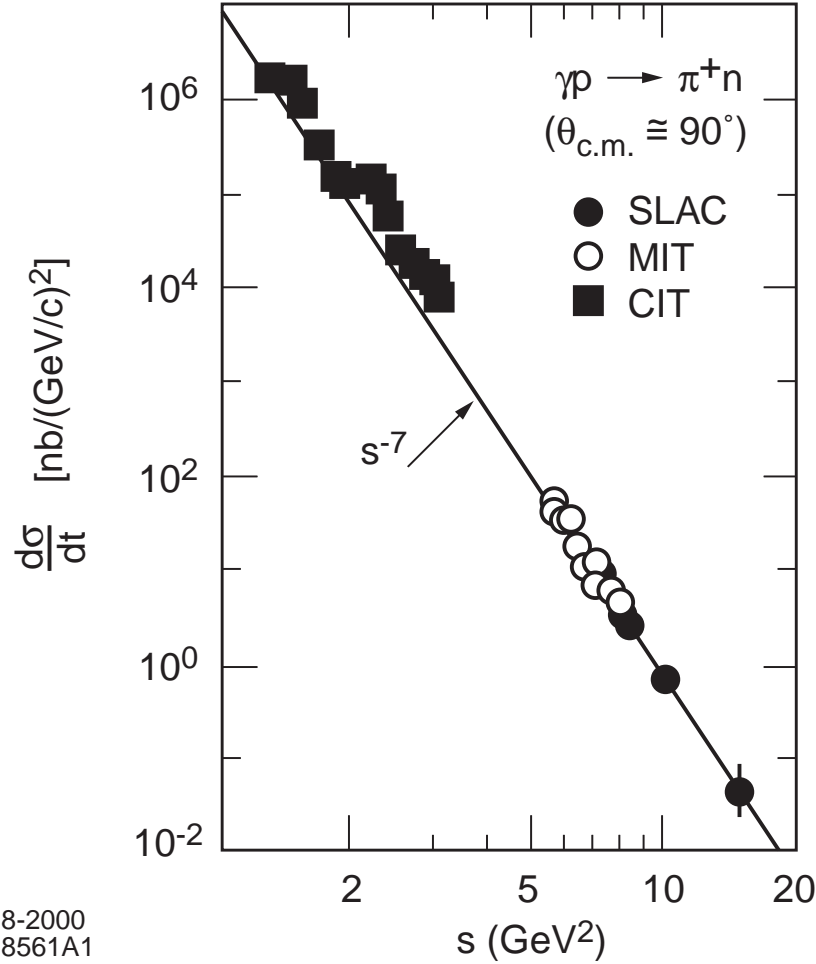}
\caption[ *]{Comparison of photoproduction data with the
dimensional counting power-law prediction.  The data are
summarized in Anderson {\em et al.}~\cite{Arr}} \label{figzkpic}
\end{figure}
The approach to scaling of $s^7 d\sigma/dt(\gamma p \to \pi^+ n)$
shown in Fig. \ref{figzkpic} appears to indicate that
leading-twist PQCD is applicable at momentum transfers exceeding a
few GeV.  If anything, the scaling appears to work too well,
considering that one expects logarithmic deviations due to the
running of the QCD coupling and the logarithmic evolution of the
hadron distribution amplitudes.   The deviations from scaling at
lower energies~\cite{Besch:sx} are interesting and can be
attributed to $s$-channel resonances or perhaps heavy quark
threshold effects, merging into the fixed-angle scaling in a
similar way as one observes the approach to leading-twist
Bjorken-scaling behavior in deep inelastic scattering via
quark-hadron duality~\cite{Melnitchouk:2001zy}.

The absence of
significant corrections to leading-twist scaling suggests that the
running coupling is effectively frozen at the kinematics relevant
to the data.   If higher-twist soft processes are conspiring to
mimic leading-twist scaling $s^7 d\sigma/dt(\gamma p \to \pi^+n)$,
then we would have the strange situation of seeing two separate
kinematic domains of $s^7$ scaling of the photoproduction cross
section. It has been argued~\cite{Radyushkin:1998rt,Diehl:2001fv}
that the Compton amplitude is dominated by soft end-point
contributions of the proton wavefunctions where the two photons
both interact on a quark line carrying nearly all of the proton's
momentum.  However, a corresponding soft endpoint explanation of
the observed $s^7 d\sigma/dt(\gamma p \to \pi^+ n)$ scaling of the
pion photoproduction data is not apparent;  there is no endpoint
contribution which could explain the success of dimensional
counting in large-angle pion photoproduction apparent in Fig.
\ref{figzkpic}.

Exclusive two-photon processes where two photons annihilate into
hadron pairs $\gamma \gamma \to H \bar H$ at high transverse
momentum  provide highly valuable probes of coherent effects in
quantum chromodynamics.  For example, in the case of exclusive
final states at high momentum transfer and fixed $\theta_{cm}$
such as $\gamma \gamma \rightarrow p \bar p $ or meson pairs,
photon-photon collisions provide a time-like microscope for
testing fundamental scaling laws of PQCD and for measuring
distribution amplitudes. Counting rules predict asymptotic
fall-off $s^4 d\sigma/dt \sim f(t/s)$ for meson pairs and  $s^6
d\sigma/dt \sim f(t/s)$ for baryon pairs.  Hadron-helicity
conservation predicts dominance of final states with $\lambda_H +
\lambda_{\bar H} = 0.$ The angular dependence reflects the
distribution amplitudes.  One can also study $\gamma^* \gamma \to
$ hadron pairs in $ e^\pm e^-$ collisions as a function of photon
virtuality, the time-like analog of deeply virtual Compton
scattering which is sensitive to the two hadron distribution
amplitude.  One can also study the interference of the time-like
Compton amplitude with the bremsstrahlung amplitude $e^\pm e \to B
B e^\pm e^-$ where a time-like photon produces the pair.  The
$e^\pm$ asymmetry measures the relative phase of the time-like
hadron form factor with that of the virtual Compton amplitude.

\begin{figure}[htb]
\centering
\includegraphics[width=4.3in]
{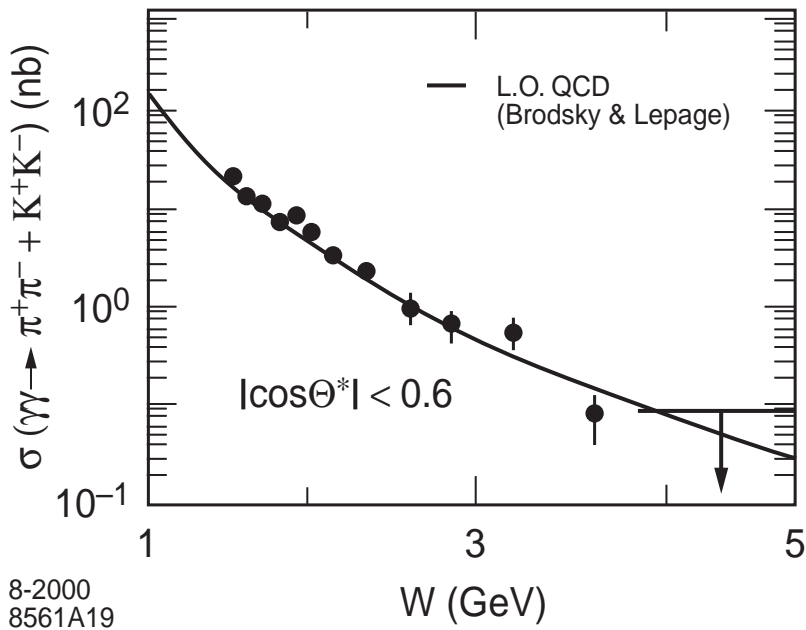} \caption[*]{Comparison of the sum of $\gamma \gamma
\rightarrow \pi^+ \pi^-$ and $\gamma \gamma \rightarrow K^+ K^-$
meson pair production cross sections with the perturbative QCD
prediction~\cite{Brodsky:1981rp} normalized to the timelike pion
form factor. The data are from the CLEO
collaboration~\cite{Savinov:2001wj}.} \label{Fig:CLEO}
\end{figure}

The PQCD predictions for the two-photon production of charged
pions and kaons is insensitive to the shape of the meson
distribution amplitudes.  In fact, the ratio of the $\gamma \gamma
\to \pi^+ \pi^-$ and  $e^+ e^- \to \mu^+ \mu^-$ amplitudes at
large $s$ and fixed $\theta_{CM}$ can be predicted since the ratio
is nearly insensitive to the running coupling and the shape of the
pion distribution amplitude:
\begin{equation}
{{d\sigma \over dt }(\gamma \gamma \to \pi^+ \pi^-) \over {d\sigma
\over dt }(\gamma \gamma \to \mu^+ \mu^-)} \sim {4 \vert F_\pi(s)
\vert^2 \over 1 - \cos^2 \theta_{\rm c.m.} } .\end{equation} The
comparison of the PQCD prediction for the sum of $\pi^+ \pi^-$
plus $K^+ K^-$ channels with CLEO data~\cite{Savinov:2001wj} is
shown in Fig. \ref{Fig:CLEO}. Results for separate pion and kaon
channels have been given by the TPC/$2\gamma$
collaboration~\cite{Boyer}.
The angular distribution of meson pairs is also predicted by PQCD
at large momentum transfer. The CLEO data for charged pion and
kaon pairs show a clear transition to the angular distribution
predicted by PQCD for $W = \sqrt s_{\gamma \gamma} > 2$ GeV.
Similarly in $\gamma \gamma \to p \bar p$ one can see a dramatic
change in the fixed angle distribution as one enters the hard
scattering domain. It is clearly important to measure the
two-photon production of neutral pions and $\rho^+ \rho^-$ cross
sections in view of their strong sensitivity to the shape of meson
distribution amplitudes. Furthermore, the ratio of $\pi^+ \pi^-$
to $\pi^0 \pi^0$ cross sections is highly sensitive to the
production dynamics. The ratio ${\sigma(\gamma \gamma \to \pi^0
\pi^0)/\sigma(\gamma \gamma \to \pi^+ \pi^-)}$ at fixed
angles is predicted to be very small in PQCD; in contrast, this ratio is
of
$\cal O$
$1$ in soft handbag models.

An interesting contribution to $K^+ p \to K^+ p$ scattering comes
from the exchange of the common $u$ quark.  The  quark interchange
amplitude for $A + B \to  C + D$ can be written as a convolution
of the four light-cone wavefunctions multiplied by a factor
$\Delta^- = P^-_A + P^-_B - \sum_i k^-_i,$  the inverse of the
central propagator~\cite{Gunion:1973ex}. The interchange amplitude
is consistent with constituent counting rule scaling, and often
provides a phenomenologically accurate representation of the
$\theta_{c.m.}$ angular distribution at large momentum transfer.
For example, the angular distribution of processes such as $K^+ p
\to K^+ p$ appear to follow the predictions based on quark interchange,
\eg, $T_H((u_1
\bar s) (u_2 u_3 d) \to (u_2 \bar s) (u_1 u_3
d)$~\cite{Gunion:1973ex}. This mechanism also provides constraints
on Regge intercepts $\alpha_R(t)$ for  meson exchange trajectories
at large momentum transfer~\cite{Blankenbecler:1973kt}.  An
extensive review of this phenomenology is given in the review by
Sivers {\em et al.}~\cite{Sivers:1976dg}

it is also interesting to study
amplitudes where a nuclear wavefunction has to absorb large
momentum transfer. For example, the helicity-conserving deuteron
form factor is predicted to scale as $F_d(Q^2) \propto (Q^2)^{-5}$
reflecting the minimal six quark component of nuclear
wavefunction. The deuteron form factor at high $Q^2$ is sensitive
to wavefunction configurations where all six quarks overlap within
an impact separation $b_{\perp i} < \mathcal{O} (1/Q).$  The
leading power-law fall off predicted by QCD is $F_d(Q^2) =
f(\alpha_s(Q^2))/(Q^2)^5$, where,
asymptotically~\cite{Brodsky:1976rz,Brodsky:1983vf},
$f(\alpha_s(Q^2)) \propto \alpha_s(Q^2)^{5+2\gamma}$.  In general,
the six-quark wavefunction of a deuteron is a mixture of five
different color-singlet states.  The dominant color configuration
at large distances corresponds to the usual proton-neutron bound
state. However, at small impact space separation, all five Fock
color-singlet components eventually acquire equal weight, \ie, the
deuteron wavefunction evolves to 80\%\ ``hidden
color''\cite{Brodsky:1983vf}. The relatively large normalization
of the deuteron form factor observed at large $Q^2$ hints at
sizable hidden-color contributions~\cite{Farrar:1991qi}.

\begin{figure}[htb]
\centering
\includegraphics[width=4.3in]
{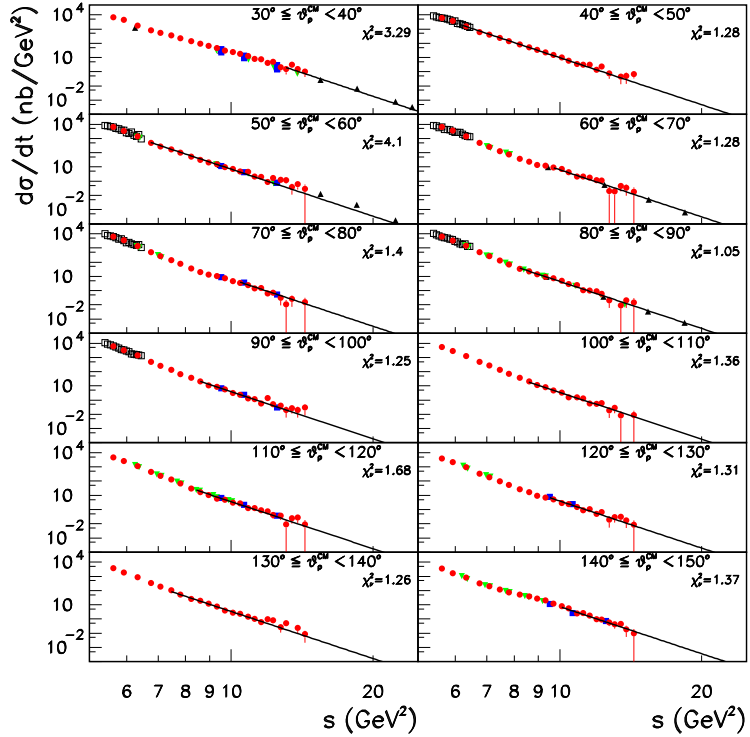} \caption{Fits of the cross sections $d\sigma/dt$ to
$s^{-11}$ for $P_T \ge P_T^{th}$ and proton angles between
$30^{\circ}$ and $150^{\circ}$ (solid lines). Data are from CLAS
(full/red circles), Mainz(open/black squares), SLAC
(full-down/green triangles), JLab Hall~A (full/blue squares) and
Hall~C (full-up/black triangles). Also shown in each panel is the
$\chi^2_\nu$ value of the fit. From Ref.~\cite{Rossi:2004qm}.}
\label{fig:fitfin}
\end{figure}

Hidden color components can also play a predominant role in the
reaction $\gamma d \to J/\psi p n$ at threshold if it is dominated
by the multi-fusion process $\gamma g g \to J/\psi$.  In the case
of nuclear structure functions beyond the single nucleon kinematic
limit, $1 < x_{bj} < A$, the nuclear light-cone momentum must be
transferred to a single quark, requiring quark-quark correlations
between quarks of different nucleons in a compact, far-off-shell
regime. This physics is also sensitive to the part of the nuclear
wavefunction which contains hidden-color components in distinction
from a convolution of separate color-singlet nucleon
wavefunctions. One also sees the onset of the predicted
perturbative QCD scaling behavior for exclusive nuclear amplitudes
such as deuteron photodisintegration (Here $n = 1+ 6 + 3 + 3 = 13 .$)
$s^{11}{ d\sigma\over dt}(\gamma d \to p n) \sim $ constant at
fixed CM angle. The measured deuteron form factor and the deuteron
photodisintegration cross section appear to follow the
leading-twist QCD predictions at large momentum transfers in the
few GeV region~\cite{Holt:1990ze,Bochna:1998ca,Rossi:2004qm}. A
comparison of the data with the QCD predictions is shown in
Fig.~\ref{fig:fitfin}.

In the case of the deuteron form factor, the proton and neutron
share the deuteron's momentum equally to first approximation.
Since the deuteron form factor contains the probability amplitudes
for the proton and neutron to scatter from $p/2$ to $p/2+q/2$, it
is natural to define the reduced deuteron form
factor~\cite{Brodsky:1976rz,Brodsky:1983vf}
\begin{equation}
f_d(Q^2) \equiv \frac{F_d(Q^2)}{F_{1N} \left(Q^2\over 4\right)\,
F_{1N}\,\left(Q^2\over
4\right)}\ .
\end{equation}
The effect of nucleon compositeness is
removed from the reduced form factor. QCD then predicts the
scaling
\begin{equation}
f_d(Q^2) \sim {1\over Q^2} ;
\end{equation}
\ie\ the same scaling law as a meson form factor. This scaling is
consistent with experiment for $Q \gsim$ 1 GeV. In fact as seen in
Fig.~\ref{reduced}, the deuteron reduced form factor contains two
components: (1) a fast falling component characteristic of nuclear
binding with probability $85\%$, and (2) a hard contribution
falling as a monopole with a scale of order $0.5~{\rm GeV}$  with
probability $15\%.$

\begin{figure}[htb]
\centering
\includegraphics[width=4.3in]
{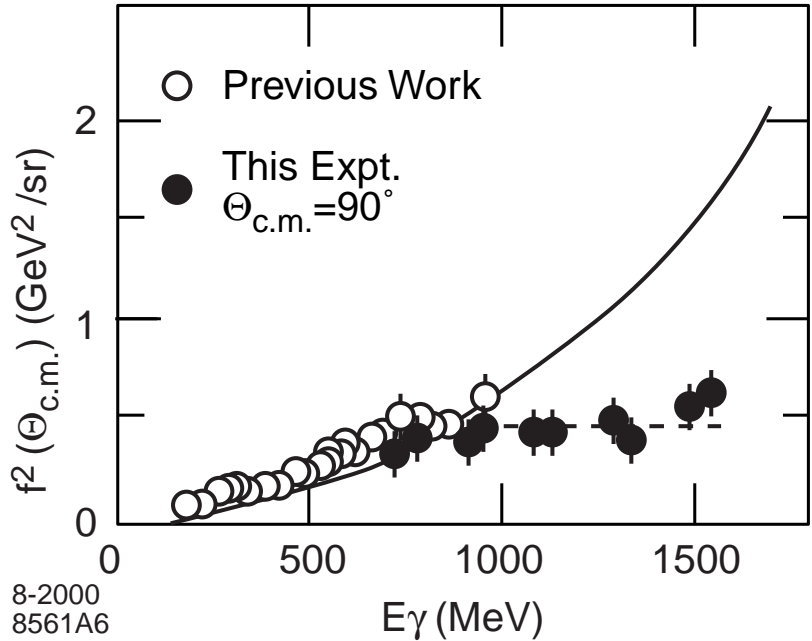} \caption[*]{ Comparison of deuteron
photodisintegration data with the scaling prediction which
requires $f^2(\theta_{cm})$ to be at most logarithmically
dependent on energy at large momentum transfer.  The data in are
from Belz {\em et al.}\cite{Belz:1995ge} The solid curve is a
nuclear physics prediction~\cite{Lee:1988pi}.} \label{figyqpic}
\end{figure}

In the case of deuteron photodisintegration $\gamma d \to p n$ the
amplitude requires the scattering of each nucleon at $t_N =
t_d/4$.  The perturbative QCD scaling is~\cite{Brodsky:1983kb}
\begin{equation}
{d\sigma\over d\Omega_{c.m.}} (\gamma d \to n p) = {1\over \sqrt {s (s-M_d^2)}}
 {F^2_n(t_d/4)  F^2_p(t_d/4) f^2_{red}(\theta_{c.m})\over p^2_\perp} .
\end{equation}
The predicted scaling of the reduced photodisintegration amplitude
$f_{red}(\theta_{c.m.})  \simeq $ const  is also consistent with
experiment~\cite{Brodsky:1983kb,Holt:1990ze,Bochna:1998ca}.  See
Fig. \ref{figyqpic}.

The observation of conformal scaling
behavior~\cite{Brodsky:1976rz} in exclusive deuteron processes
such as deuteron photoproduction~\cite{Rossi:2004qm}  and the
deuteron form factor~\cite{Arnold:1975dd} is particularly
interesting. For example, at high $Q^2$ the deuteron form factor
is sensitive to wavefunction configurations where all six quarks
overlap within an impact separation $b_{\perp i} < {\cal O}
(1/Q).$ In general, the six-quark wavefunction of a deuteron is a
mixture of five different color-singlet states. The dominant color
configuration at large distances corresponds to the usual
proton-neutron bound state. However at small impact space
separation, all five Fock color-singlet components eventually
acquire equal weight, \ie, the deuteron wavefunction evolves to
80\%\ ``hidden color.'' The derivation of the evolution equation
for the deuteron distribution amplitude and its leading anomalous
dimension $\gamma$ is given in Ref.~\cite{Brodsky:1983vf}. As
emphasized in the introduction, the relatively large normalization
of the deuteron form factor observed at large
$Q^2$~\cite{Farrar:1991qi}, as well as the presence of two mass
scales in the scaling behavior of the reduced deuteron form
factor~\cite{Brodsky:1976rz} $f_d(Q^2)= F_d(Q^2)/F^2(Q^2/4)$
suggests sizable hidden-color contributions in the deuteron
wavefunction. See Fig. \ref{reduced}.

The postulate that the QCD coupling has an infrared fixed-point
provides additional understanding of the applicability of
conformal scaling and constituent counting rules to physical QCD
processes~\cite{Brodsky:1974vy,Matveev:1973ra}. The general
success of dimensional counting rules implies that the effective
coupling $\alpha_V(Q^*)$ controlling the gluon exchange
propagators in $T_H$ are frozen in the infrared, since the
effective momentum transfers $Q^*$ exchanged by the gluons are
often a small fraction of the overall momentum
transfer~\cite{Brodsky:1998dh}. In this case, the pinch
contributions are suppressed by a factor decreasing faster than a
fixed power~\cite{Brodsky:1974vy}. The effective coupling
$\alpha_\tau(s)$ extracted from $\tau$ decays displays a flat
behavior at low mass scales~\cite{Brodsky:2002nb}.

The field of analyzable exclusive processes has been expanded to a
wide range of QCD processes, such as electroweak decay amplitudes,
highly virtual diffractive processes such as $\gamma^* p \to \rho
p$~\cite{Brodsky:1994kf,Collins:1997sr}, and semi-exclusive
processes such as $\gamma^* p \to \pi^+ X$~\cite{Brodsky:1998sr}
where the $\pi^+$ is produced in isolation at large $p_T$.  An
important new application is the recent analysis of hard exclusive
$B$ decays by Beneke {\em et al.}~\cite{Beneke:2000ry} and Keum
{\em et al.}~\cite {Keum:2000wi}

Deeply virtual Compton amplitude $\gamma^* p \to \gamma p$ has
emerged as one of the most important exclusive QCD
reactions~\cite{Ji:1997nm,Radyushkin:1997ki,Diehl:1999tr,Diehl:1999kh}.
The process factorizes into a hard amplitude representing Compton
scattering on the quark convoluted with the skewed parton distributions.
The resulting amplitudes can be represented as
diagonal and off-diagonal convolutions of light-cone
wavefunctions, as in semileptonic $B$ decay~\cite{Brodsky:2000xy}.
New sum rules can be constructed which correspond to gravitons
coupling to the quarks of the proton~\cite{Ji:1997nm}.  It is
possible that the handbag approximation to DVCS may be modified by
corrections to the quark propagator similar to those which appear
in the final state interaction corrections to deep inelastic
scattering~\cite{Brodsky:2002ue,Brodsky:2000ii}.  In particular,
one can expect that the propagator corrections will give
single-spin asymmetries correlating the spin of the proton with
the normal to the production plane in DVCS~\cite{Brodsky:2002cx}.

The hard diffraction of vector mesons $\gamma^* p \to V^0 p$ at
high $Q^2$ and high energies for longitudinally polarized vector
mesons factorizes into a skewed parton distribution convoluted with the
hard scale $\gamma^* g \to g V^0$ amplitude, where the physics of the
vector meson is contained in its distribution
amplitude~\cite{Brodsky:1994kf,Muller:1994cn,Collins:1997fb}. The
data appears consistent with the $s,t$ and $Q^2$ dependence
predicted by  theory.  Ratios of these processes for different
mesons are sensitive to the ratio  of $1/x$ moments of the $V^0$
distribution amplitudes.

The virtual two-photon annihilation process $\gamma^* \gamma \to $
hadrons, which is measurable in single-tagged $e^+ e^- \to e^+ e^-
{\rm hadrons}$ events, provides a semi-local probe of even charge
conjugation $C=^+$ hadron systems $\pi^0, \eta^0, \eta^\prime,
\eta_c, \pi^+ \pi^-$, etc.   The   $\gamma^* \gamma \to \pi^+
\pi^-$ hadron pair process is related to virtual Compton
scattering on a pion target by crossing.  Hadron pair production
is of particular interest since the leading-twist amplitude is
sensitive to the $1/x - 1/(1-x)$ moment of the two-pion
distribution amplitude coupled to two valence
quarks~\cite{Muller:1994cn,Diehl:2000uv}.   This type of
measurement can also constrain the parameters of the effective
chiral theory, which is needed for example to constrain the
hadronic light-by-light contribution to the muon magnetic
moment~\cite{Ramsey-Musolf:2002cy}.

One can also study hard ``semi-exclusive''
processes~\cite{Brodsky:1998sr} of the form $A+B \to C + Y$ which
are characterized by a large momentum transfer between the
particles $A$ and $C$ and a large rapidity gap between the final
state particle $C$ and the inclusive system $Y$. Such reactions
are in effect generalizations of deep inelastic lepton scattering,
providing novel currents which probe specific quark distributions
of the target $B$ at fixed momentum fraction and novel
spin-dependent parton distributions.

\section{Heavy Quark Components of the Proton Structure Function}

In the simplest treatment of deep inelastic scattering, nonvalence
quarks are produced via gluon splitting and DGLAP evolution.
However, in a full theory heavy quarks are multiply-connected to
the valence quarks~\cite{Brodsky:1980pb}. For example, the
asymmetry of the strange and anti-strange distributions in the
nucleon is due to their different interactions with the other
quark constituents.   The probability for  Fock states of a light
hadron such as the proton to have an extra heavy quark pair
decreases as $1/m^2_Q$ in non-Abelian gauge
theory~\cite{Franz:2000ee,Brodsky:1984nx}.  The relevant matrix
element is the cube of the QCD field strength $G^3_{\mu nu}.$ This
is in contrast to abelian gauge theory where the relevant operator
is $F^4_{\mu \nu}$ and the probability of intrinsic heavy leptons
in QED bound state is suppressed as $1/m^4_\ell.$ The intrinsic
Fock state probability is maximized at minimal off shellness. The
maximum probability occurs at $x_i = { m^i_\perp / \sum^n_{j = 1}
m^j_\perp}$; \ie, when the constituents have equal rapidity. Thus
the heaviest constituents have the highest momentum fractions and
highest $x$. Intrinsic charm thus predicts that the charm
structure function has support at large $x_{bj}$  in excess of
DGLAP extrapolations~\cite{Brodsky:1980pb}; this is in agreement
with the EMC measurements~\cite{Harris:1995jx}.  It predicts
leading charm hadron production and fast charmonium production in
agreement with measurements~\cite{Anjos:2001jr}.   The production
cross section for the double charmed $\Xi_{cc}^+$
baryon~\cite{Ocherashvili:2004hi} and the production of double
$J/\psi's$ appears to be consistent with the dissociation and
coalescence of double IC Fock states~\cite{Vogt:1995tf}. Intrinsic
charm can also explain the $J/\psi \to \rho \pi$
puzzle~\cite{Brodsky:1997fj}. It also affects the extraction of
suppressed CKM matrix elements in $B$
decays~\cite{Brodsky:2001yt}. It is thus critical for new
experiments (HERMES, HERA, COMPASS) to definitively establish the
phenomenology of  the charm structure function at large $x_{bj}.$

Since the intrinsic charm quarks have a relatively hard
distribution in the nucleon, one expects enhanced open and hidden
charm production near the kinematic threshold. This will be
particularly interesting to study in the new GSI antiproton
facility.

\section{The Strange Quark Asymmetry}

Although the strange and antistrange distributions in the nucleon
are identical when they derive from gluon-splitting $g \to s \bar
s$, this is not the case when the strange quarks are part of the
intrinsic structure of the nucleon.  There is a simple analog in
QED: Consider the $\tau^\pm$ distributions in the (rare!)
$\ket{e^- \mu^+ \tau^+ \tau^-}$ Fock state of muonium $(\mu^+
e^-).$  The $\tau^- $ is attracted by Coulomb interactions to the
high rapidity $\mu+.$ Thus the $\tau^-$ will tend to have higher
rapidity than the $\tau^+.$

Similar effects will  happen in QCD. If we use the diquark model
$\ket p  \sim \ket{u_{3_c} (ud)_{\bar 3_C}},$ then the $Q_{3_C}$
in the $\ket{u (ud) Q \bar Q }$ Fock state will be attracted to
the heavy diquark and thus have higher rapidity than the $\bar Q$.

An alternative model is the $\ket{K \Lambda}$ fluctuation model
for the $\ket{uud s \bar s}$ Fock state of the
proton~\cite{Brodsky:1996hc}. The $s$ quark tends to have higher
$x$.

The experimentally observed asymmetry~\cite{Portheault:2004xy}
appears to be small but positive:\break $\int dx x [s(x)- \bar
s(x)]$. The $\bar s(x)-s(x)$ asymmetry can be studied in detail in
$p \bar p$ collisions by searching for antisymmetric
forward-backward strange quark distributions in the $\bar p-p$ CM
frame.

\section{The Infrared Behavior of  Effective QCD Couplings}

It is often assumed that color confinement in QCD can be traced to
the singular behavior of the running coupling in the infrared,
{\em i.e.} ``infrared slavery." For example, if $\alpha_s(q^2) \to
{1}/{q^2}$ at $q^2 \to 0$, then one-gluon exchange leads to a
linear potential at large distances.  However,
theoretical~\cite{vonSmekal:1997is,Zwanziger:2003cf,Howe:2002rb,Howe:2003mp,%
Furui:2003mz} and
phenomenological~\cite{Mattingly:ej,Brodsky:2002nb,Baldicchi:2002qm}
evidence is now accumulating that the QCD coupling becomes
constant at small virtuality; {\em i.e.}, $\alpha_s(Q^2)$ develops
an infrared fixed point in contradiction to the usual assumption
of singular growth in the infrared.  Since all observables are
related by commensurate scale relations, they all should have an
IR fixed point~\cite{Howe:2003mp}.  A recent study of the QCD
coupling using lattice gauge theory in Landau gauge in fact shows
an infrared fixed point~\cite{Furui:2004bq}. This result is also
consistent with Dyson-Schwinger equation studies of the physical
gluon propagator~\cite{vonSmekal:1997is,Zwanziger:2003cf}.  The
relationship of these results to the infrared-finite coupling for
the vector interaction defined in the quarkonium potential has
recently been discussed by Badalian and
Veselov~\cite{Badalian:2004ig}.

One can define the fundamental coupling of QCD from virtually any
physical observable~\cite{Grunberg:1980ja}. Such couplings, called
``effective charges", are all-order resummations of perturbation
theory, so they  correspond to the complete theory of QCD. Unlike
the $\barMS$ coupling, a physical coupling is analytic across
quark flavor thresholds~\cite{Brodsky:1998mf,Brodsky:1999fr}.  In
particular, heavy particles will contribute to physical
predictions even at energies below their threshold. This is in
contrast to mathematical renormalization schemes such as ${\bar
MS},$ where mass thresholds are treated as step functions. In
addition, since the QCD running couplings defined from observables
are bounded, integrations over  effective charges are well defined
and the arguments requiring renormalon resummations do apply. The
physical couplings satisfy the standard renormalization group
equation for its logarithmic derivative, ${{\rm d}\alpha_{\rm
phys}/{\rm d}\ln k^2} = \widehat{\beta}_{\rm phys}[\alpha_{\rm
phys}(k^2)]$, where the first two terms in the perturbative
expansion of $\widehat{\beta}_{\rm phys}$ are scheme-independent
at leading twist; the higher order terms have to be calculated for
each observable separately using perturbation theory.

Menke, Merino, and Rathsman~\cite{Brodsky:2002nb} and I have
considered a physical coupling for QCD which is defined from the
high precision measurements of the hadronic decay channels of the
$\tau^- \to \nu_\tau {\rm h}^-$.  Let $R_{\tau}$ be the ratio of
the hadronic decay rate to the leptonic rate.  Then
$R_{\tau}\equiv R_{\tau}^0\left[1+\frac{\alpha_\tau}{\pi}\right]$,
where $R_{\tau}^0$ is the zeroth order QCD prediction, defines the
effective charge $\alpha_\tau$.  The data for $\tau$ decays is
well-understood channel by channel, thus allowing the calculation
of the hadronic decay rate and the effective charge as a function
of the $\tau$ mass below the physical mass.  The vector and
axial-vector decay modes which can be studied separately. Using an
analysis of the $\tau$ data from the OPAL
collaboration~\cite{Ackerstaff:1998yj}, we have found that the
experimental value of the coupling $\alpha_{\tau}(s)=0.621 \pm
0.008$ at $s = m^2_\tau$ corresponds to a value of
$\alpha_{\MSbar}(M^2_Z) = (0.117$-$0.122) \pm 0.002$, where the
range corresponds to three different perturbative methods used in
analyzing the data.  This result is in good agreement with the
world average $\alpha_{\MSbar}(M^2_Z) = 0.117 \pm 0.002$.
However, from the figure we also see that the effective charge
only reaches $\alpha_{\tau}(s) \sim 0.9 \pm 0.1$ at $s=1\,{\rm
GeV}^2$, and it even stays within the same range down to
$s\sim0.5\,{\rm GeV}^2$. This result is in good agreement with the
estimate of Mattingly and Stevenson~\cite{Mattingly:ej} for the
effective coupling $\alpha_R(s) \sim 0.85 $ for $\sqrt s <
0.3\,{\rm GeV}$ determined from ${\rm e}^+{\rm e}^-$ annihilation,
especially if one takes into account the perturbative commensurate
scale relation, $\alpha_{\tau}(m_{\tau^\prime}^2)= \alpha_R(s^*),$
where $s^* \simeq 0.10\,m_{\tau^\prime}^2.$ This behavior is not
consistent with the coupling having a Landau pole, but rather
shows that the physical coupling is close to constant at low
scales, suggesting that physical QCD couplings are effectively
constant or ``frozen" at low scales.

\begin{figure}[htb]
\centering
\includegraphics[width=4.3in]   
{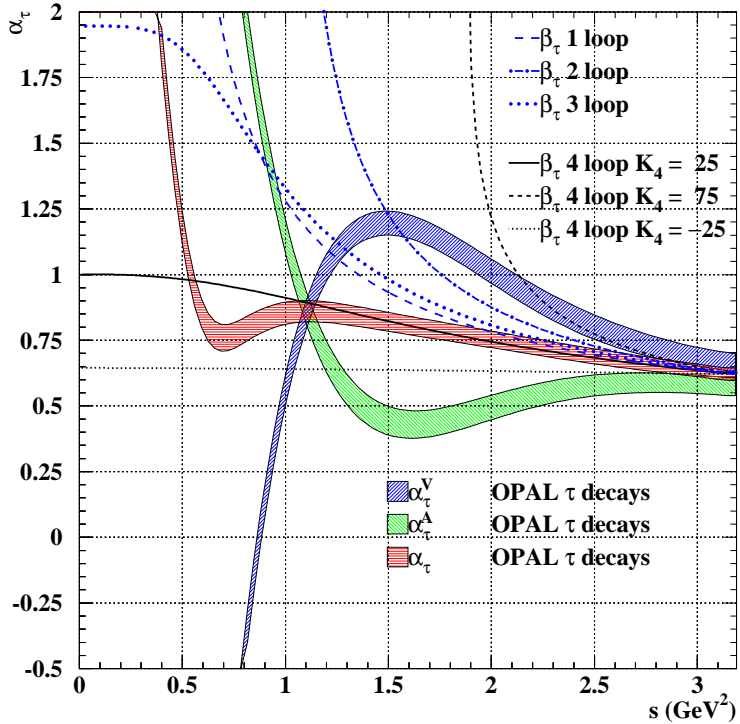} \caption[*]{The effective charge $\alpha_{\tau}$ for
non-strange hadronic decays of a hypothetical $\mathit{\tau}$
lepton with $\mathit{m_{\tau'}^2 = s}$ compared to solutions of
the fixed order evolution equation for $\alpha_{\tau}$ at two-,
three-, and four-loop order.  The error bands include statistical
and systematic errors. \label{fig:fopt_comp}}
\end{figure}

Figure~\ref{fig:fopt_comp} shows a comparison of the
experimentally determined effective charge $\alpha_{\tau}(s)$ with
solutions to the evolution equation for $\alpha_{\tau}$ at two-,
\hbox{three-,} and four-loop order normalized at $m_\tau$.  At
three loops the behavior of the perturbative solution drastically
changes, and instead of diverging, it freezes to a value
$\alpha_{\tau}\simeq 2$ in the infrared. The infrared behavior is
not perturbatively stable since the evolution of the coupling is
governed by the highest order term.  This is illustrated by the
widely different results obtained for three different values of
the unknown four loop term $\beta_{\tau,3}$ which are also shown.
The values of $\beta_{\tau,3}$ used are obtained from the estimate
of the four loop term in the perturbative series of $R_\tau$,
$K_4^{\overline{\rm MS}} = 25\pm 50$~\cite{LeDiberder:1992fr}. It
is interesting to note that the central four-loop solution is in
good agreement with the data all the way down to $s\simeq1\,{\rm
GeV}^2$.

The results for $\alpha_{\tau}$ resemble the behavior of the one-loop
``time-like"
effective coupling~\cite{Beneke:1994qe,Ball:1995ni,Dokshitzer:1995qm}
\begin{equation}\label{eq:alphaeff}
\alpha_{\rm eff}(s)=\frac{4\pi}{\beta_0} \left\{\frac{1}{2} -
\frac{1}{\pi}\arctan\left[\frac{1}{\pi}\ln\frac{s}{\Lambda^2}\right]\right\}
\end{equation}
which is finite in the infrared and freezes to the value $\alpha_{\rm
eff}(s)={4\pi}/{\beta_0}$ as $s\to 0$.  It is instructive to expand the
``time-like"
effective coupling for large $s$,
\begin{eqnarray}
\alpha_{\rm eff}(s) &=&\frac{4\pi}{\beta_0\ln\left(s/\Lambda^2\right)} \left\{1
-\frac{1}{3}\frac{\pi^2}{\ln^2\left(s/\Lambda^2\right)}
+\frac{1}{5}\frac{\pi^4}{\ln^4\left(s/\Lambda^2\right)} +\ldots \right\}
\nonumber\\
&=&\alpha_{\rm s}(s)\left\{1 -\frac{\pi^2\beta_0^2}{3}\left(\frac{\alpha_{\rm
s}(s)}{4\pi}\right)^2 +\frac{\pi^4\beta_0^4}{5}\left(\frac{\alpha_{\rm
s}(s)}{4\pi}\right)^4 +\ldots \right\}.
\end{eqnarray}
This shows that the ``time-like" effective coupling is a
resummation of $(\pi^2\beta_0^2\alpha_{\rm s}^2)^n$-corrections to
the usual running couplings.  The finite coupling $\alpha_{\rm
eff}$ given in Eq.~(\ref{eq:alphaeff}) obeys standard PQCD
evolution at LO.  Thus one can have a solution for the
perturbative running of the QCD coupling which obeys asymptotic
freedom but does not have a Landau singularity.

The near constancy of the effective QCD coupling at small scales
illustrates the near-conformal behavior of QCD.  It helps explain
the empirical success of dimensional counting rules for the power
law fall-off of form factors and fixed angle scaling.  As shown in
the references~\cite{Brodsky:1997dh,Melic:2001wb}, one can
calculate the hard scattering amplitude $T_H$ for such
processes~\cite{Lepage:1980fj} without scale ambiguity in terms of
the effective charge $\alpha_\tau$ or $\alpha_R$ using
commensurate scale relations.  The effective coupling is evaluated
in the regime where the coupling is approximately constant, in
contrast to the rapidly varying behavior from powers of
$\alpha_{\rm s}$ predicted by perturbation theory (the universal
two-loop coupling).  For example, the nucleon form factors are
proportional at leading order to two powers of $\alpha_{\rm s}$
evaluated at low scales in addition to two powers of $1/q^2$; The
pion photoproduction amplitude at fixed angles is proportional at
leading order to three powers of the QCD coupling.  The essential
variation from leading-twist counting-rule behavior then only
arises from the anomalous dimensions of the hadron distribution
amplitudes.

\section{The Role of Conformal Symmetry in QCD Phenomenology}

The classical Lagrangian of QCD for massless quarks is conformally
symmetric.  Since it has no intrinsic mass scale, the classical
theory is invariant under the $SO(4,2)$ translations, boosts, and
rotations of the Poincare  group, plus the dilatations and other
transformations of the conformal group.  Scale invariance, and
therefore conformal symmetry is destroyed in the quantum theory by
the renormalization procedure which introduces a renormalization
scale as well as by quark masses.  Conversely,
Parisi~\cite{Parisi:zy} has shown that perturbative QCD becomes a
conformal theory  for $\beta \to 0$ and zero quark mass.
Conformal symmetry is thus broken in physical QCD; nevertheless,
we can still recover the underlying features of the conformally
invariant theory by evaluating any expression in QCD in the
analytic limit of zero quark mass and zero $\beta$ function:
\begin{equation}
\lim_{m_q \to 0, \beta \to 0} \mathcal{O}_{QCD} = \mathcal{ O}_{\rm
conformal\ QCD} \ .
\end{equation} This conformal correspondence limit is analogous
to Bohr's correspondence principle where one recovers predictions
of classical theory from quantum theory in the limit of zero
Planck constant.  The contributions to an expression in QCD from
its nonzero $\beta$-function can be systematically
identified~\cite{Brodsky:2000cr,Rathsman:2001xe,Grunberg:2001bz}
order-by-order in perturbation theory using the Banks-Zaks
procedure~\cite{Banks:1981nn}.

There are a number of useful phenomenological consequences of near
conformal behavior of QCD: the conformal approximation with zero
$\beta$ function can be used as template for QCD
analyses~\cite{Brodsky:1985ve,Brodsky:1984xk} such as the form of
the expansion polynomials for distribution
amplitudes~\cite{Braun:2003rp,Braun:1999te}.  The near-conformal
behavior of QCD is the basis for commensurate scale
relations~\cite{Brodsky:1994eh} which relate observables to each
other without renormalization scale or scheme
ambiguities~\cite{Brodsky:2000cr,Rathsman:2001xe}.  By definition,
all contributions from the nonzero $\beta$ function can be
incorporated into the QCD running coupling $\alpha_s(Q)$ where $Q$
represents the set of physical invariants.  Conformal symmetry
thus provides a template for physical QCD expressions. For
example, perturbative expansions in QCD for massless quarks must
have the form
\begin{equation}
 \mathcal{O} = \sum_{n=0} C_n \alpha^n_s(Q^*_n)
 \end{equation}
where the $C_n$ are identical to the expansion coefficients in the
conformal theory, and $Q^*_n$ is the scale chosen to resum all of
the contributions from the nonzero $\beta$ function at that order
in perturbation theory.  Since the conformal theory does not
contain renormalons, the $C_n$ do not have the divergent $n\!$
growth characteristic of conventional PQCD expansions evaluated at
a fixed scale.

\section{The AFS/CFT Correspondence and Conformal Properties of Hadronic
Light-Front Wavefunctions}

As shown by Maldacena~\cite{Maldacena:1997re}, there is a
remarkable correspondence between large $N_C$ supergravity theory
in a higher dimensional  anti-de Sitter space and supersymmetric
QCD in 4-dimensional space-time.  String/gauge duality provides a
framework for predicting QCD phenomena based on the conformal
properties of the AdS/CFT correspondence.

The AdS/CFT correspondence is based on the fact that the generators of conformal and
Poincare transformations have representations on the five-dimensional anti-deSitter space
$AdS_5$  as well as Minkowski spacetime.  For example, Polchinski and
Strassler~\cite{Polchinski:2001tt} have shown that the power-law fall-off of hard
exclusive hadron-hadron scattering amplitudes at large momentum transfer can be derived
without the use of perturbation theory by using the scaling properties of the hadronic
interpolating fields in the large-$r$ region of  AdS space.  Thus one can use the
Maldacena correspondence to compute the leading power-law behavior of exclusive processes
such as high-energy fixed-angle scattering of gluonium-gluonium scattering in
supersymmetric QCD. The resulting predictions for hadron physics effectively
coincide~\cite{Polchinski:2001tt,Brower:2002er,Andreev:2002aw} with QCD dimensional
counting rules for form factors and hard scattering
amplitudes~\cite{Brodsky:1973kr,Matveev:ra,Brodsky:1974vy,Brodsky:2002st}. Polchinski and
Strassler~\cite{Polchinski:2001tt} have also derived counting rules for deep inelastic
structure functions at $x \to 1$ in agreement with perturbative QCD
predictions~\cite{Lepage:1980fj,Brodsky:1994kg} as well as Bloom-Gilman
exclusive-inclusive duality~\cite{Bloom:1971ye}.

The supergravity analysis is based on an extension of classical
gravity theory in higher dimensions and is nonperturbative.  Thus
analyses of exclusive processes~\cite{Lepage:1980fj} which were
based on perturbation theory can be extended by the Maldacena
correspondence to all orders.  An important point is that the hard
scattering amplitudes which are normally or order $\alpha_s^p$ in
PQCD appear as order $\alpha_s^{p/2}$ in the supergravity
predictions.  This can be understood as an all-orders resummation
of the effective potential~\cite{Maldacena:1997re,Rey:1998ik}.

The superstring theory results are derived in the limit of a large
$N_C$~\cite{'tHooft:1973jz}.  For gluon-gluon scattering, the
amplitude scales as ${1}/{N_C^2}$.   For color-singlet bound
states of quarks, the amplitude scales as ${1}/{N_C}$.  This large
$N_C$-counting, in fact, corresponds to the quark interchange
mechanism~\cite{Gunion:1973ex}.  For example, for $K^+ p \to K^+
p$ scattering, the $u$-quark exchange amplitude scales
approximately as $\frac{1}{u}$\ $\frac{1}{t^2},$ which agrees
remarkably well with the measured large $\theta_{CM}$ dependence
of the $K^+ p$ differential cross section~\cite{Sivers:1975dg}.
This implies that the nonsinglet Reggeon trajectory asymptotes to
a negative integer~\cite{Blankenbecler:1973kt}, in this case,
$\lim_{-t \to \infty}\alpha_R(t) \to -1.$

De Teramond and I have extended the Polchinski-Strassler analysis
to hadron-hadron scattering~\cite{Brodsky:2003px}.  We have also
shown how to compute the form and scaling of light-front hadronic
wavefunctions using the AdS/CFT correspondence in quantum field
theories which have an underlying conformal structure, such as
${\mathcal N} = 4$ super-conformal QCD. For example, baryons are
included in the theory by adding an open string sector in $AdS_5
\times S^5$ corresponding to quarks in the fundamental
representation of  the $SU(4)$ symmetry defined on $S^5$ and the
fundamental and higher representations of $SU(N_C).$ The hadron
mass scale is introduced by imposing boundary conditions at the
$AdS_5$ coordinate  $r= r_0 = \Lambda_{QCD} R^2.$ The quantum
numbers of the lowest Fock state of each hadron, including its
internal orbital angular momentum and spin-flavor symmetry, are
identified by matching the fall-off of the string wavefunction
$\Psi(x,r)$ at the asymptotic $3+1$ boundary.  Higher Fock states
are identified with conformally invariant quantum fluctuations of
the bulk geometry about the AdS background.  The eigenvalues of
the 10-dimensional Dirac and Rarita-Schwinger equations have also
been used to determine the  nucleon and $\Delta$ spectrum in
conformal QCD.  The results are in surprising agreement with the
empirical spectra~\cite{deTeramond:2004qd}.

One can also use the scaling properties of the hadronic
interpolating operator in the extended AdS/CFT space-time theory
to determine the scaling of light-front hadronic wavefunctions at
high relative transverse momentum. De Teramond and
I~\cite{Brodsky:2003px} have also shown how the angular momentum
dependence of the light-front wavefunctions also follow from the
conformal properties of the AdS/CFT correspondence. where $g_s$ is
the string scale and $\Lambda_o$ represents the basic QCD mass
scale.  Quantum fluctuations of the strings in the AdS radial
direction correspond to the quantum fluctuations of the hadron
wavefunctions due to orbital angular momentum and radial nodes in
the 3+1 theory.

The scaling and conformal properties of the AdS/CFT correspondence
leads to a hard component of light-front wavefunctions of the
form~\cite{Brodsky:2003px}:
\begin{eqnarray}
\psi_{n/h} (x_i, \vec k_{\perp i} , \lambda_i, l_{z i}) &\sim&
\frac{(g_s~N_C)^{\frac{1}{2} (n-1)}}{\sqrt {N_C}} ~\prod_{i =1}^{n
- 1} (k_{i \perp}^\pm)^{\vert l_{z i}\vert} ~ \nonumber \\[1ex]
&\times&\left[\frac{ \Lambda_o}{ {M}^2 - \sum _i\frac{\vec k_{\perp i}^2 +
m_i^2}{x_i} +
\Lambda_o^2}  \right] ^{n +\vert l_z \vert -1}, \label{eq:lfwfR}
\end{eqnarray}
where $g_s$ is the string scale and $\Lambda_o$ represents the
basic QCD mass scale.  The scaling predictions agree with
perturbative QCD analyses~\cite{Ji:bw,Lepage:1980fj}, but the
AdS/CFT analysis is performed at strong coupling without the use
of perturbation theory.  The near-conformal scaling properties of
light-front wavefunctions lead to a number of other predictions
for QCD which are normally discussed in the context of
perturbation theory, such as constituent counting scaling laws for
structure functions at $x \to 1$, as well as the leading power
fall-off of form factors and hard exclusive scattering amplitudes
for QCD processes.

\section{Applicability of PQCD and Conformal Symmetry to Hard Exclusive
Processes}

The PQCD/conformal symmetry predictions for hadron form factors
are leading-twist predictions.  The only mass parameter is the QCD
scale, so the power-law predictions must be relevant---up to
logarithms---even in the few GeV domain.  Note also that the same
PQCD couplings which enter hard exclusive reactions are tested in
DGLAP evolution even at small $Q^2.$ As noted above, the
dimensional counting rules for form factors and exclusive
processes have also been derived for conformal QCD using the
AdS/CFT correspondence~\cite{Polchinski:2001tt,Brodsky:2003px}.

In fact, there have been a remarkable number of empirical
successes of PQCD predictions, including the scaling and angular
dependence of $\gamma \gamma \to \pi^+ \pi^-$, pion
photoproduction, vector meson electroproduction, and the
photon-to-pion transition form factor.   A particularly dramatic
example is deuteron photodisintegration  which satisfies the
predicted scaling law $[s^{11} {d\sigma\over dt}(\gamma d \to p
n)\sim {\rm const}]$ at large $p_\perp$ and fixed CM
angle~\cite{Rossi:2004qm} to remarkable high precision.
Perturbative QCD predicts that only the small compact part of the
light-front wavefunctions enter exclusive hard scattering
processes, and that these hadronic fluctuations have diminished
interactions in a nuclear target~\cite{Brodsky:1988xz}. Evidence
for QCD color transparency has been observed for quasi-elastic
photoproduction~\cite{Dutta:2003mk} and proton-proton
scattering~\cite{Aclander:2004zm}. In general, the PQCD scaling
behavior can be modulated by resonances and heavy quark threshold
phenomena~\cite{Brodsky:1987xw} which can cause dramatic spin
correlations~\cite{Court:1986dh} as well as novel color
transparency effects~\cite{Brodsky:1988xz,Aclander:2004zm}. The
approach to scaling in pion photoproduction: $[s^{7} {d\sigma\over
dt}(\gamma p \to n \pi^+)\sim {\rm const}]$ and evidence for
structure due to the strangeness threshold have recently been
studied at Jefferson Laboratory~\cite{Zhu:2004dy}.

Leading-order perturbative QCD predicts the empirical scaling of
form factors and other hard exclusive amplitudes, but it typically
underestimate the normalization.  The normalization of theoretical
prediction involves questions of the shape of the hadron
distribution amplitudes, the proper scale for the running
coupling~\cite{Brodsky:1997dh} as well as higher order
corrections. In fact, as noted above, in the AdS/CFT analysis,
hard scattering amplitudes which are normally of order
$\alpha_s^p$ in PQCD appear as order $\alpha_s^{p/2}$ in the
nonperturbative theory~\cite{Maldacena:1997re,Rey:1998ik}.

Further experimental studies, particularly measurements of $\bar p
p$ annihilation into two hadrons at GSI, electroproduction at
Jefferson Laboratory and the study of two-photon exclusive
channels at CLEO and the B-factories have the potential of
providing critical information on the hadron wavefunctions as well
as testing the dominant dynamical processes at short distances.

\section{Color Transparency}

One of the most distinctive tests for the underlying gauge theory
basis for the strong interactions is color transparency: the
small transverse size fluctuations of a hadron wavefunction with a
small color dipole moment have minimal interactions in a
nucleus~\cite{Bertsch:1981py,Brodsky:1988xz}.  Each hadron entering
or emitted from a hard exclusive reaction initially emerges with
high momentum and small transverse size $b_\perp.$
A fundamental feature of gauge
theory is that soft gluons decouple from the small color-dipole
moment of the compact fast-moving color-singlet wavefunction
configurations of the incident and final-state hadrons. The
transversely compact color-singlet configurations can effectively
persist over a distance of order $\ell_{\rm Ioffe} = \mathcal{O}
(E_{\rm lab}/Q^2)$, the Ioffe coherence length. Thus if we study
hard quasi-elastic processes in a nuclear target such as $e A \to
e' p' (A-1)$ or $p A \to p' (A-1)$, the outgoing and ingoing
hadrons will have minimal absorption in a nucleus.  The diminished
absorption of hadrons produced in hard exclusive reactions implies
additivity of the nuclear cross section in nucleon number $A$ and
is the theoretical basis for the ``color transparency" of hard
quasi-elastic
reactions~\cite{Brodsky:1988xz,Frankfurt:1988nt,Jain:1996dd}.  In
contrast, in conventional Glauber scattering, one predicts strong,
nearly energy-independent initial and final state attenuation.
Similarly, in hard diffractive processes such as $\gamma^*(Q^2) p
\to \rho p$~\cite{Brodsky:1994kf}, only the small transverse
configurations $b_\perp \sim 1/Q$ of the longitudinally polarized
vector meson distribution amplitude are involved. Its hadronic
interactions as it exits the nucleus will be minimal, and thus the
$\gamma^*(Q^2) N \to \rho N$ reaction can occur coherently
throughout a nuclear target in reactions without absorption or
shadowing.

\begin{figure}[htb]
\centering
\includegraphics[width=4.3in]{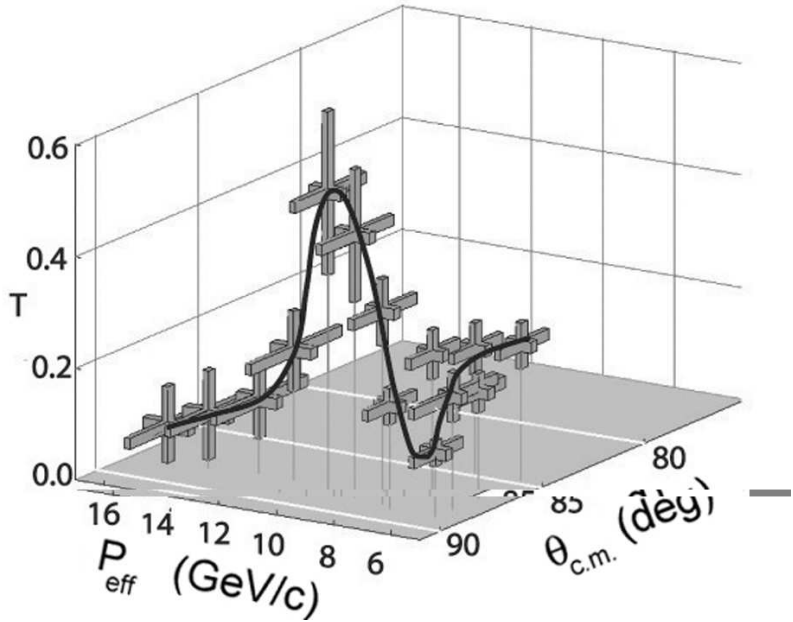}
\caption{The dependence of Carbon transparency on effective
incident beam momentum ($P_{\rm eff}$) and on center of mass
scattering angle ($\theta_{\rm c.m.}$). The data are from the E850
C(p,2p)
experiments~\cite{Carroll:1988rp,Mardor:1998fz,%
Leksanov:2001ui,Aclander:2004zm}.} \label{TvsEthcm}
\end{figure}

Color transparency has also been tested in large angle
quasi-elastic $p A \to p p A-1$
scattering~\cite{Carroll:1988rp,Mardor:1998fz,Leksanov:2001ui,Aclander:2004zm}
where only the small size fluctuations of the hadron wavefunction
enters the hard exclusive scattering amplitude. There is  evidence
for the onset of color transparency  in the regime $6< s < 25$
GeV$^2$, indicating that small wavefunction configurations are
indeed controlling this exclusive reaction at moderate momentum
transfers. However at $p_{\rm lab} \simeq 12$ GeV, $E_{cm} \simeq
5$ GeV, color transparency dramatically fails. See
Fig.~\ref{TvsEthcm}. It is noteworthy that in the same energy
range, the normal-normal spin asymmetry $A_{NN}$ in elastic $pp
\to pp$ scattering at $\theta_{cm} = 90^0$ increases dramatically
to $A_{NN} \simeq 0.6$; it is about four times more probable that
the protons scatter with helicity normal to the scattering plane
than anti-normal~\cite{Court:1986dh}.

The unusual spin and color transparency effects seen in elastic
proton-proton scattering at $E_{CM} \sim 5$ GeV and large angles
could be related to the charm threshold and the effects of a
$\ket{ uud uud c \bar c }$ resonance which would appear as in the
$J=L=S=1$ $p p $ partial
wave~\cite{Brodsky:1987xw,deTeramond:1998ny}. The intermediate
state $\vert u u d u u d c \bar c \rangle$ has odd intrinsic
parity and couples to the $J=S=1$ initial state, thus strongly
enhancing scattering when the incident projectile and target
protons have their spins parallel and normal to the scattering
plane.  A similar enhancement of $A_{NN}$ is observed at the
strangeness threshold.  The physical protons coupling at the charm
threshold will have normal Glauber interactions, thus explaining
the anomalous change in color transparency observed at the same
energy in quasi-elastic $ p p$ scattering.  A crucial test of this
hypothesis is the observation of open charm production near
threshold with a cross section of order of $1
\mu$b~\cite{Brodsky:1987xw,deTeramond:1998ny}. A similar cross
section is expected for the second threshold for open charm
production from $p \bar p \to {\rm charm} ~ p \bar p.$

An alternative explanation of the color transparency and spin
anomalies in $pp$ elastic scattering has been postulated by
Ralston, Jain, and Pire~\cite{Ralston:1986zn,Jain:1996dd}. The
oscillatory effects in the large-angle $pp \to pp$ cross section
and spin structure are postulated to be due to the interference of
Landshoff pinch and perturbative QCD amplitudes.  In the case of
quasi-elastic reactions, the nuclear medium absorbs and filters
out the non-compact pinch contributions, leaving the additive hard
contributions unabsorbed.  It is clearly important that these two
alternative explanations be checked by experiment.

In general, one can expect strong effects whenever heavy quarks
are produced at low relative velocity with respect to each other
or the other quarks in the reaction since the QCD van der Waals
interactions become maximal in this domain. The opening of the
strangeness and charm threshold in intermediate states can become
most apparent in large angle reactions such as $pp$ scattering and
pion photoproduction since the competing perturbative QCD
amplitudes are power-suppressed. Charm and bottom production near
threshold such as $J/\psi$ photoproduction is also sensitive to
the multi-quark, gluonic, and hidden-color correlations of
hadronic and nuclear wavefunctions in QCD since all of the
target's constituents must act coherently within the small
interaction volume of the heavy quark production
subprocess~\cite{Brodsky:2000zc}. Although such multi-parton
subprocess cross sections are suppressed by powers of $1/m^2_Q$,
they have less phase-space suppression and can dominate the
contributions of the leading-twist single-gluon subprocesses in
the threshold regime.

\section{Measuring Light-Front Wavefunctions in QCD and Testing Color
Transparency using Diffractive Dissociation}

Diffractive multi-jet production in heavy nuclei provides a novel
way to measure the shape of light-front Fock state wave functions
and test color transparency~\cite{Brodsky:1988xz}. For example,
consider the reaction~\cite{Bertsch:1981py,Frankfurt:1999tq} $\pi
A \rightarrow {\rm Jet}_1 + {\rm Jet}_2 + A^\prime$ at high energy
where the nucleus $A^\prime$ is left intact in its ground state.
The transverse momenta of the jets balance so that $ \vec k_{\perp
i} + \vec k_{\perp 2} = \vec q_\perp < {R^{-1}}_A \ . $ The
light-cone longitudinal momentum fractions also need to add to
$x_1+x_2 \sim 1$ so that $\Delta p_L < R^{-1}_A$.  The process can
then occur coherently in the nucleus.  Because of color
transparency, the valence wave function of the pion with small
impact separation will penetrate the nucleus with minimal
interactions, diffracting into jet pairs~\cite{Bertsch:1981py}.
The $x_1=x$, $x_2=1-x$ dependence of the di-jet distributions will
thus reflect the shape of the pion valence light-cone wave
function in $x$; similarly, the $\vec k_{\perp 1}- \vec k_{\perp
2}$ relative transverse momenta of the jets gives key information
on the derivative of the underlying shape of the valence pion
wavefunction~\cite{Frankfurt:1999tq,Nikolaev:2000sh}. The
diffractive nuclear amplitude extrapolated to $t = 0$ should be
linear in nuclear number $A$ if color transparency is correct. The
integrated diffractive rate should then scale as $A^2/R^2_A \sim
A^{4/3}.$

These predictions have been verified by the E791 experiment at
Fermilab for 500 GeV incident pions on nuclear
targets~\cite{Aitala:2000hc}. The high energy pion diffracts into
dijets $\pi A \to q \bar q A^\prime$ which balance in transverse
momentum and leave the nucleus in its ground
state~\cite{Ashery:2002jx}. The measured momentum fraction
distribution of the jets is consistent with the shape of the pion
asymptotic distribution amplitude, $\phi^{\rm asympt}_\pi (x) =
\sqrt 3 f_\pi x(1-x)$~\cite{Aitala:2000hb}. Data from
CLEO~\cite{Gronberg:1998fj} for the $\gamma \gamma^* \rightarrow
\pi^0$ transition form factor also favor a form for the pion
distribution amplitude close to the asymptotic solution to its
perturbative QCD evolution
equation~\cite{Lepage:1979zb,Efremov:1978rn,Lepage:1980fj}.

These ``self-resolving" diffractive processes thus provide direct
experimental information on the light-cone wavefunctions of the
photon and proton in terms of their QCD degrees of freedom, as
well as the composition of nuclei in terms of their nucleon and
mesonic degrees of freedom. When the hadronic jets have balancing
but high transverse momentum, one studies the small size
fluctuation of the incident pion. The forward diffractive
amplitude is observed to grow in proportion to the total number of
nucleons in the nucleus, in agreement with the color transparency
prediction (see below), but in strong contrast to standard Glauber
theory which predicts that only the front surface of the nucleus
should be effective.

The diffractive dissociation of a hadron or nucleus can also occur
via the Coulomb dissociation of a beam particle on an electron
beam (\eg\ at HERA or eRHIC) or on the strong Coulomb field of a
heavy nucleus (\eg\ at RHIC or nuclear collisions at the
LHC)~\cite{BHDP}. The amplitude for Coulomb exchange at small
momentum transfer is proportional to the first derivative $\sum_i
e_i {\partial \over \vec k_{T i}} \psi$ of the light-cone
wavefunction, summed over the charged constituents.  The Coulomb
exchange reactions fall off less fast at high transverse momentum
compared to pomeron exchange reactions since the light-cone
wavefunction is effective differentiated twice in two-gluon
exchange reactions. It is also interesting to study diffractive
tri-jet production using proton beams $ p A \rightarrow {\rm
Jet}_1 + {\rm Jet}_2 + {\rm Jet}_3 + A^\prime $ to determine the
fundamental shape of the 3-quark structure of the valence
light-cone wavefunction of the nucleon at small transverse
separation~\cite{MillerFrankfurtStrikman}.

There has been an important debate whether diffractive jet
production faithfully measures the light-front wavefunctions of
the projectile. Braun {\em et al.}~\cite{Braun:2002wu} and
Chernyak~\cite{Chernyak:2002jc} have argued that one should
systematically iterate the gluon exchange kernel from all sources,
including final state interactions. Thus if the hard momentum
exchange which produces the high transverse momentum di-jets
occurs in the final state, then the $x$ and $k_\perp$
distributions will reflect the gluon exchange kernel, not the
pion's wavefunction. However, it should be noted that the
measurements of pion diffraction by the E791
experiment~\cite{Aitala:2000hb} are performed on a platinum
target. Only the part of the pion wavefunction with small impact
separation can give the observed color transparency; {\em i.e.},
additivity of the amplitude on nuclear number. Thus the nucleus
automatically selects events where the jets are produced at high
transverse momentum in the initial state before the pion reaches
the nucleus~\cite{Bertsch:1981py}.

The debate~\cite{Ivanov:2002hp,Braun:2002wu,Chernyak:2002jc}
concerning the nature of diffractive dijet dissociation also
applies to the simpler analysis of diffractive dissociation via
Coulomb exchange. The one-photon exchange matrix element can be
identified with the spacelike electromagnetic form factor for $\pi
\to q \bar q$; $ \VEV{\pi; P -q | j^+(0)|q \bar q; P}$. Here the
state $\ket{q \bar q}$ is the eigenstate of the QCD Hamiltonian;
it is effectively an `out' state.  If we choose the $q^+=0$ frame
where $q^2 = - \vec q\,^2_\perp$, then the form factor is exactly
the overlap integral in transverse momentum of the pion and $\bar
q q$ LFWFs summed over Fock States. The form factor vanishes at
$Q^2= 0$ because it is the matrix element of the total charge
operator and the pion and jet-jet eigenstates are orthogonal.  The
$n = 2$ contribution to the form factor is the convolution
$\psi_\pi(x,k_\perp-(1-x)q_\perp)$ with $\psi_{\bar q
q}(x,k_\perp)$.  This can be expanded at small $q^2$ in terms of
the transverse momentum derivatives of the pion wavefunction. The
final-state wavefunction represents an outgoing wave of free
quarks with momentum $y, \ell_\perp$ and $1-y, -\ell_\perp$. To
first approximation, the wavefunction $\psi_{\bar q q}(x,k_\perp)$
peaks strongly at $x = y$ and $k_\perp = \ell_\perp.$ Using this
approximation, the form factor at small $Q^2$ is proportional to
the derivative of the pion light-cone wavefunction $[e_q (1-x)-
e_{\bar q} x] {\partial \over d k_\perp} \psi_\pi(x,k_\perp)$
evaluated at $x = y$ and $k_\perp = \ell_\perp.$ One can also
consider corrections to the final state wavefunction from gluon
exchange.  However, the final quarks are already moving in the
correct direction at zeroth order, so these corrections would be
expected to be of higher order.

\section{The Generalized Crewther Relation}

A central principle of renormalization theory is that predictions
which relate physical observables to each other cannot depend on
theoretical conventions.  For example, one can use any
renormalization scheme, such as the modified minimal subtraction
scheme, and any choice of renormalization scale $\mu$ to compute
perturbative series relating observables $A$ and $B$.  However,
all traces of the choices of the renormalization scheme and scale
must disappear when one algebraically eliminates the
$\alpha_{\overline{\mbox{\tiny MS}}}(\mu)$ and directly relates $A$
to $B$.  This is the principle underlying ``commensurate scale
relations" (CSR)~\cite{Brodsky:1995ht}, which are general
leading-twist QCD predictions relating physical observables to
each other at their respective scales. An important example is the
generalized Crewther relation~\cite{Brodsky:1995tb}:
\begin{equation}
\left[1 + \frac{\alpha_R(s^*)}{\pi} \right] \left[1 -
\frac{\alpha_{g_1}(Q^2)}{\pi}\right] = 1
\end{equation}
where the underlying form at zero $\beta$ function is dictated by
conformal symmetry~\cite{Crewther:1972kn}. Here $\alpha_R(s)/\pi$
and $-\alpha_{g_1}(Q^2)/\pi$ represent the entire radiative
corrections to $R_{e^+ e^-}(s)$ and the Bjorken sum rule for the
$g_1(x,Q^2)$ structure function measured in spin-dependent deep
inelastic scattering, respectively. The relation between $s^*$ and
$Q^2$ can be computed order by order in perturbation theory, as in
the BLM method~\cite{Brodsky:1982gc}. The ratio of physical scales
guarantees that the effect of new quark thresholds is
commensurate. Commensurate scale relations are
renormalization-scheme independent and satisfy the group
properties of the renormalization group.  Each observable can be
computed in any convenient renormalization scheme such as
dimensional regularization. The $\barMS$ coupling can then be
eliminated; it becomes only an intermediary~\cite{Brodsky:1994eh}.
In such a procedure there are no further renormalization scale
($\mu$) or scheme ambiguities.

The generalized Crewther relation~\cite{Brodsky:1995tb} can be
derived by calculating the QCD radiative corrections to the deep
inelastic sum rules and $R_{e^+ e^-}$ in a convenient
renormalization scheme such as the modified minimal subtraction
scheme $\overline{\rm MS}$.  One then algebraically eliminates
$\alpha_{\overline{\mbox{\tiny MS}}}(\mu)$. Finally, BLM
scale-setting~\cite{Brodsky:1982gc} is used to eliminate the
$\beta$-function dependence of the coefficients.  The form of the
resulting relation between the observables thus matches the result
which would have been obtained had QCD been a conformal theory
with zero $\beta$ function.  The final result relating the
observables is independent of the choice of intermediate
$\overline{\rm MS}$ renormalization scheme.

The Crewther relation was originally derived assuming that the
theory is conformally invariant; \ie, for zero $\beta$ function.
In the physical case, where the QCD coupling runs, all
non-conformal effects are resummed into the energy and momentum
transfer scales of the effective couplings $\alpha_R$ and
$\alpha_{g1}$.    The coefficients are independent of color and
are the same in Abelian, non-Abelian, and conformal gauge theory.
The non-Abelian structure of the theory is reflected in the
expression for the scale ${Q}^{*}$.

Fits~\cite{Mattingly:ej} to the experimental measurements of the
$R$-ratio above the threshold for the production of
$c\overline{c}$ bound states provide the empirical constraint:
$\alpha_{R}({\sqrt s} =5.0~{\rm GeV})/\pi \simeq 0.08\pm 0.03.$
The prediction for the effective coupling for the deep inelastic
sum rules at the commensurate momentum transfer $Q$ is then
$\alpha_{g_1}(Q=12.33\pm 1.20~{\rm GeV})/\pi \simeq \alpha_{\rm
GLS}(Q=12.33\pm 1.20~{\rm GeV})/\pi \simeq 0.074\pm 0.026.$
Measurements of the Gross-Llewellyn Smith sum rule have so far
only been carried out at relatively small values of
$Q^2$~\cite{CCFRL1,CCFRL2}; however, one can use the results of
the theoretical extrapolation~\cite{KS} of the experimental data
presented in~\cite{CCFRQ} to obtain $ \alpha_{\rm GLS}^{\rm
extrapol}(Q=12.25~{\rm GeV})/\pi\simeq 0.093\pm 0.042.$ This range
overlaps with the prediction from the generalized Crewther
relation. It is clearly important to have higher precision
measurements to fully test this fundamental QCD prediction.

The ratio of commensurate scales $\Lambda_{BA}$ is determined by
the requirement that all terms involving the $\beta$ function are
incorporated into the arguments of the running couplings, as in
the original BLM procedure~\cite{Brodsky:1982gc}.  Physically, the
ratio of scales corresponds to the fact that the physical
observables have different quark threshold and distinct
sensitivities to fermion loops.  More generally, the differing
scales are in effect relations between mean values of the physical
scales which appear in loop integrations. Commensurate scale
relations are transitive; \ie, given the relation between
effective charges for observables $A$ and $C$ and $C$ and $B$, the
resulting relation between $A$ and $B$ is independent of $C$.  In
particular, transitivity implies $\Lambda_{AB} = \Lambda_{AC}
\times \Lambda_{CB}$. The shift in scales which gives conformal
coefficients in effect pre-sums the large and strongly divergent
terms in the PQCD series which grow as $n!  (\beta_0 \alpha_s)^n$,
\ie, the infrared renormalons associated with coupling-constant
renormalization~\cite{tHooft,Mueller,LuOneDim,BenekeBraun}.

Similarly, commensurate scale relations obey the ``conformal
correspondence principle": the CSRs reduce to correct conformal
relations when $N_C$ and $N_F$ are tuned to produce zero $\beta$
function.  Thus conformal symmetry provides a {\it template} for
QCD predictions, providing relations between observables which are
present even in theories which are not scale invariant.  All
effects of the nonzero beta function are encoded in the
appropriate choice of relative scales $\Lambda_{AB} = Q_A/Q_B$.

In the case of QED,  the heavy lepton potential (in the limit of
vanishing external charge) is conventionally used to define the
effective charge $\alpha_{qed}(q^2)$.  This definition, the Dyson
Goldberger-Low effective charge, resums all lepton pair vacuum
polarization contributions in the photon propagator, and it is
analytic in the lepton masses.  The scale of the QCD coupling is
thus the virtuality of the exchanged photon. The extension of this
concept to non-abelian gauge theories is non-trivial due to the
self interactions of the gauge bosons which make the usual
self-energy gauge dependent. However, by systematically
implementing the Ward identities of the theory, one can project
out the unique self-energy of each {\it physical} particle.  This
results in a gluonic self-energy which is gauge independent and
which can be resummed to define an effective charge that is
related through the optical theorem to differential cross
sections. The algorithm for performing the calculation at the
diagrammatic level is called the ``pinch
technique"~\cite{Cornwall:1981zr,Degrassi:1992ue,watson,prw}. The
generalization of the pinch technique to higher loops has recently
been
investigated~\cite{Watson:1998vw,watson2b,rafael2,%
Binosi:2002ft,Binosiqcdall,Binosi:2004qe}. Binosi and Papavassiliou
have shown the consistency of the pinch technique to all orders in
perturbation theory, thus allowing a systematic application to the
QCD and electroweak effective charges at higher orders. The pinch
scheme is in fact used to define the evolution of the couplings in
the electroweak theory. The pinch scheme thus provides an ideal
scheme for QCD couplings as well.

\section{Effective Charges and Unification}

\begin{figure}[htb]
 \centering
 \includegraphics[height=3in]{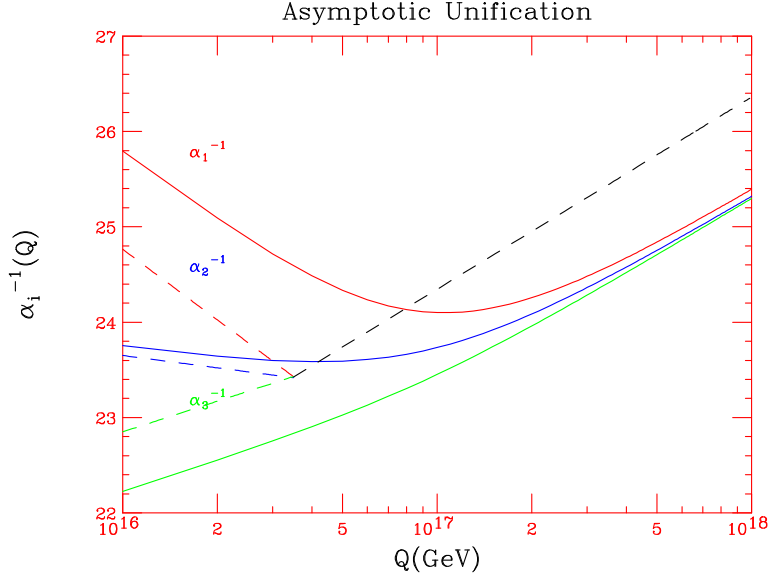}
 \caption[*]{
{\bf Asymptotic Unification.}  An illustration of strong and
electroweak coupling unification in an $SU(5)$ supersymmetric
model based on the pinch scheme effective charge.  The solid lines
are the analytic pinch scheme $\overline{PT}$ effective couplings,
while the dashed lines are the $\overline{DR}$ couplings.  For
illustrative purposes, $a_3(M_Z)$ has been chosen  so that
unification occurs at a finite scale for $\overline{DR}$ and
asymptotically for the $\overline{PT}$ couplings.  Here
$M_{SUSY}=200{\rm GeV}$ is the mass of all light superpartners
except the wino and gluino which have values $ \frac{1}{2}{mgx} =
M_{SUSY} = 2 m_{wx}$.} \label{fig:4}
\end{figure}

Recently Michael Binger and I have analyzed a supersymmetric grand
unification model in the context of physical renormalization
schemes~\cite{Binger:2003by}.  Our essential assumption is that
the underlying {\it forces} of the theory become unified at the
unification scale. We have found a number of qualitative differences and
improvements in precision over conventional approaches.  There is
no need to assume that the particle spectrum has any specific
structure; the effect of heavy particles is included both below
and above the physical threshold.  Unlike mathematical schemes
such as dimensional reduction, $\overline{DR}$, the evolution of
the coupling is analytic and  unification is approached
continuously rather than at a fixed scale. The effective charge
formalism thus provides a template for calculating all mass
threshold effects for any given grand unified theory. These new
threshold corrections  are important in making the measured values
of the gauge couplings consistent with unification. A comparison
with the conventional scheme based on $\overline{DR}$ dimensional
regularization scheme is summarized in Fig.~\ref{fig:4}.

\section{Conclusions on Commensurate Scale Relations}

Commensurate scale relations have a number of attractive properties:
\begin{enumerate}
\item The ratio of physical scales $Q_A/Q_B$ which appears in
commensurate scale relations reflects the relative position of
physical thresholds, \ie\, quark anti-quark pair production. \item
The functional dependence and perturbative expansion of the CSR
are identical to those of a conformal scale-invariant theory where
$\beta_A(\alpha_A)=0$ and $\beta_B(\alpha_B)=0$. \item In the case
of theories approaching fixed-point behavior
$\beta_A(\bar\alpha_A)=0$ and $\beta_B(\bar\alpha_B)=0$, the
commensurate scale relation relates both the ratio of fixed point
couplings $\bar\alpha_A/\bar\alpha_B$ and the ratio of scales as
the fixed point is approached. \item Commensurate scale relations
satisfy the Abelian correspondence
principle~\cite{Brodsky:1997jk}; \ie\ the non-Abelian gauge theory
prediction reduces to Abelian theory for $N_C \to 0$ at fixed $
C_F\alpha_s$ and fixed $N_F/C_F$. \item The perturbative expansion
of a commensurate scale relation has the same form as a conformal
theory, and thus has no $n!$ renormalon growth arising from the
$\beta$-function.  It is an interesting conjecture whether the
perturbative expansion relating observables to observable are in
fact free of all $n!$ growth.  The generalized Crewther relation,
where the commensurate relation's perturbative expansion forms a
geometric series to all orders, has convergent behavior.
\end{enumerate}

Virtually any perturbative QCD prediction can be written in the
form of a commensurate scale relation, thus eliminating any
uncertainty due to renormalization scheme or scale dependence.

\section{Inclusive Reactions: Complications from Final-State Interactions}

It is usually assumed---following the parton model---that the
leading-twist structure functions measured in deep  inelastic
lepton-proton scattering are simply the probability distributions for
finding quarks and gluons in the target nucleon.  In fact, gluon  exchange
between the fast, outgoing quarks and the target spectators
effects the leading-twist structure functions in a profound way,
leading to  diffractive leptoproduction processes, shadowing of
nuclear structure  functions, and target spin asymmetries.  In
particular, the final-state  interactions from gluon exchange lead
to single-spin asymmetries in  semi-inclusive deep inelastic
lepton-proton scattering which are not  power-law suppressed in
the Bjorken limit.

\begin{figure}[htb]
\centering
\includegraphics[width=3.3in,height=3in]{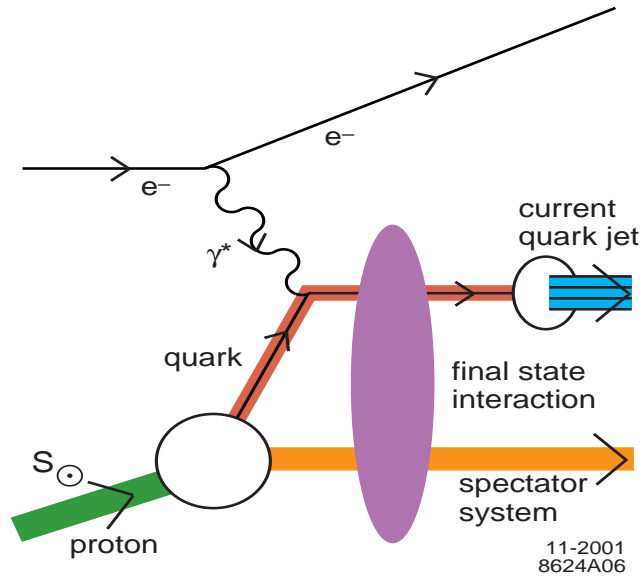}
\caption{A final-state interaction from gluon exchange in deep
inelastic lepton scattering. The difference of the QCD
Coulomb-like  phases in different orbital states of the proton
produces a single proton spin asymmetry.} \label{SSA}
\end{figure}

A new understanding of the role of final-state interactions in
deep inelastic scattering has recently
emerged~\cite{Brodsky:2002ue}.  The final-state interactions from
gluon exchange between the outgoing quark and the target spectator
system lead to single-spin asymmetries in semi-inclusive deep
inelastic lepton-proton scattering at leading twist in
perturbative QCD; {\em i.e.}, the rescattering corrections of the
struck quark with the target spectators are not power-law
suppressed at large photon virtuality $Q^2$ at fixed
$x_{bj}$~\cite{Brodsky:2002cx} See Fig. \ref{SSA}. The final-state
interaction from gluon exchange occurring immediately after the
interaction of the current also produces a leading-twist
diffractive component to deep inelastic scattering $\ell p \to
\ell^\prime p^\prime X$ corresponding to color-singlet exchange
with the target system; this in turn produces shadowing and
anti-shadowing of the nuclear structure
functions~\cite{Brodsky:2002ue,Brodsky:1989qz}.  In addition, one
can show that the pomeron structure function derived from
diffractive DIS has the same form as the quark contribution of the
gluon structure function~\cite{Brodsky:2004hi}.

The final-state
interactions occur at a light-cone time $\Delta\tau \simeq 1/\nu$
after the virtual photon interacts with the struck quark,
producing a nontrivial phase.  Thus none of the above phenomena is
contained in the target light-front wave functions computed in
isolation.   In particular, the shadowing of nuclear structure
functions is due to destructive interference effects from
leading-twist diffraction of the virtual photon, physics not
included in the nuclear light-front wave functions.  Thus the
structure functions measured in deep inelastic lepton scattering
are affected by final-state rescattering, modifying their
connection to light-front probability distributions.  Some of
these results can be understood by augmenting the light-front wave
functions with a gauge link, but with a gauge potential created by
an external field created by the virtual photon $q \bar q$ pair
current~\cite{Belitsky:2002sm}.  The gauge link is also process
dependent~\cite{Collins:2002kn}, so the resulting augmented LFWFs
are not universal.

\section{Single-Spin Asymmetries in Drell-Yan Processes}

Single-spin asymmetries in hadronic reactions provide a remarkable
window to QCD mechanisms at the amplitude level.  In general,
single-spin asymmetries measure the correlation of the spin
projection of a hadron with a production or scattering
plane~\cite{Sivers:1990fh}.  Such correlations are odd under time
reversal, and thus they can arise in a time-reversal invariant
theory only when there is a phase difference between different
spin amplitudes. Specifically, a nonzero correlation of the proton
spin normal to a production plane measures the phase difference
between two amplitudes coupling the proton target with $J^z_p =
\pm {1\over 2}$ to the same final-state.  The calculation requires
the overlap of target light-front wavefunctions with different
orbital angular momentum: $\Delta L^z = 1;$ thus a single-spin
asymmetry (SSA) provides a direct measure of orbital angular
momentum in the QCD bound state.

Consider the SSA produced in semi-inclusive deep inelastic
scattering $\ell p^\uparrow \to \ell^\prime \pi X$.   In the
target rest frame, such a single target spin correlation
corresponds to the $T$-odd triple product $ \vec S_p \cdot \vec
p_\pi \times \vec q.$ (The covariant form of this correlation is
$\epsilon_{\mu \nu \sigma \tau} S^\mu_p p^\nu q^\sigma
p^\tau_\pi.)$   Significant asymmetries $A_{UL}$ and $A_{UT}$ of
this type have in fact been observed for targets polarized
parallel to or transverse to the lepton beam
direction~\cite{hermes0001,smc99}.

Dae Sung Hwang and Ivan Schmidt~\cite{Brodsky:2002cx} and I have
shown that the QCD final-state interactions (gluon exchange)
between the struck quark and the proton spectator system in
semi-inclusive deep inelastic lepton scattering can produce
single-spin asymmetries which survive in the Bjorken limit.  Such
effects are proportional to the matrix element of a higher-twist
quark-quark-gluon correlator in the target hadron, and thus it has
been assumed on dimensional grounds that any single-spin asymmetry
arising from this source must be suppressed by a power of the momentum
transfer
$Q$ in the Bjorken limit. However, another momentum scale enters into
the semi-inclusive process---the transverse momentum ${\vec
r}_{\perp} = {\vec {p_\pi}}_{\perp} - {\vec q}_{\perp}$ of the
emitted pion relative to the photon direction, and we have shown
that the power-law suppression due to the higher-twist
quark-quark-gluon correlator takes the form of an inverse power of
$r_\perp$ rather than $Q.$

Corrections from spin-one gluon exchange in the initial- or
final-state of QCD processes are not suppressed at high energies
because the coupling is vector-like. Therefore, as a consequence
of the gauge coupling of QCD, single-spin asymmetries in
semi-inclusive deep inelastic scattering survive in the Bjorken
limit of large $Q^2$ at fixed $x_{bj}$ and fixed $\vec r_\perp$.
The resulting final-state phases are analogous to the ``Coulomb"
phases to the hard subprocess which arises from gauge interactions
between outgoing charge particles in QED~\cite{Weinberg:1965nx}.
More specifically, we require the difference  between the gauge
interaction phases for the $J^z_p= \pm {1\over 2}$ amplitudes.
The phases depend on the spin because the outgoing particles
interact at different impact separation corresponding to their
different relative orbital angular momentum.

In our paper~\cite{Brodsky:2002cx}, we explicitly evaluated the
SSA for electroproduction for a specific model of a spin-half ~
proton of mass $M$ with charged spin-half ~ and spin-0
constituents of mass $m$ and $\lambda$, respectively, as in the
QCD-motivated quark-diquark model of a nucleon.  The basic
leptoproduction reaction is then $\gamma^* p \to q (qq)_0$. Our
analysis predicts a nonzero SSA for the target spin normal to the
photon to quark-jet $ \vec S_p \cdot \vec p_q \times \vec q$ which
can be determined by using a jet variable such as thrust to
determine the current quark direction;  \ie, we predict a SSA even
without final-state jet hadronization.  Our mechanism is thus
distinct from a description of SSA based on transversity and
phased fragmentation functions.

\begin{figure}[htb]
\centering
\includegraphics[width=4.3in]{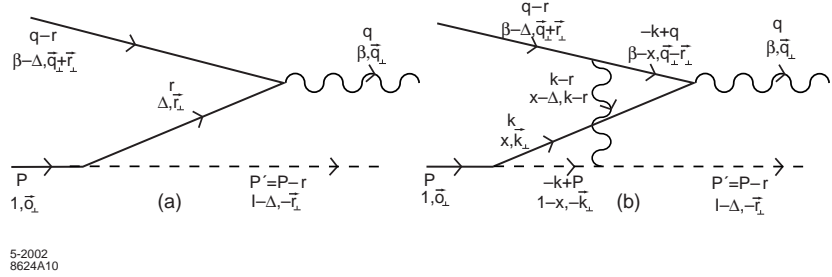}
\caption[*]{The initial-state interaction in the Drell-Yan process.}
\label{fig:1DYSSA}
\end{figure}

Recently Collins~\cite{Collins} has pointed out some important
consequences of these results for SSA in deep inelastic
scattering. In this treatment, the final-state interactions of the
struck quark are incorporated into Wilson line path-ordered
exponentials which augment the light-cone wavefunctions. [See
also~\cite{Ji:2002aa}.] Since the final-state interactions appear
at short light-cone times $\Delta x^+ = {\cal O}(1/\nu)$ after the
virtual photon acts, they can be distinguished from hadronization
processes which occur over long times. As first noted by Collins,
initial-state interactions between the annihilating antiquark and
the spectator system of the target can produce single-spin
asymmetries in the Drell-Yan process.

Dae Sung Hwang, Ivan Schmidt and I~\cite{Brodsky:2002rv} have
extended our analysis to initial- and final-state QCD effects to
predict single-spin asymmetries in hadron-induced hard QCD
processes. Specifically, we shall consider the Drell-Yan (DY) type
reactions~\cite{Drell:1970wh} such as $\bar p p^\uparrow \to
\ell^+ \ell^- X$.  Here the target particle is polarized normal to
the production plane.  The target spin asymmetry can be produced
due to the initial-state gluon-exchange interactions between the
interacting antiquark coming from one hadronic system and the
spectator system of the other. This is shown in the diagram of
Fig. \ref{fig:1DYSSA}. The importance of initial-state
interactions in the theory of massive lepton pair production,
$Q_\perp$ broadening, and energy loss in a nuclear target has been
discussed in
Refs.~\cite{Bodwin:1981fv,Bodwin:1988fs,Brodsky:1996nj}.

The orientation of the target spin $S_z = \pm 1/2$ corresponds to
amplitudes differing by relative orbital angular momentum $\Delta
L^z = 1.$ The initial-state interaction from a gluon exchanged
between the annihilating antiquark and target spectator system
depends in detail on this relative orbital angular momentum.  In
contrast, the initial or final-state interactions due to the
exchange of gauge particles between partons not participating in
the hard subprocess do not contribute to the SSA.  Such
spectator-spectator interactions occur at large impact separation
and are not sensitive to just one unit difference $\Delta L^z = 1$
of the orbital angular momentum of the target wavefunction.

Our mechanism thus depends on the interference of different
amplitudes arising from the target hadron's wavefunction and is
distinct from probabilistic measures of the target such as
transversity. It is also important to note that the target spin
asymmetries which we compute in the DY and DIS processes require
the same overlap of wavefunctions which enters the computation of
the target nucleon's magnetic moment.  In addition, by selecting
different initial mesons in the DY process, we can isolate the
flavor of the annihilating quark and antiquark.  The flavor
dependence of single-spin asymmetries thus has the potential to
provide detailed information of the spin and flavor content of
nucleons at the amplitude level.

As in our analysis of semi-inclusive DIS, we  calculated the
single-spin asymmetry in the Drell-Yan process induced by
initial-state interactions by adopting an effective theory of a
spin-${1\over 2}$ proton of mass $M$ with charged spin-${1\over
2}$ and spin-$0$ constituents of mass $m$ and $\lambda$,
respectively, as in a quark-diquark model. We will take the
initial particle to be just an antiquark.  The result for specific
meson projectiles such as $M p^\uparrow \to \ell^+ \ell^- X$ is
then obtained by convolution with the antiquark distribution of
the incoming meson.  One can also incorporate target nucleon
wavefunctions with a quark-vector diquark structure.  In a more
complete study, one should allow for a many-parton light-front
Fock state wavefunction representation of the target. The results,
however, are always normalized to the quark contribution to the
proton anomalous moment, and thus are basically model-independent.

\section{Crossing}

There is a simple diagrammatic connection between the amplitude
describing the initial-state interaction of the annihilating
antiquark, which gives a single-spin asymmetry for the Drell-Yan
process $\pi p^{\uparrow} \to \ell^+ \ell^- X$, and the
final-state rescattering amplitude of the struck quark, which
gives the single-spin asymmetries in semi-inclusive deep inelastic
leptoproduction $\ell p^{\uparrow}\to \ell' \pi X$. The crossing
of the Feynman amplitude for $\gamma^*(\tilde q) p(P) \to (\tilde
q+r) (P-r)$ in DIS gives $(-\tilde q-r) p(P) \to \gamma^*(-\tilde
q) (P-r)$ for DY by reversing the four-vectors of the photon and
quark lines.  The outgoing quark with momentum $\tilde q+r$ in DIS
becomes the incoming antiquark with momentum $-\tilde q-r$ in DY.
We can use crossing of the Lorentz invariant amplitudes for DIS as
a guide for obtaining the amplitudes for DY
amplitude~\cite{Brandenburg:1994mm}.

In general, one cannot use crossing to relate imaginary parts of
amplitudes to each other, since under crossing, real and imaginary
parts become connected. However, in our case, the relevant
one-gluon exchange diagrams in DIS and DY are both purely
imaginary at high energy, so their magnitudes are related by
crossing. Thus a crucial test of our mechanism is an exact
relation between the magnitude and flavor dependence of the SSA in
the Drell-Yan reaction and the SSA in deep inelastic scattering.
We thus predict the DY SSA of the proton spin with the normal to
the antiquark to virtual photon plane: $\vec S_p \cdot \vec
p_{\bar q} \times \vec{\tilde q}$.  It is identical -- up to a
sign -- to the SSA computed in DIS for $\vec S_p \cdot \vec p_{ q}
\times \vec q$.

The phase arising from the initial- and final-state interactions
in QCD is analogous to the Coulomb phase of Abelian QED
amplitudes. The Coulomb phase depends on the  product of charges
and relative velocity of each ingoing and outgoing charged
pair~\cite{Weinberg:1965nx}.  Thus the sign of the phase in DY and
DIS are opposite because of the different color charge of the
ingoing $\overline 3_C$ antiquark in DY and the outgoing $3_C$
quark in DIS.

\begin{figure}[htbp]
\centering
\includegraphics[width=3.3in]
{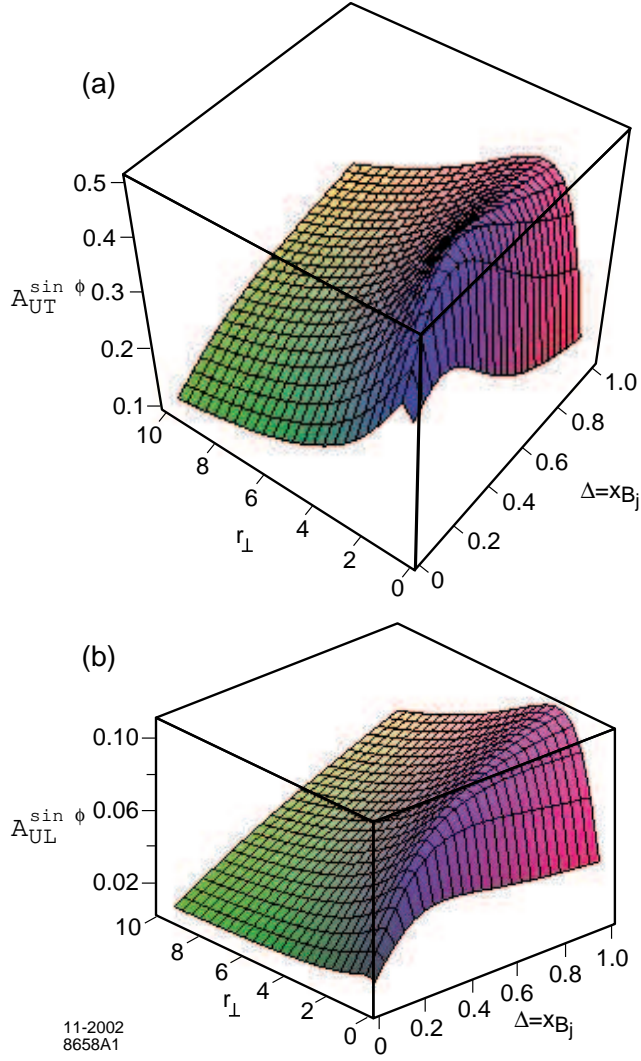}
\caption[*]{\baselineskip 13pt
 Model predictions for the target single-spin asymmetry
$A^{\sin \phi}_{UT}$ for charged and neutral current deep
inelastic scattering resulting from gluon exchange in the final
state.  Here $r_\perp$ is the magnitude of the transverse momentum
of the outgoing quark relative to the photon or vector boson
direction, and $\Delta = x_{bj}$ is the light-cone momentum
fraction of the struck quark. The parameters of the model are
given in the text.  In (a) the target polarization is transverse
to the incident lepton direction.  The asymmetry in (b) $A^{\sin
\phi}_{UL} = K A^{\sin \phi}_{UT}$ includes a kinematic factor $K
= {Q\over \nu}\sqrt{1-y}$ for the case where the target nucleon is
polarized along the incident lepton direction.  For illustration,
we have taken $K= 0.26 \sqrt x,$ with $E_{lab} = 27.6 ~{\rm GeV}$
and $y = 0.5.$} \label{fig:SSA1}
\end{figure}

The asymmetry in the Drell-Yan process is thus the same as that
obtained in DIS, with the appropriate identification of variables,
but with the opposite sign.  This has been stressed recently by
Collins~\cite{Collins}.  Therefore the single-spin asymmetry
transverse to the production plane in the Drell-Yan process can be
obtained from the results of our recent
paper~\cite{Brodsky:2002cx}:
\begin{eqnarray}
{\cal P}_y &=& -\ {e_1e_2\over 8\pi} \ {2\ \Bigl(\ \Delta\, M+m\ \Bigr)\
r^1\over \Big[\
\Bigl( \ \Delta\, M+m\ \Bigr)^2\ +\ {\vec r}_{\perp}^2\ \Big]}\ \Big[\ {\vec
r}_{\perp}^2+\Delta (1-\Delta)(-M^2+{m^2\over\Delta} +{\lambda^2\over
1-\Delta})\ \Big]
\nonumber\\ &\times& \ {1\over {\vec r}_{\perp}^2}\ {\rm ln}{{\vec
r}_{\perp}^2 +\Delta
(1-\Delta)(-M^2+{m^2\over\Delta}+{\lambda^2\over 1-\Delta})\over \Delta
(1-\Delta)(-M^2+{m^2\over\Delta}+{\lambda^2\over 1-\Delta})}\ . \label{sa2b}
\end{eqnarray}
Here $\Delta  = {q^2\over 2P\cdot q } = {q^2 \over 2 M \nu} $
where $\nu$ is the energy of the lepton pair in the target rest
frame. An explicit calculation is given in
Ref.~\cite{Brodsky:2002rv}.  An illustration of the predictions
for electroproduction is shown in Fig.~\ref{fig:SSA1}.

The natural framework for the wavefunctions which appear in the
SSA calculations is the light-front Fock
expansion~\cite{Lepage:1980fj,Brodsky:1989pv}. In principle, the
light-front wavefunctions for hadrons can be obtained by solving
for the eigen-solutions of the light-front QCD Hamiltonian. Such
wavefunctions are real and include all interactions up to a given
light-front time.  The final-state gluon exchange corrections
which provide the SSA for semi-inclusive DIS occur immediately
after the virtual photon strikes the active quark. Such
interactions are not included in the light-front wavefunctions,
just as Coulomb final-state interactions are not included in the
Schr\"odinger bound state wavefunctions in QED.  As noted above
Collins~\cite{Collins} has argued that since the relevant
rescattering interactions of the struck quark occur very close in
light-cone time to the hard interaction, one can augment the
light-front wavefunctions by a Wilson line factor which
incorporates the effects of the final-state interactions in
semi-inclusive DIS. However, such augmented wavefunctions are not
universal and process independent; for example, in the case of the
DY process, an incoming Wilson line of opposite phase must be
used.

It should be emphasized that the same overlap of light-front
wavefunctions with $\Delta L^z = 1$  which gives single-spin
asymmetries also yields the Pauli form factor $F_2(t)$ and the
generalized parton distribution $E(x,\zeta,t)$ entering deeply
virtual Compton
scattering~\cite{Muller:1998fv,Ji:1996ek,Radyushkin:1997ki,%
Diehl:2000xz,Brodsky:2000xy}. Each quark of the target
wavefunction appears additively, weighted linearly by the quark
charge in the case of the Pauli form factor and weighted
quadratically in the case of deep inelastic scattering, the
Drell-Yan reaction and deeply virtual Compton scattering.

Thus the same physical mechanism which produces a leading-twist
single-spin asymmetry in semi-inclusive deep inelastic lepton scattering,
also leads to a leading twist single-spin asymmetry in the Drell-Yan
process. The initial-state interaction between the annihilating antiquark
with the spectator of the target produces the required phase
correlation.  The equality in magnitude, but opposite sign, of the
single-spin asymmetries in semi-inclusive DIS and the
corresponding Drell-Yan processes is an important check of our
mechanism.

It has been conventional to assume that the effects of initial-
and final-state interactions are always power-law suppressed for
hard processes in QCD. In fact, this is not in general correct, as
can be seen from our analyses of leading-twist single-spin
asymmetries in the Drell-Yan process and semi-inclusive deep
inelastic scattering.  The initial- and final-state interactions
which survive in the scaling limit occur in light-cone time $\tau
= {\cal O}(1/Q)$ immediately before or after the hard subprocess.
Other initial- and final-state interactions, such as those between
the spectator of the incident hadron and the spectator of the
target hadron in the DY process, take place over long time scales,
and they only provide inconsequential unitary phase corrections to
the process. This is in accord with our intuition that
interactions which occur at distant times cannot affect the
primary reaction.

Our formalism can be adopted to single-spin asymmetries in more
general hard inclusive reactions, such as $\bar p p^\uparrow \to
\pi X,$ where the pion is detected at high transverse
momentum~\cite{E70496,lambda}. In such cases, one must identify the
relevant hard quark-gluon subprocess and analyze a set of gluon exchange
corrections which connect the spectators of the polarized hadron
with the active quarks and gluons participating in the hard
subprocess. However, this type of final-state interaction cannot be
readily identified as an augmented target wavefunction. It is also clear
from our analyses that there are potentially important corrections
to the hard quark propagator in hard exclusive subprocesses such
as deeply virtual Compton scattering or exclusive meson
electroproduction.  These rescattering interactions of the
propagating quark can provide new single-spin observables and will
correct analyses based on the handbag approximation.

The empirical study of single-spin asymmetries in hard inclusive
and exclusive processes thus provides a new window to the
investigation of hadron spin, angular momentum, and flavor
structure of hadrons.

\section{The Origin of Nuclear Shadowing and Antishadowing}

\begin{figure}[htbp]
\centering
\includegraphics[width=3.3in]{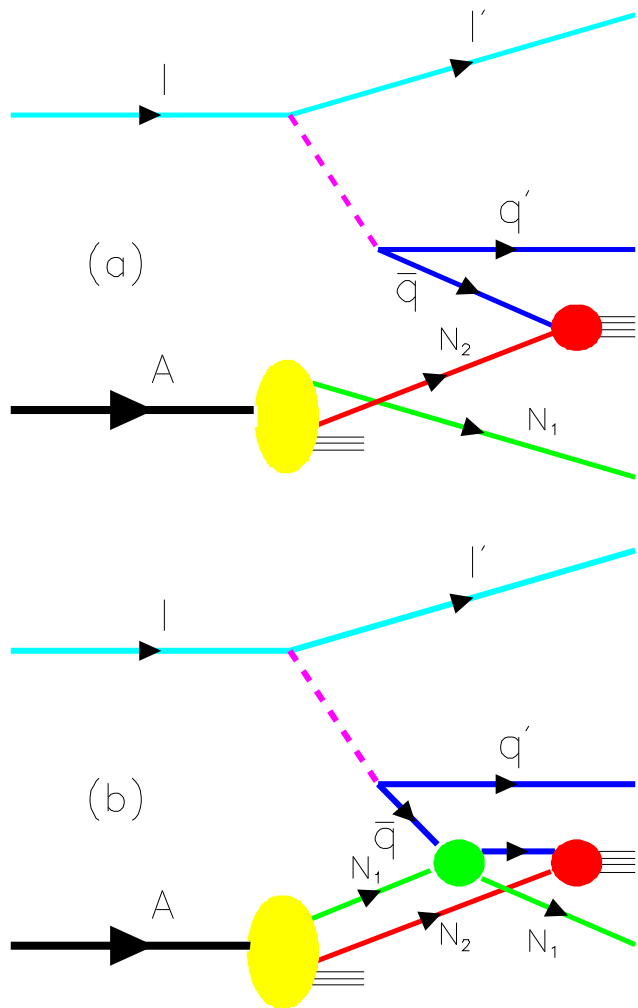}
\caption[*]{\baselineskip 13pt The one-step and two-step processes
in DIS on a nucleus. If the scattering on nucleon $N_1$ is via
pomeron exchange, the one-step and two-step amplitudes are
opposite in phase, thus diminishing the $\bar q$ flux reaching
$N_2.$ This causes shadowing of the charged and neutral current
nuclear structure functions. \label{bsy1f1}}
\end{figure}

The Bjorken-scaling diffractive interactions on nucleons in a
nucleus also lead to shadowing and anti-shadowing of the nuclear
structure functions~\cite{Brodsky:2002ue,Brodsky:1989qz}.
The physics of nuclear shadowing in deep inelastic scattering can
be most easily understood in the laboratory frame using the
Glauber-Gribov picture~\cite{Glauber:1955qq,Gribov:1968gs}.  The
virtual photon, $W$ or $Z^0,$  produces a quark-antiquark
color-dipole pair which can interact diffractively or
inelastically on the nucleons in the nucleus. The destructive
interference of diffractive amplitudes from pomeron exchange on
the upstream nucleons then causes shadowing of the virtual photon
interactions on the back-face
nucleons~\cite{Stodolsky:1966am,Brodsky:1969iz,Ioffe:1969kf,
Frankfurt:1988zg,Kopeliovich:1998gv,Kharzeev:2002fm}. As
emphasized by Ioffe~\cite{Ioffe:1969kf}, the coherence between
processes which occur on different nucleons at separation $L_A$
requires small Bjorken $x_{B}:$ $1/M x_B = {2\nu/ Q^2}  \ge L_A .$
The coherence between different quark processes is also the basis
of saturation phenomena in DIS and other hard QCD reactions at
small $x_B$~\cite{Mueller:2004se}, and coherent multiple parton
scattering has been used in the analysis of $p+A$ collisions in
terms of the perturbative QCD factorization
approach~\cite{Qiu:2004da}. An example of the interference of one-
and two-step processes in deep inelastic lepton-nucleus scattering
illustrated in Fig.~\ref{bsy1f1}.

An important aspect of the shadowing phenomenon is that the
diffractive contribution $\gamma^* N \to X N^\prime$ to deep
inelastic scattering (DDIS) where the nucleon $N_1$ in
Fig.~\ref{bsy1f1} remains intact is a constant fraction of the
total DIS rate, confirming that it is a leading-twist
contribution. The Bjorken scaling of DDIS has been observed at
HERA~\cite{Martin:2004xw,Adloff:1997sc,Ruspa:2004jb}. As shown in
Ref.~\cite{Brodsky:2002ue}, the leading-twist contribution to DDIS
arises in QCD in the usual parton model frame when one includes
the nearly instantaneous gluon exchange final-state interactions
of the struck quark with the target spectators. The same final
state interactions also lead to leading-twist single-spin
asymmetries in semi-inclusive DIS~\cite{Brodsky:2002cx}. Thus the
shadowing of nuclear structure functions is also a leading-twist
effect.

It was shown in Ref.~\cite{Brodsky:1989qz}  that if one allows for
Reggeon exchanges which leave a nucleon intact,  then one can
obtain {\it constructive} interference among the multi-scattering
amplitudes in the nucleus.   A Bjorken-scaling contribution to
DDIS from Reggeon exchange has in fact also been observed at
HERA~\cite{Adloff:1997sc,Ruspa:2004jb}. The strength and energy
dependence of the $C=+$ Reggeon $t-$channel exchange contributions
to virtual Compton scattering is constrained by the
Kuti-Weisskopf~\cite{Kuti:1971ph} behavior $F_2(x) \sim
x^{1-\alpha_R}$ of the non-singlet electromagnetic structure
functions at small $x$.  The phase of the Reggeon exchange
amplitude is determined by its signature factor.  Because of this
complex phase structure~\cite{Brodsky:1989qz}, one obtains
constructive interference and {\it antishadowing} of the nuclear
structure functions in the range $0.1 < x < 0.2$---a pronounced
excess of the nuclear cross section with respect to nucleon
additivity~\cite{Arneodo:1992wf}.

In the case where the diffractive amplitude on $N_1$ is imaginary,
the two-step process has the phase $i \times i = -1 $ relative to
the one-step amplitude, producing destructive interference. (The
second factor of $i$ arises from integration over the quasi-real
intermediate state.)  In the case where the diffractive amplitude
on $N_1$ is due to $C=+$ Reggeon exchange with intercept
$\alpha_R(0) = 1/2$, for example, the phase of the two-step
amplitude is ${1\over \sqrt 2}(1-i) \times i = {1\over \sqrt 2}
(i+1)$ relative to the one-step amplitude, thus producing
constructive interference and antishadowing.

\begin{figure}[htb]
\centering
\includegraphics[width=4.3in]
{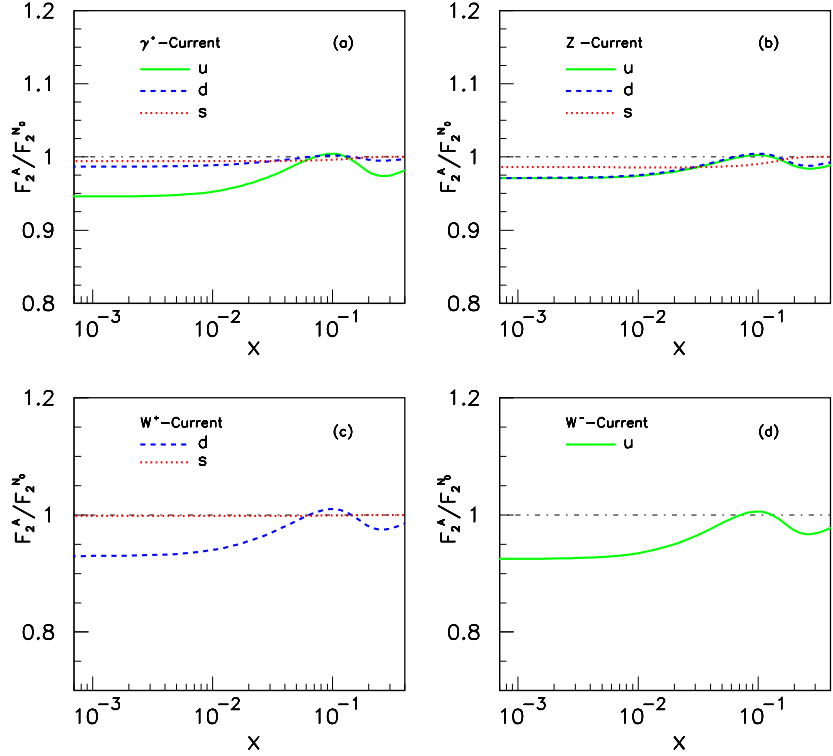} \caption{The  quark contributions to the ratios of
structure functions at $ Q^2 = 1~\rm{GeV}^2$. The solid, dashed
and dotted curves correspond to the $u$, $d$ and $s$ quark
contributions, respectively. This corresponds in our model to the
nuclear dependence of the $\sigma(\bar u-A)$, $\sigma(\bar d-A)$,
$\sigma(\bar s-A)$ cross sections, respectively. In order to
stress the individual contribution of quarks, the numerator of the
ratio $F_2^{A} / F_2^{N_0}$ shown in these two figures is obtained
from the denominator by a replacement $q^{N_0}$ into $q^{A}$ for
only the considered quark. As a result, the effect of
antishadowing appears diminished.
 \label{bsy1f5}}
\end{figure}

The effective quark-nucleon scattering amplitude includes Pomeron
and Odderon contributions from multi-gluon exchange as well as
Reggeon quark-exchange contributions~\cite{Brodsky:1989qz}.  The
coherence of these multiscattering nuclear processes leads to
shadowing and antishadowing of the electromagnetic nuclear
structure functions in agreement with measurements. The Reggeon
contributions to the quark scattering amplitudes depend
specifically on the quark flavor; for example the isovector Regge
trajectories couple differently to $u$ and $d$ quarks. The $s$ and
$\bar s$ couple to yet different Reggeons. This implies distinct
anti-shadowing effects for each quark and antiquark component of
the nuclear structure function. Ivan Schmidt, Jian-Jun Yang, and
I~\cite{Brodsky:2004bg} have shown that this picture leads to
substantially different antishadowing for charged and neutral
current reactions.

\begin{figure}[htb]
\centering
\includegraphics[width=4.3in]{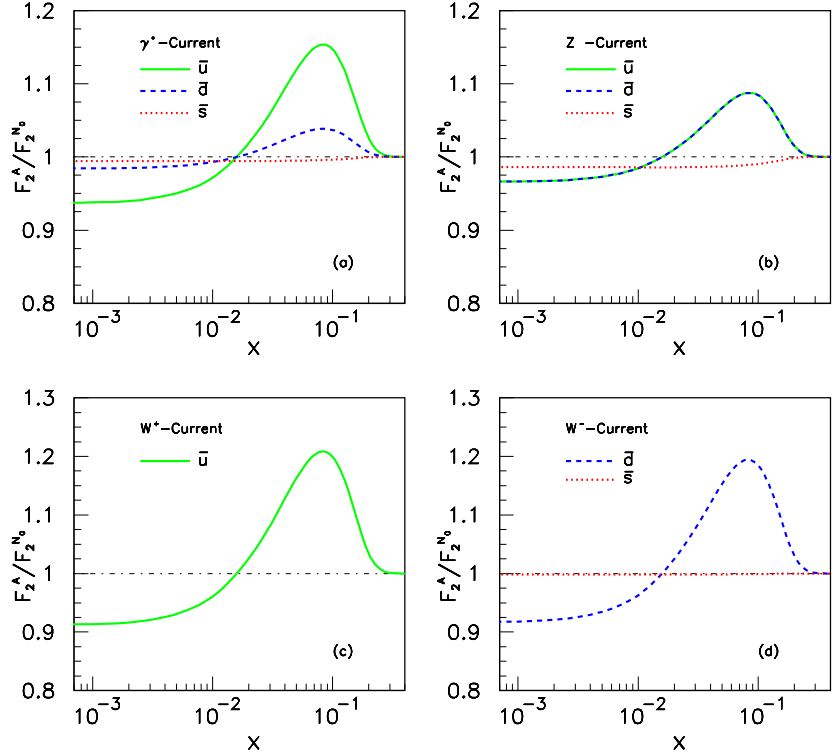}
\caption{ The anti-quark contributions to ratios of the structure
functions at $ Q^2 = 1~\rm{GeV}^2$. The solid, dashed and dotted
curves correspond to $\bar{u}$, $\bar{d}$ and $\bar{s}$ quark
contributions, respectively. This corresponds in our model to the
nuclear dependence of the $\sigma(u-A)$, $\sigma(d-A)$,
$\sigma(s-A)$ cross sections, respectively. In order to stress the
individual contribution of quarks, the numerator of the ratio
$F_2^{A} / F_2^{N_0}$ shown in these two figures is obtained from
the denominator by a replacement $\bar{q}^{N_0}$ into
$\bar{q}^{A}$ for only the considered anti-quark. \label{bsy1f6} }
\end{figure}

Figures~\ref{bsy1f5}--\ref{bsy1f6} illustrate the individual quark
$q$ and anti-quark $\bar{q}$ contributions to the ratio of nuclear
$^{56} Fe$ (structure functions $R=F_2^{A} / F_2^{N_0}$ in a model
calculation where the Reggeon contributions are constrained by the
Kuti-Weisskopf behavior~\cite{Kuti:1971ph} of the nucleon
structure functions at small $x_{bj}.$  Because the strange quark
distribution is much smaller than $u$ and $d$ quark distributions,
the strange quark contribution to the ratio is very close to 1
although $s^{A}/s^{N_0}$ may significantly deviate from 1.

Our analysis leads to substantially different nuclear
antishadowing for charged and neutral current reactions; in fact,
the neutrino and antineutrino DIS cross sections are each modified
in different ways due to the various allowed Regge exchanges. The
non-universality of nuclear effects will modify the extraction of
the weak-mixing angle $\sin^2\theta_W$, particularly because of
the strong nuclear effects for the $F_3$ structure function. The
shadowing and antishadowing of the strange quark structure
function in the nucleus can also be considerably different than
that of the light quarks. We thus find that part of the anomalous
NuTeV result~\cite{McFarland:2003gx} for $\sin^2\theta_W$ could be
due to the nonuniversality  of nuclear antishadowing for charged
and neutral currents. Our picture also implies non-universality
for the nuclear modifications of spin-dependent structure
functions.  A new determination of $\sin^2\theta_W$ is also
expected from the neutrino scattering experiment NOMAD at
CERN~\cite{Petti}. A systematic program of measurements of the
nuclear effects in charged and neutral current reactions could
also be carried out in high energy electron-nucleus colliders such
as HERA and eRHIC, or by using high intensity neutrino
beams~\cite{Geer}.

Thus the antishadowing of nuclear structure functions depends in
detail on quark flavor. Careful measurements of the nuclear
dependence of charged, neutral, and electromagnetic DIS processes
are thus needed to establish the distinctive phenomenology of
shadowing and antishadowing and to make the NuTeV results
definitive. It is also important to map out the shadowing and
antishadowing of each quark component of the nuclear structure
functions to illuminate the underlying QCD mechanisms. Such
studies can be carried out in semi-inclusive deep inelastic
scattering for the electromagnetic current at Hermes and at
Jefferson Laboratory by tagging the flavor of the current quark.
It is also important to measure antishadowing and test
non-universality in antiproton reactions at GSI such as $\bar p A
\to \ell^+ \ell^- X$ as well in pion- and kaon-induced Drell-Yan
reactions~\cite{Berger:1986dd,Accardi:2003be}.

\section{Conclusions}

New theoretical developments in QCD, together with a number of key
experiments, have brought new perspectives to our understanding of
dynamical aspects of the strong interactions.

Color transparency, as evidenced by the Fermilab measurements of
diffractive dijet production, implies that a pion can interact
coherently throughout a nucleus with minimal absorption, in
dramatic contrast to traditional Glauber theory based on a fixed
$\sigma_{\pi n}$ cross section.  Color transparency gives direct
validation of the gauge interactions of QCD.

The observation that $\simeq 10\%$ of the positron-proton deep
inelastic cross section at HERA is diffractive points to the
importance of final-state gauge interactions as well as a new
perspective to the nature of the hard pomeron.  The same
interactions are responsible for nuclear shadowing and
Sivers-type single-spin asymmetries in semi-inclusive deep
inelastic scattering and in Drell-Yan reactions.  These new
observations are in contradiction to parton model and light-cone
gauge based arguments that final state interactions can be ignored
at leading twist.  The  modifications of the deep inelastic
lepton-proton cross section due to final state interactions are
consistent with color-dipole based scattering models and imply
that the traditional identification of structure functions with
the quark probability distributions computed from the
wavefunctions of the target hadron computed in isolation must be
modified.

Empirical evidence continues to accumulate that the
strange-antistrange quark distributions are not symmetric in the
proton, and that the proton wavefunction contains charm quarks
with large light-cone momentum fractions $x$.  The recent
observation by SELEX that doubly-charmed quarks are produced at
large $x_F$ and small $p_T$ in hadron-nucleus collisions is
evidence for the diffractive dissociation of complex off-shell
Fock states of the projectile.  These observations contradict the
traditional view that sea quarks and gluons are always produced
perturbatively via DGLAP evolution.   The existence of intrinsic
glue and charm has strong consequences for lepton-pair and open
charm production Measurable with antiproton beams.

The dynamical origins of the antishadowing of nuclear structure
functions in the domain $0.1 < x_{Bj} < 1$  is now becoming
understood.  An important conclusion is that antishadowing is
nonuniversal -- different for quarks and antiquarks and different
for strange quarks versus light quarks.  This has important
consequences for antiproton--nucleus Drell-Yan and other inclusive
reactions.

The most dramatic spin correlation ever observed is the large
$4:1$ $A_{NN}$ asymmetry measured in elastic proton-proton
scattering at large CM angles at $\sqrt s \simeq 5 ~ {\rm Gev}$.
If this effect is due to intermediate $uud uud c \bar c$
states---corresponding to the formation of ``octoquark"
resonances---then there should be comparable effects in elastic
$\bar p  p$ scattering.

Dimensional counting rules for hard exclusive processes have now been derived in the
context of nonperturbative QCD using the AdS/CFT correspondence.  The recent measurements
of the predicted $s^{11}$ scaling behavior in deuteron photodisintegration adds further
evidence for the dominance of leading-twist quark-gluon subprocesses. Lowest-order
perturbation theory appears to give the proper scaling, but not the magnitude of the
measured hard exclusive cross sections, suggesting the importance of higher-order
corrections or even a nonperturbative resummation.  I have given evidence that the
running coupling has constant fixed-point behavior, which together with BLM scale fixing,
helps explains the near conformal scaling behavior of the fixed-CM angle cross sections.
The angular distribution of hard exclusive processes is generally consistent with quark
interchange, as predicted from large $N_C$ considerations.  Other important tests of hard
exclusive processes can be carried out with antiproton beams, including timelike proton
form factors.  I have also noted the importance of testing  for the presence of $J=0$
behavior in two-photon reactions such as $\bar p p \to \gamma \gamma$ as a test of the
near-local two-photon couplings to quarks.

A rigorous prediction of QCD is the ``hidden color" of nuclear
wavefunctions at short distances.  I have noted that the two-scale
behavior, as well as the large magnitude of the hard component of
the reduced deuteron form factor at high $Q^2$, gives importance
evidence for this essential feature of the non-Abelian theory.
This points to the importance of studying hard $\bar p$ deuteron
reactions.

I have emphasized several theoretical tools: light-front
wavefunctions as a representation of hadrons at the amplitude
level, the Abelian correspondence principle, and the conformal
template.  It is important to note that there is no
renormalization scale ambiguity in QCD when one uses effective
charges to define the perturbative expansions or when one related
observable to observable, as in commensurate scale relations such
as the generalized Crewther relation.

I have also discussed how the remarkable AdS/CFT duality, which has been established
between supergravity string theory in 10 dimensions and conformal extensions of QCD, is
now providing to a new understanding of QCD at strong coupling and a  new examination at
its nearly-conformal structure.

The antiproton storage ring HESR to be constructed at GSI will
open up a new range of perturbative and nonperturbative tests of
QCD in exclusive and inclusive reactions. I have discussed 21
tests of QCD using antiproton beams which can illuminate these
novel features of QCD.  The proposed experiments include the
formation of exotic hadrons, measurements of timelike generalized
parton distributions, the production of charm at threshold,
transversity measurements in Drell-Yan reactions, and searches for
single-spin asymmetries.  The interactions of antiprotons in
nuclear targets will allow tests of exotic nuclear phenomena such
as color transparency, hidden color, reduced nuclear amplitudes,
and the non-universality of nuclear antishadowing.

\section*{Acknowedgements}

I wish to thank Uli Wiedner and his colleagues for hosting this
outstanding Fermi School of Physics at Varenna. This talk is based
on collaborations with Carl Carlson, Markus Diehl, Guy de Teramond,
Rikard Enberg, John Hiller, Paul Hoyer, Dae Sung Hwang, Gunnar
Ingelman, Volodya Karmanov, Sven Menke, Carlos Merino, Joerg
Raufeisen, Johan Rathsman, and Ivan Schmidt.


\begin{thebibliography}{000}

\bibitem{Greensite:2003bk}
J.~Greensite,
Prog.\ Part.\ Nucl.\ Phys.\  {\bf 51}, 1 (2003) [arXiv:hep-lat/0301023].


\bibitem{Gross:1973id}
D.~J.~Gross and F.~Wilczek,
Phys.\ Rev.\ Lett.\  {\bf 30}, 1343 (1973).

\bibitem{Politzer:1973fx}
H.~D.~Politzer,
Phys.\ Rev.\ Lett.\  {\bf 30}, 1346 (1973).

\bibitem{Klempt:2000ud}
E.~Klempt,
arXiv:hep-ex/0101031.

\bibitem{Rajagopal:2000wf}
K.~Rajagopal and F.~Wilczek,
arXiv:hep-ph/0011333.

\bibitem{Rischke:2003mt}
D.~H.~Rischke,
Prog.\ Part.\ Nucl.\ Phys.\  {\bf 52}, 197 (2004) [arXiv:nucl-th/0305030].

\bibitem{Schwarz:2003du}
D.~J.~Schwarz,
Annalen Phys.\  {\bf 12}, 220 (2003) [arXiv:astro-ph/0303574].

\bibitem{Collins:1989gx}
J.~C.~Collins, D.~E.~Soper and G.~Sterman,
Adv.\ Ser.\ Direct.\ High Energy Phys.\  {\bf 5}, 1 (1988).

\bibitem{Bodwin:1984hc}
G.~T.~Bodwin,
Phys.\ Rev.\ D {\bf 31}, 2616 (1985) [Erratum-ibid.\ D {\bf 34}, 3932 (1986)].

\bibitem{Brodsky:1989pv}
S. J. Brodsky and G. P. Lepage,
in: \textit{Perturbative Quantum Chromodynamics}, edited by A. H.~Mueller (World
Scientific, Singapore 1989).

\bibitem{Maldacena:1997re}
J.~M.~Maldacena,
{\em Adv.\ Theor.\ Math.\ Phys.}\  {\bf 2}, 231 (1998) [{\em Int.\ J.\
Theor.\ Phys.}\
{\bf 38}, 1113 (1999)] [arXiv:hep-th/9711200].

\bibitem{Polchinski:2001tt}
J.~Polchinski and M.~J.~Strassler,
{\em Phys.\ Rev.\ Lett.}\  {\bf 88}, 031601 (2002) [arXiv:hep-th/0109174].

\bibitem{Brower:2002er}
R.~C.~Brower and C.~I.~Tan,
{\em Nucl.\ Phys.\ B} {\bf 662}, 393 (2003) [arXiv:hep-th/0207144].

\bibitem{Andreev:2002aw}
O.~Andreev,
{\em Phys.\ Rev.\ D} {\bf 67}, 046001 (2003) [arXiv:hep-th/0209256].

\bibitem{Fritzsch:1973pi}
H.~Fritzsch, M.~Gell-Mann and H.~Leutwyler,
Phys.\ Lett.\ B {\bf 47}, 365 (1973).

\bibitem{'tHooft:1974hx}
G.~'t Hooft,
Nucl.\ Phys.\ B {\bf 75}, 461 (1974).

\bibitem{Pauli:1985pv}
H.~C.~Pauli and S.~J.~Brodsky,
Phys.\ Rev.\ D {\bf 32}, 1993 (1985).

\bibitem{Hornbostel:1988fb}
K.~Hornbostel, S.~J.~Brodsky and H.~C.~Pauli,
Phys.\ Rev.\ D {\bf 41}, 3814 (1990).

\bibitem{burkardt89}
M. Burkardt, Nucl. Phys. A {\bf 504}, 762 (1989).

\bibitem{Brodsky:1997de}
S.~J.~Brodsky, H.~C.~Pauli and S.~S.~Pinsky,
Phys.\ Rept.\  {\bf 301}, 299 (1998) [arXiv:hep-ph/9705477].


\bibitem{Wilson:1976zj}
K.~G.~Wilson,
Phys.\ Rept.\  {\bf 23}, 331 (1976).

\bibitem{DeGrand:2003xu}
T.~DeGrand,
Int.\ J.\ Mod.\ Phys.\ A {\bf 19}, 1337 (2004) [arXiv:hep-ph/0312241].

\bibitem{Maris:2003vk}
P.~Maris and C.~D.~Roberts,
Int.\ J.\ Mod.\ Phys.\ E {\bf 12}, 297 (2003) [arXiv:nucl-th/0301049].

\bibitem{Dalley:2004rq}
S.~Dalley,
arXiv:hep-ph/0409139.

\bibitem{Dirac:1949cp}
P.~A.~M.~Dirac,
Rev.\ Mod.\ Phys.\  {\bf 21}, 392 (1949).

\bibitem{cdkm}
J.~Carbonell, B.~Desplanques, V.~A.~Karmanov, and J.~F.~Mathiot,
Phys.\ Rep.\  {\bf 300}, 215 (1998) [arXiv:nucl-th/9804029].

\bibitem{Brodsky:2004sb}
S.~J.~Brodsky, {\it Presented at Workshop on QCD Down Under, Barossa Valley
and Adelaide,
Australia, 10-19 Mar 2004.}
arXiv:hep-ph/0408071.

\bibitem{Brodsky:1996hc}
S.~J.~Brodsky and B.~Q.~Ma,
Phys.\ Lett.\ B {\bf 381}, 317 (1996) [arXiv:hep-ph/9604393].


\bibitem{Bjorken:1968dy}
J.~D.~Bjorken,
Phys.\ Rev.\  {\bf 179}, 1547 (1969).

\bibitem{Bloom:1969kc}
E.~D.~Bloom {\it et al.},
Phys.\ Rev.\ Lett.\  {\bf 23}, 930 (1969).

\bibitem{Brodsky:2002cx}
S.~J.~Brodsky, D.~S.~Hwang and I.~Schmidt,
Phys.\ Lett.\ B {\bf 530}, 99 (2002) [arXiv:hep-ph/0201296].

\bibitem{Belitsky:2002sm}
A.~V.~Belitsky, X.~Ji and F.~Yuan,
Nucl.\ Phys.\ B {\bf 656}, 165 (2003) [arXiv:hep-ph/0208038].

\bibitem{Collins:2002kn}
J.~C.~Collins,
Phys.\ Lett.\ B {\bf 536}, 43 (2002) [arXiv:hep-ph/0204004].

\bibitem{Brodsky:2002ue}
S.~J.~Brodsky, P.~Hoyer, N.~Marchal, S.~Peigne and F.~Sannino,
Phys.\ Rev.\ D {\bf 65}, 114025 (2002) [arXiv:hep-ph/0104291].

\bibitem{Brodsky:2004hi}
S.~J.~Brodsky, R.~Enberg, P.~Hoyer and G.~Ingelman,
arXiv:hep-ph/0409119.


\bibitem{Brodsky:1983vf}
S.~J.~Brodsky, C.~R.~Ji and G.~P.~Lepage,
Phys.\ Rev.\ Lett.\  {\bf 51}, 83 (1983).

\bibitem{Brodsky:1988xz}
S.~J.~Brodsky and A.~H.~Mueller,
Phys.\ Lett.\ B {\bf 206}, 685 (1988).


\bibitem{Brodsky:1989qz}
S.~J.~Brodsky and H.~J.~Lu,
Phys.\ Rev.\ Lett.\  {\bf 64}, 1342 (1990).

\bibitem{Brodsky:2004qa}
S.~J.~Brodsky, I.~Schmidt and J.~J.~Yang,
arXiv:hep-ph/0409279.

\bibitem{Zeller:2001hh}
G.~P.~Zeller {\it et al.}  [NuTeV Collaboration],
Phys.\ Rev.\ Lett.\  {\bf 88}, 091802 (2002) [Erratum-ibid.\  {\bf
90}, 239902 (2003)] [arXiv:hep-ex/0110059].


\bibitem{Brodsky:2003px}
S.~J.~Brodsky and G.~F.~de Teramond,
Phys.\ Lett.\ B {\bf 582}, 211 (2004) [arXiv:hep-th/0310227].

\bibitem{Brodsky:1973kr}
S.~J.~Brodsky and G.~R.~Farrar,
{\em Phys.\ Rev.\ Lett.}\  {\bf 31}, 1153 (1973).

\bibitem{Matveev:ra}
V.~A.~Matveev, R.~M.~Muradian and A.~N.~Tavkhelidze,
{\em Lett.\ Nuovo Cim.}\  {\bf 7}, 719 (1973).

\bibitem{Brodsky:1974vy}
S.~J.~Brodsky and G.~R.~Farrar,
{\em Phys.\ Rev.\ D} {\bf 11}, 1309 (1975).

\bibitem{Brodsky:2004qb}
S.~J.~Brodsky,
arXiv:hep-ph/0408069.

\bibitem{Brodsky:2003dn}
S.~J.~Brodsky,
SLAC-PUB-10206
{\it Invited talk at International Conference on Color Confinement
and Hadrons in Quantum Chromodynamics - Confinement 2003, Wako,
Japan, 21-24 Jul 2003}

\bibitem{Parisi:zy}
G.~Parisi,
{\em Phys.\ Lett.\ B} {\bf 39}, 643 (1972).

\bibitem{Brodsky:1980ny}
S. J.~Brodsky, Y.~Frishman, G. P.~Lepage and C.~Sachrajda, {Phys.
Lett.} {\bf 91B}, 239 (1980).

\bibitem{Brodsky:1984xk}
S.~J.~Brodsky, P.~Damgaard, Y.~Frishman and G.~P.~Lepage,
{\em Phys.\ Rev.\ D} {\bf 33}, 1881 (1986).

\bibitem{Brodsky:1985ve}
S.~J.~Brodsky, Y.~Frishman and G.~P.~Lepage,
{\em Phys.\ Lett.\ B} {\bf 167}, 347 (1986).

\bibitem{Braun:2003rp}
V.~M.~Braun, G.~P.~Korchemsky and D.~Muller,
Prog.\ Part.\ Nucl.\ Phys.\  {\bf 51}, 311 (2003)
[arXiv:hep-ph/0306057].

\cite{Lepage:1979zb}
\bibitem{Lepage:1979zb}
G.~P.~Lepage and S.~J.~Brodsky,
Phys.\ Lett.\ B {\bf 87}, 359 (1979).

\bibitem{Brodsky:2000dr}
S.~J.~Brodsky and G.~P.~Lepage,
SLAC-PUB-4947 {\it  In *A.H. Mueller, (ed): Perturbative Quantum
Chromodynamics, 1989, p. 93-240,} and  S.~J.~Brodsky,
SLAC-PUB-8649 {\it  In *Shifman, M. (ed.): At the frontier of
particle physics, vol. 2* 1343-1444.}

\bibitem{Brodsky:1994eh}
S.~J.~Brodsky and H.~J.~Lu,
{\em Phys.\ Rev.\ D} {\bf 51}, 3652 (1995) [arXiv:hep-ph/9405218].

\bibitem{Rathsman:2001xe}
J.~Rathsman,
in {\it Proc. of the 5th International Symposium on Radiative
Corrections (RADCOR 2000) } ed. Howard E. Haber,
arXiv:hep-ph/0101248.

\bibitem{Brodsky:2000cr}
S.~J.~Brodsky, E.~Gardi, G.~Grunberg and J.~Rathsman,
Phys.\ Rev.\ D {\bf 63}, 094017 (2001) [arXiv:hep-ph/0002065].

\bibitem{Grunberg:2001bz}
G.~Grunberg,
JHEP {\bf 0108}, 019 (2001) [arXiv:hep-ph/0104098].

\bibitem{Banks:1981nn}
T.~Banks and A.~Zaks,
Nucl.\ Phys.\ B {\bf 196}, 189 (1982).

\bibitem{Braun:1999te}
V.~M.~Braun, S.~E.~Derkachov, G.~P.~Korchemsky and A.~N.~Manashov,
Nucl.\ Phys.\  {\bf B553}, 355 (1999) [arXiv:hep-ph/9902375].

\bibitem{Brodsky:1995tb}
S.~J.~Brodsky, G.~T.~Gabadadze, A.~L.~Kataev and H.~J.~Lu,
Phys.\ Lett. {\bf B372}, 133 (1996) [arXiv:hep-ph/9512367].

\bibitem{Brodsky:1982gc}
S.~J.~Brodsky, G.~P.~Lepage and P.~B.~Mackenzie,
{\em Phys.\ Rev.\ D} {\bf 28}, 228 (1983).

\bibitem{Brodsky:1997dh}
S.~J.~Brodsky, C.~R.~Ji, A.~Pang and D.~G.~Robertson,
Phys.\ Rev.\ D {\bf 57}, 245 (1998) [arXiv:hep-ph/9705221].

\bibitem{Cornwall:1981zr}
J.~M.~Cornwall,
Phys.\ Rev.\ D {\bf 26}, 1453 (1982).

\bibitem{Brodsky:2002nb}
S.~J.~Brodsky, S.~Menke, C.~Merino and J.~Rathsman,
{\em Phys.\ Rev.\ D} {\bf 67}, 055008 (2003)
[arXiv:hep-ph/0212078].

\bibitem{Baldicchi:2002qm}
M.~Baldicchi and G.~M.~Prosperi,
{\em Phys.\ Rev.\ D} {\bf 66}, 074008 (2002)
[arXiv:hep-ph/0202172].

\bibitem{Furui:2004bq}
S.~Furui and H.~Nakajima,
arXiv:hep-lat/0403021.

\bibitem{Badalian:2004ig}
A.~M.~Badalian and A.~I.~Veselov,
arXiv:hep-ph/0407082.

\bibitem{Ackerstaff:1998yj}
K.~Ackerstaff {\it et al.}  [OPAL Collaboration],
{\em Eur.\ Phys.\ J.\ C} {\bf 7}, 571 (1999)
[arXiv:hep-ex/9808019].

\bibitem{Brodsky:1997jk}
Phys.\ Lett.\ B {\bf 417}, 145 (1998) [arXiv:hep-ph/9707543].

\bibitem{Ellis:2001xc}
J.~R.~Ellis and M.~Karliner,
New J.\ Phys.\  {\bf 4}, 18 (2002) [arXiv:hep-ph/0108259].

\bibitem{Bjorken:1969wi}
J.~D.~Bjorken and S.~J.~Brodsky,
Phys.\ Rev.\ D {\bf 1}, 1416 (1970).

\bibitem{Abe:2002rb}
K.~Abe {\it et al.}  [Belle Collaboration],
Phys.\ Rev.\ Lett.\  {\bf 89}, 142001 (2002)
[arXiv:hep-ex/0205104].

\bibitem{Brodsky:2003hv}
S.~J.~Brodsky, A.~S.~Goldhaber and J.~Lee,
Phys.\ Rev.\ Lett.\  {\bf 91}, 112001 (2003)
[arXiv:hep-ph/0305269].

\bibitem{Brodsky:1987xw}
S.~J.~Brodsky and G.~F.~de Teramond,
Phys.\ Rev.\ Lett.\  {\bf 60}, 1924 (1988).

\bibitem{Matveev:1973ra}
V.~A.~Matveev, R.~M.~Muradian and A.~N.~Tavkhelidze,
Lett.\ Nuovo Cim.\  {\bf 7}, 719 (1973).


\bibitem{Lepage:1980fj}
G.~P.~Lepage and S.~J.~Brodsky,
{\em Phys.\ Rev.\ D} {\bf 22}, 2157 (1980).


\bibitem{Blankenbecler:1973kt}
R.~Blankenbecler, S.~J.~Brodsky, J.~F.~Gunion and R.~Savit,
Phys.\ Rev.\ D {\bf 8}, 4117 (1973).

\bibitem{Andreev:2004sy}
O.~Andreev and W.~Siegel,
arXiv:hep-th/0410131.

\bibitem{Brodsky:1981kj}
S.~J.~Brodsky and G.~P.~Lepage,
Phys.\ Rev.\ D {\bf 24}, 2848 (1981).

\bibitem{Gunion:1973ex}
J.~F.~Gunion, S.~J.~Brodsky and R.~Blankenbecler,
Phys.\ Rev.\ D {\bf 8}, 287 (1973).

\bibitem{Brodsky:1981rp}
S.~J.~Brodsky and G.~P.~Lepage,
Phys.\ Rev.\ D {\bf 24}, 1808 (1981).


\bibitem{Brodsky:2001hv}
For a review of QCD tests in photon-photon collisions see,
S.~J.~Brodsky,
in {\it Proc. of the $e^+ e^-$ Physics at Intermediate Energies
Conference } ed. Diego Bettoni, eConf {\bf C010430}, W01 (2001)
[arXiv:hep-ph/0106294].


\bibitem{Brooks:2000nb}
T.~C.~Brooks and L.~J.~Dixon,
Phys.\ Rev.\ D {\bf 62}, 114021 (2000) [arXiv:hep-ph/0004143].

\bibitem{Brodsky:2000xy}
S.~J.~Brodsky, M.~Diehl and D.~S.~Hwang,
Nucl.\ Phys.\ B {\bf 596}, 99 (2001) [arXiv:hep-ph/0009254].

\bibitem{Diehl:2000xz}
M.~Diehl, T.~Feldmann, R.~Jakob and P.~Kroll,
Nucl.\ Phys.\ B {\bf 596}, 33 (2001) [Erratum-ibid.\ B {\bf 605},
647 (2001)] [arXiv:hep-ph/0009255].

\bibitem{Brodsky:1972vv}
S.~J.~Brodsky, F.~E.~Close and J.~F.~Gunion,
Phys.\ Rev.\ D {\bf 6}, 177 (1972).

\cite{Ji:2003fw}
\bibitem{Ji:2003fw}
X.~d.~Ji, J.~P.~Ma and F.~Yuan,
Phys.\ Rev.\ Lett.\  {\bf 90}, 241601 (2003)
[arXiv:hep-ph/0301141].

\bibitem{Belitsky:2002kj}
A.~V.~Belitsky, X.~D.~Ji and F.~Yuan,
Phys.\ Rev.\ Lett.\  {\bf 91}, 092003 (2003)
[arXiv:hep-ph/0212351].

\bibitem{Brodsky:2003pw}
S.~J.~Brodsky, J.~R.~Hiller, D.~S.~Hwang and V.~A.~Karmanov,
Phys.\ Rev.\ D {\bf 69}, 076001 (2004) [arXiv:hep-ph/0311218].

\bibitem{Jones:1999rz}
M.~K.~Jones {\it et al.}  [Jefferson Lab Hall A Collaboration],
Phys.\ Rev.\ Lett.\  {\bf 84}, 1398 (2000)
[arXiv:nucl-ex/9910005].

\bibitem{Chen:2004tw}
 Y.~C.~Chen, A.~Afanasev, S.~J.~Brodsky, C.~E.~Carlson and
 M.~Vanderhaeghen,
 arXiv:hep-ph/0403058.

\bibitem{Brodsky:2003gs}
S.~J.~Brodsky, C.~E.~Carlson, J.~R.~Hiller and D.~S.~Hwang,
Phys.\ Rev.\ D {\bf 69}, 054022 (2004) [arXiv:hep-ph/0310277].

\bibitem{Brodsky:2004ck}
S.~J.~Brodsky, C.~E.~Carlson, J.~R.~Hiller and D.~S.~Hwang,
arXiv:hep-ph/0408131.

\bibitem{Bertsch:1981py}
G.~Bertsch, S.~J.~Brodsky, A.~S.~Goldhaber and J.~F.~Gunion,
Phys.\ Rev.\ Lett.\  {\bf 47}, 297 (1981).

\bibitem{Brodsky:1980pb}
S.~J.~Brodsky, P.~Hoyer, C.~Peterson and N.~Sakai,
Phys.\ Lett.\ B {\bf 93}, 451 (1980).

\bibitem{Franz:2000ee}
M.~Franz,~V.~Polyakov and K.~Goeke,
Phys.\ Rev.\ D {\bf 62}, 074024 (2000) [arXiv:hep-ph/0002240].

\bibitem{Brodsky:1984nx}
S.~J.~Brodsky, J.~C.~Collins, S.~D.~Ellis, J.~F.~Gunion and
A.~H.~Mueller,
DOE/ER/40048-21 P4
{\it Proc. of 1984 Summer Study on the SSC, Snowmass, CO, Jun 23 -
Jul 13, 1984}

\bibitem{Harris:1995jx}
B.~W.~Harris, J.~Smith and R.~Vogt,
Nucl.\ Phys.\ B {\bf 461}, 181 (1996) [arXiv:hep-ph/9508403].

\bibitem{Brodsky:2000zc}
S.~J.~Brodsky, E.~Chudakov, P.~Hoyer and J.~M.~Laget,
Phys.\ Lett.\ B {\bf 498}, 23 (2001) [arXiv:hep-ph/0010343].

\bibitem{Gittelman:1975ix}
B.~Gittelman, K.~M.~Hanson, D.~Larson, E.~Loh, A.~Silverman and
G.~Theodosiou,
Phys.\ Rev.\ Lett.\  {\bf 35}, 1616 (1975).

\bibitem{Arnold:1975dd}
R.~G.~Arnold {\it et al.},
Phys.\ Rev.\ Lett.\  {\bf 35}, 776 (1975).

\bibitem{Brodsky:1976rz}
S.~J.~Brodsky and B.~T.~Chertok,
Phys.\ Rev.\ D {\bf 14}, 3003 (1976).

\bibitem{Farrar:1991qi}
G.~R.~Farrar, K.~Huleihel and H.~y.~Zhang,
Phys.\ Rev.\ Lett.\  {\bf 74}, 650 (1995).

\bibitem{Miyama:1999yp}
M.~Miyama,
Nucl.\ Phys.\ Proc.\ Suppl.\  {\bf 79}, 620 (1999)
[arXiv:hep-ph/9905559].

\bibitem{Brodsky:2002rv}
S.~J.~Brodsky, D.~S.~Hwang and I.~Schmidt,
Nucl.\ Phys.\ B {\bf 642}, 344 (2002) [arXiv:hep-ph/0206259].

\bibitem{Boer:2002ju}
D.~Boer, S.~J.~Brodsky and D.~S.~Hwang,
Phys.\ Rev.\ D {\bf 67}, 054003 (2003) [arXiv:hep-ph/0211110].

\bibitem{Brodsky:1994kg}
S.~J.~Brodsky, M.~Burkardt and I.~Schmidt,
Nucl.\ Phys.\ B {\bf 441}, 197 (1995) [arXiv:hep-ph/9401328].


\bibitem{Berger:1979du}
E.~L.~Berger and S.~J.~Brodsky,
Phys.\ Rev.\ Lett.\  {\bf 42}, 940 (1979).


\bibitem{Brodsky:1999mz}
S.~J.~Brodsky, J.~Rathsman and C.~Merino,
Phys.\ Lett.\ B {\bf 461}, 114 (1999) [arXiv:hep-ph/9904280].

\bibitem{Tomboulis:jn}
E.~Tomboulis,
Phys.\ Rev.\ D {\bf 8}, 2736 (1973).

\bibitem{Srivastava:2000cf}
P.~P.~Srivastava and S.~J.~Brodsky,
Phys.\ Rev.\ D {\bf 64}, 045006 (2001) [arXiv:hep-ph/0011372].

\bibitem{Karmanov:1991fv}
V.~A.~Karmanov and A.~V.~Smirnov,
Nucl.\ Phys.\ A {\bf 546}, 691 (1992).

\bibitem{Dalley:ug}
S.~Dalley,
Nucl.\ Phys.\ B (Proc.\ Suppl.)  {\bf 108}, 145 (2002).

\bibitem{Brodsky:2003gk}
S.~J.~Brodsky,
in {\it Nagoya 2002, Strong coupling gauge theories and effective
field theories, 1-18.} [arXiv:hep-th/0304106].

\bibitem{Raufeisen:2004dg}
J.~Raufeisen and S.~J.~Brodsky,
arXiv:hep-th/0408108.

\bibitem{Brodsky:1980zm}
S.~J.~Brodsky and S.~D.~Drell,
Phys.\ Rev.\ D {\bf 22}, 2236 (1980).

\bibitem{Brodsky:1998hn}
S.~J.~Brodsky and D.~S.~Hwang,
Nucl.\ Phys.\ B {\bf 543}, 239 (1999) [arXiv:hep-ph/9806358].


\bibitem{Brodsky:1971zh}
 S.~J.~Brodsky, F.~E.~Close and J.~F.~Gunion,
 Phys.\ Rev.\ D {\bf 5}, 1384 (1972).


\bibitem{Brodsky:1973hm}
S.~J.~Brodsky, F.~E.~Close and J.~F.~Gunion,
Phys.\ Rev.\ D {\bf 8}, 3678 (1973).

\bibitem{Brodsky:2000ii}
S.~J.~Brodsky, D.~S.~Hwang, B.~Q.~Ma and I.~Schmidt,
Nucl.\ Phys.\ B {\bf 593}, 311 (2001) [arXiv:hep-th/0003082].

\bibitem{Jaffe:1989jz}
R.~L.~Jaffe and A.~Manohar,
Nucl.\ Phys.\ B {\bf 337}, 509 (1990).

\bibitem{Ji:2002qa}
X.~d.~Ji,
Nucl.\ Phys.\ Proc.\ Suppl.\  {\bf 119}, 41 (2003)
[arXiv:hep-lat/0211016].

\bibitem{Landshoff:ew}
P.~V.~Landshoff,
Phys.\ Rev.\ D {\bf 10}, 1024 (1974).

\bibitem{Brodsky:1979qm}
S.~J.~Brodsky and G.~P.~Lepage,
SLAC-PUB-2294 {\em Workshop on Current Topics in High Energy
Physics}, Cal Tech., Pasadena, Calif., Feb 13-17, 1979.

\bibitem{Lepage:1979za}
G.~P.~Lepage and S.~J.~Brodsky,
Phys.\ Rev.\ Lett.\  {\bf 43}, 545 (1979).

\bibitem{Efremov:1980rn}
A. V.~Efremov and A. V.~Radyushkin,
Theor.\ Math.\ Phys.\  {\bf 42} (1980) 97;

\bibitem{Muller:1994cn}
D.~Muller,
Phys.\ Rev.\ D {\bf 51}, 3855 (1995) [arXiv:hep-ph/9411338].

\bibitem{Ball:1998ff}
P.~Ball and V.~M.~Braun,
Nucl.\ Phys.\ B {\bf 543}, 201 (1999) [arXiv:hep-ph/9810475].

\bibitem{Beneke:2000ry}
M.~Beneke, G.~Buchalla, M.~Neubert and C.~T.~Sachrajda,
Nucl.\ Phys.\ B {\bf 591}, 313 (2000) [arXiv:hep-ph/0006124].



\bibitem{Keum:2000wi}
Y.~Y.~Keum, H.~N.~Li and A.~I.~Sanda,
Phys.\ Rev.\ D {\bf 63}, 054008 (2001) [arXiv:hep-ph/0004173].

\bibitem{Szczepaniak:1990dt}
A.~Szczepaniak, E.~M.~Henley and S.~J.~Brodsky,
Phys.\ Lett.\ B {\bf 243}, 287 (1990).

\bibitem{Brodsky:2001jw}
A review of QCD analyses of exclusive $B$ decays is given in
S.~J.~Brodsky,
Published in {\it Ise-Shima 2001, B physics and CP violation,
229-234} [arXiv:hep-ph/0104153].

\bibitem{Chua:2002wn}
C.~K.~Chua, W.~S.~Hou and S.~Y.~Tsai,
Phys.\ Rev.\ D {\bf 66}, 054004 (2002) [arXiv:hep-ph/0204185].

\bibitem{Drell:1970km}
S. D.~Drell and T.~Yan,
Phys.\ Rev.\ Lett.\  {\bf 24}, 181 (1970).

\bibitem{West:1970av}
G. B.~West,
Phys.\ Rev.\ Lett.\  {\bf 24}, 1206 (1970).

\bibitem{Chernyak:1977fk}
V.~L.~Chernyak, A.~R.~Zhitnitsky and V.~G.~Serbo,
JETP Lett.\  {\bf 26}, 594 (1977).

\bibitem{Chernyak:1980dk}
V.~L.~Chernyak, V.~G.~Serbo and A.~R.~Zhitnitsky,
Sov.\ J.\ Nucl.\ Phys.\  {\bf 31}, 552 (1980).

\bibitem{Farrar:1979aw}
G.~R.~Farrar and D.~R.~Jackson,
Phys.\ Rev.\ Lett.\  {\bf 43}, 246 (1979).

\bibitem{Duncan:1980hi}
A.~Duncan and A.~H.~Mueller,
Phys.\ Rev.\  {\bf D21}, 1636 (1980).

\bibitem{Brodsky:1998dh}
S.~J.~Brodsky, C.~R.~Ji, A.~Pang and D.~G.~Robertson,
Phys.\ Rev.\ D {\bf 57}, 245 (1998) [arXiv:hep-ph/9705221].

\bibitem{Beneke:2002bs}
M.~Beneke,
Nucl.\ Phys.\ Proc.\ Suppl.\  {\bf 111}, 62 (2002)
[arXiv:hep-ph/0202056].

\bibitem{Muller:1994fv}
D.~Muller, D.~Robaschik, B.~Geyer, F.~M.~Dittes and J.~Horejsi,
Fortsch.\ Phys.\ {\bf 42}, 101 (1994) [arXiv:hep-ph/9812448].

\bibitem{Melic:1998qr}
B.~Melic, B.~Nizic and K.~Passek,
Phys.\ Rev.\ D {\bf 60}, 074004 (1999) [arXiv:hep-ph/9802204].

\bibitem{Szczepaniak:1998sa}
A.~Szczepaniak, A.~Radyushkin and C.~Ji, { Phys. Rev.} {\bf D57},
2813 (1998) [arXiv:hep-ph/9708237].

\bibitem{Peskin:1979mn}
M.~E.~Peskin,
Phys.\ Lett.\  {\bf B88}, 128 (1979).


\bibitem{jsl}
C-R Ji, A. F. Sill and R. M. Lombard-Nelsen, Phys. Rev. {\bf D36},
165 (1987).


\bibitem{Chernyak:1984bm}
V.~L.~Chernyak and I.~R.~Zhitnitsky,
Nucl.\ Phys.\  {\bf B246}, 52 (1984).

\bibitem{Chernyak:1989nv}
V.~L.~Chernyak, A.~A.~Ogloblin and I.~R.~Zhitnitsky,
Z.\ Phys.\  {\bf C42}, 583 (1989).

\bibitem{Stefanis:1999wy}
N.~G.~Stefanis,
Eur.\ Phys.\ J.\  {\bf C7}, 1 (1999) [arXiv:hep-ph/9911375].

\bibitem{Brodsky:1983st}
S.~J.~Brodsky, J.~R.~Ellis, J.~S.~Hagelin and C.~T.~Sachrajda,
Nucl.\ Phys.\ B {\bf 238}, 561 (1984).

\bibitem{Kuramashi:2000hw}
Y.~Kuramashi  [JLQCD Collaboration],
[arXiv:hep-ph/0103264].

\bibitem{Brodsky:1981sx}
S.~J.~Brodsky, G.~P.~Lepage and S.~A.~Zaidi,
Phys.\ Rev.\  {\bf D23}, 1152 (1981).

\bibitem{ks}
I. D. King and C. T. Sachrajda, Nucl. Phys. {\bf B279}, 785
(1987).

\bibitem{gs}
M. Gari and N. Stefanis, Phys. Lett. {\bf B175}, 462 (1986), M.
Gari and N. Stefanis, Phys. Lett. {\bf 187B}, 401 (1987).


\bibitem{Carlson:1986mm}
C.~E.~Carlson,
Phys.\ Rev.\  {\bf D34}, 2704 (1986).

\bibitem{Stoler:1993yk}
P.~Stoler,
Phys.\ Rept.\  {\bf 226}, 103 (1993).

\bibitem{Pobylitsa:2001cz}
P.~V.~Pobylitsa,~V.~Polyakov and M.~Strikman,
Phys.\ Rev.\ Lett.\  {\bf 87}, 022001 (2001)
[arXiv:hep-ph/0101279].

\bibitem{perdrisat} M.~K.~Jones {\it et al.}  [Jefferson Lab Hall A
Collaboration],
Phys.\ Rev.\ Lett.\  {\bf 84}, 1398 (2000).
[arXiv:nucl-ex/9910005].
O.~Gayou {\it et al.}  [Jefferson Lab Hall A Collaboration],
Phys.\ Rev.\ Lett.\  {\bf 88}, 092301 (2002).
[arXiv:nucl-ex/0111010].

\bibitem{walecka} R. G. Sachs, Phys. Rev. {\bf 126}, 2256 (1962); J. D.
Walecka, Nuovo Cim.\ {\bf 11}, 821 (1959).

\bibitem{Guichon:2003qm}
P.~A.~M.~Guichon and M.~Vanderhaeghen,
Phys.\ Rev.\ Lett.\  {\bf 91}, 142303 (2003)
[arXiv:hep-ph/0306007].

\bibitem{Blunden:2003sp}
P.~G.~Blunden, W.~Melnitchouk and J.~A.~Tjon,
Phys.\ Rev.\ Lett.\  {\bf 91}, 142304 (2003)
[arXiv:nucl-th/0306076].

\bibitem{acg81} R.~G.~Arnold, C.~E.~Carlson, and F.~Gross,
Phys.\ Rev.\ C {\bf 23}, 363 (1981), and references cited therein.

\bibitem{baldini} R.~Baldini, S.~Dubnicka, P.~Gauzzi, S.~Pacetti,
E.~Pasqualucci, and Y.~Srivastava,
Eur.\ Phys.\ J.\ C {\bf 11}, 709 (1999);
R.~Baldini {\it et al.},
{\it Proc. of the $e^+ e^-$ Physics at Intermediate Energies
Conference } ed. Diego Bettoni, eConf {\bf C010430}, T20 (2001)
[hep-ph/0106006].

\bibitem{Geshkenbein74}
For a discussion on the validity of continuing spacelike form
factors to the timelike region, see, B.~V.~Geshkenbein,
B.~.L.~Ioffe, and M.~A.~Shifman, Sov. J. Nucl. Phys. {\bf 20}, 66
(1975) [Yad.\ Fiz.\ {\bf 20}, 128 (1974)].

\bibitem{seealso}  See also
R.~Calabrese,
in {\it Proc. of the $e^+ e^-$ Physics at Intermediate Energies
Conference } ed. Diego Bettoni, eConf {\bf C010430}, W07 (2001);
H.~W.~Hammer,
{\it ibid.}, W08 (2001) [arXiv:hep-ph/0105337];
Carl~E.~Carlson,
{\it ibid.}, W09 (2001) [arXiv:hep-ph/0106290];
M.~Karliner,
{\it ibid.}, W10 (2001) [arXiv:hep-ph/0108106].

\bibitem{Brodsky:xz}
S.~J.~Brodsky and A.~H.~Mueller,
Phys.\ Lett.\ B {\bf 206}, 685 (1988).

\bibitem{Bjorken:kk}
J.~D.~Bjorken,
Nucl.\ Phys.\ Proc.\ Suppl.\  {\bf 11}, 325 (1989).

\bibitem{Beneke:2001ev}
M.~Beneke, G.~Buchalla, M.~Neubert and C.~T.~Sachrajda,
Nucl.\ Phys.\ B {\bf 606}, 245 (2001) [arXiv:hep-ph/0104110].

\bibitem{belitsky02} A.~V.~Belitsky, X.~Ji, and F.~Yuan,
Phys.\ Rev.\ Lett.\  {\bf 91}, 092003 (2003).
arXiv:hep-ph/0212351.

\bibitem{Ralston:2003mt}
J.~P.~Ralston and P.~Jain,
Phys.\ Rev.\ D {\bf 69}, 053008 (2004) [arXiv:hep-ph/0302043].

\bibitem{Miller:2002qb}
G.~A.~Miller and M.~R.~Frank,
Phys.\ Rev.\ C {\bf 65}, 065205 (2002) [arXiv:nucl-th/0201021].

\bibitem{d} A.~Z.~Dubnickova, S.~Dubnicka, and M.~P.~Rekalo,
Nuovo Cim.\ A {\bf 109}, 241 (1996);
S.~Rock,
{\it Proc. of the $e^+ e^-$ Physics at Intermediate Energies
Conference } ed. Diego Bettoni, eConf {\bf C010430}, W14 (2001)
[hep-ex/0106084].

\bibitem{ijl} F.~Iachello, A.~D.~Jackson, and A.~Lande,
Phys.\ Lett.\ B {\bf 43}, 191 (1973).

\bibitem{Kronfeld:1991kp}
A.~S.~Kronfeld and B.~Nizic,
Phys.\ Rev.\ {\bf D44}, 3445 (1991).

\bibitem{Guichon:1998xv}
P.~A.~Guichon and M.~Vanderhaeghen,
Prog.\ Part.\ Nucl.\ Phys.\  {\bf 41}, 125 (1998)
[arXiv:hep-ph/9806305].

\bibitem{Shupe:vg}
M.~A.~Shupe {\it et al.},
Phys.\ Rev.\ D {\bf 19}, 1921 (1979).


\bibitem{Weisberger:1972hk}
 W.~I.~Weisberger,
 Phys.\ Rev.\ D {\bf 5}, 2600 (1972).

\bibitem{Damashek:1969xj}
 M.~Damashek and F.~J.~Gilman,
 Phys.\ Rev.\ D {\bf 1}, 1319 (1970).



\bibitem{Isgur:1989iw}
N.~Isgur and C. H.~Llewellyn Smith, { Phys. Lett.} {\bf B217}, 535
(1989).

\bibitem{Radyushkin:1998rt}
A. V.~Radyushkin, {Phys. Rev.} {\bf D58}, 114008 (1998)
hep-ph/9803316.

\bibitem{Bolz:1996sw}
J.~Bolz and P.~Kroll, { Z. Phys.} {\bf A356}, 327 (1996)
hep-ph/9603289.


\bibitem{Diehl:1998kh}
M.~Diehl, T.~Feldmann, R.~Jakob and P.~Kroll,
Eur.\ Phys.\ J.\ C {\bf 8}, 409 (1999) [arXiv:hep-ph/9811253].

\bibitem{Huang:2001ej}
H.~W.~Huang, P.~Kroll and T.~Morii,
Eur.\ Phys.\ J.\ C {\bf 23}, 301 (2002) [arXiv:hep-ph/0110208].

\bibitem{Lepage:1982gd}
G.~P.~Lepage, S.~J.~Brodsky, T.~Huang and P.~B.~Mackenzie,
and S.~J.~Brodsky, T.~Huang and G.~P.~Lepage,
{\em  In *Banff 1981, Proceedings, Particles and Fields 2*,
143-199}.

\bibitem{Li:1992nu}
H.~N.~Li and G.~Sterman,
Nucl.\ Phys.\ B {\bf 381}, 129 (1992).

\bibitem{Weiss:2002ec}
C.~Weiss,
[arXiv:hep-ph/0206295].

\bibitem{Chernyak:1999cj}
V.~Chernyak,
[arXiv:hep-ph/9906387].

\bibitem{Jones:uu}
M.~K.~Jones  [Jefferson Lab Hall A Collaboration],
Nucl.\ Phys.\ A {\bf 699}, 124 (2002).
%
\bibitem{Ambrogiani:1999bh}
M.~Ambrogiani {\it et al.}  [E835 Collaboration],
Phys.\ Rev.\ D {\bf 60}, 032002 (1999).

\bibitem{Brodsky:1997fj}
S.~J.~Brodsky and M.~Karliner,
Phys.\ Rev.\ Lett.\  {\bf 78}, 4682 (1997) [arXiv:hep-ph/9704379].

\bibitem{Brodsky:2001yt}
S.~J.~Brodsky and S.~Gardner,
Phys.\ Rev.\ D {\bf 65}, 054016 (2002) [arXiv:hep-ph/0108121].

\bibitem{Chang:2001yf}
C.~H.~Chang and W.~S.~Hou,
Phys.\ Rev.\ D {\bf 64}, 071501 (2001).

\bibitem{Arr}
R.~L.~Anderson \etal, Phys.~Rev.~Lett. {\bf 30}, 627 (1973)

\bibitem{Besch:sx}
H.~J.~Besch, F.~Krautschneider, K.~P.~Sternemann and W.~Vollrath,
Z.\ Phys.\ C {\bf 16}, 1 (1982).

\bibitem{Melnitchouk:2001zy}
W.~Melnitchouk,
Nucl.\ Phys.\ A {\bf 699}, 278 (2002) [arXiv:hep-ph/0106262].

\bibitem{Diehl:2001fv}
M.~Diehl, P.~Kroll and C.~Vogt,
Phys.\ Lett.\ B {\bf 532}, 99 (2002) [arXiv:hep-ph/0112274].

\bibitem{Savinov:2001wj}
V.~Savinov,
in {\it Proc. of the $e^+ e^-$ Physics at Intermediate Energies
Conference } ed. Diego Bettoni, eConf {\bf C010430}, W03 (2001)
[arXiv:hep-ex/0106013].

\bibitem{Boyer}
J.~Boyer \etal, Phys. Rev. Lett. {\bf 56}, 207 (1980); TPC/Two
Gamma Collaboration (H. Aihara \etal), Phys. Rev. Lett. {\bf 57},
404 (1986).

\bibitem{Sivers:1976dg}
D.~Sivers, S.~J.~Brodsky and R.~Blankenbecler,
Phys.\ Rept.\ {\bf 23}, 1 (1976).

\bibitem{Holt:1990ze}
R.~J.~Holt,
Phys.\ Rev.\  {\bf C41}, 2400 (1990).

\bibitem{Bochna:1998ca}
C.~Bochna {\it et al.}  [E89-012 Collaboration],
Phys.\ Rev.\ Lett.\  {\bf 81}, 4576 (1998)
[arXiv:nucl-ex/9808001].

\bibitem{Rossi:2004qm}
P.~Rossi {\it et al.}  [CLAS Collaboration],
arXiv:hep-ph/0405207.

\bibitem{Brodsky:1983kb}
S.~J.~Brodsky and J.~R.~Hiller,
Phys.\ Rev.\  {\bf C28}, 475 (1983).

\bibitem{Belz:1995ge}
J.~E.~Belz {\em et al.},
Phys.\ Rev.\ Lett.\ {\bf 74}, 646 (1995).

\bibitem{Lee:1988pi}
T.~S.~Lee,
CONF-8805140-10.

\bibitem{Brodsky:1994kf}
S.~J.~Brodsky, L.~Frankfurt, J.~F.~Gunion, A.~H.~Mueller and
M.~Strikman,
Phys.\ Rev.\  {\bf D50}, 3134 (1994), [arXiv:hep-ph/9402283].

\bibitem{Collins:1997sr}
J.~C.~Collins,
Phys.\ Rev.\ D {\bf 57}, 3051 (1998) [Erratum-ibid.\ D {\bf 61},
019902 (2000)] [arXiv:hep-ph/9709499].

\bibitem{Brodsky:1998sr}
S.~J.~Brodsky, M.~Diehl, P.~Hoyer and S.~Peigne,
Phys.\ Lett.\ B {\bf 449}, 306 (1999) [arXiv:hep-ph/9812277].

\bibitem{Ji:1997nm}
X.~Ji,
Phys.\ Rev.\  {\bf D55}, 7114 (1997), [arXiv:hep-ph/9609381].

\bibitem{Radyushkin:1997ki}
A.~V.~Radyushkin,
Phys.\ Rev.\  {\bf D56}, 5524 (1997) [hep-ph/9704207].

\bibitem{Diehl:1999tr}
M.~Diehl, T.~Feldmann, R.~Jakob and P.~Kroll,
Phys.\ Lett.\  {\bf B460}, 204 (1999) [arXiv:hep-ph/9903268].

\bibitem{Diehl:1999kh}
M.~Diehl, T.~Feldmann, R.~Jakob and P.~Kroll,
Eur.\ Phys.\ J.\  {\bf C8}, 409 (1999), [arXiv:hep-ph/9811253].

\bibitem{Collins:1997fb}
J.~C.~Collins, L.~Frankfurt and M.~Strikman,
Phys.\ Rev.\  {\bf D56}, 2982 (1997) [arXiv:hep-ph/9611433].

\bibitem{Diehl:2000uv}
M.~Diehl, T.~Gousset, and B.~Pire,
[arXiv:hep-ph/0003233].

\bibitem{Ramsey-Musolf:2002cy}
M.~Ramsey-Musolf and M.~B.~Wise,
Phys.\ Rev.\ Lett.\  {\bf 89}, 041601 (2002)
[arXiv:hep-ph/0201297].

\bibitem{Anjos:2001jr}
J.~C.~Anjos, J.~Magnin and G.~Herrera,
Phys.\ Lett.\ B {\bf 523}, 29 (2001) [arXiv:hep-ph/0109185].

\bibitem{Ocherashvili:2004hi}
A.~Ocherashvili {\it et al.}  [SELEX Collaboration],
arXiv:hep-ex/0406033.

\bibitem{Vogt:1995tf}
R.~Vogt and S.~J.~Brodsky,
Phys.\ Lett.\ B {\bf 349}, 569 (1995) [arXiv:hep-ph/9503206].


\bibitem{Portheault:2004xy}
B.~Portheault,
arXiv:hep-ph/0406226.

\bibitem{vonSmekal:1997is}
L.~von Smekal, R.~Alkofer and A.~Hauck,
Phys.\ Rev.\ Lett.\  {\bf 79}, 3591 (1997) [arXiv:hep-ph/9705242].

\bibitem{Zwanziger:2003cf}
D.~Zwanziger,
Phys.\ Rev.\ D {\bf 69}, 016002 (2004) [arXiv:hep-ph/0303028].

\bibitem{Howe:2002rb}
D.~M.~Howe and C.~J.~Maxwell,
Phys.\ Lett.\ B {\bf 541}, 129 (2002) [arXiv:hep-ph/0204036].

\bibitem{Howe:2003mp}
D.~M.~Howe and C.~J.~Maxwell,
Phys.\ Rev.\ D {\bf 70}, 014002 (2004) [arXiv:hep-ph/0303163].

\bibitem{Furui:2003mz}
S.~Furui and H.~Nakajima,
arXiv:hep-lat/0309166.

\bibitem{Mattingly:ej}
A.~C.~Mattingly and P.~M.~Stevenson,
Phys.\ Rev.\ D {\bf 49}, 437 (1994) [arXiv:hep-ph/9307266].

\bibitem{Grunberg:1980ja}
G.~Grunberg,
{\em Phys.\ Lett.} {\bf B95}, 70 (1980) [Erratum-ibid. {\bf B110},
501 (1982)].
%
G.~Grunberg,
{\em Phys.\ Rev.} {\bf D29}, 2315 (1984).

\bibitem{Brodsky:1998mf}
S.~J.~Brodsky, M.~S.~Gill, M.~Melles and J.~Rathsman,
{\em Phys.\ Rev.\ D} {\bf 58}, 116006 (1998)
[arXiv:hep-ph/9801330].

\bibitem{Brodsky:1999fr}
S.~J.~Brodsky, M.~Melles and J.~Rathsman,
{\em Phys.\ Rev.\ D} {\bf 60}, 096006 (1999)
[arXiv:hep-ph/9906324].



\bibitem{LeDiberder:1992fr}
F.~Le Diberder and A.~Pich,
Phys.\ Lett. {\bf B289}, 165 (1992).

\bibitem{Beneke:1994qe}
M.~Beneke and V.~M.~Braun,
Phys.\ Lett.\ B {\bf 348}, 513 (1995) [arXiv:hep-ph/9411229].

\bibitem{Ball:1995ni}
P.~Ball, M.~Beneke and V.~M.~Braun,
{\em Nucl.\ Phys.} {\bf B452}, 563 (1995) [arXiv:hep-ph/9502300].

\bibitem{Dokshitzer:1995qm}
Y.~L.~Dokshitzer, G.~Marchesini and B.~R.~Webber,
{\em Nucl.\ Phys.} {\bf B469}, 93 (1996) [arXiv:hep-ph/9512336].

\bibitem{Melic:2001wb}
B.~Melic, B.~Nizic and K.~Passek,
Phys.\ Rev.\ D {\bf 65}, 053020 (2002) [arXiv:hep-ph/0107295].

\bibitem{Brodsky:2002st}
S.~J.~Brodsky,
Published in {\it Newport News 2002, Exclusive processes at high momentum
transfer 1-33.}
[arXiv:hep-ph/0208158.]

\bibitem{Bloom:1971ye}
E.~D.~Bloom and F.~J.~Gilman,
Phys.\ Rev.\ D {\bf 4}, 2901 (1971).

\bibitem{Rey:1998ik}
S.~J.~Rey and J.~T.~Yee,
{\em Eur.\ Phys.\ J.\ C} {\bf 22}, 379 (2001) [arXiv:hep-th/9803001].

\bibitem{'tHooft:1973jz}
G.~'t Hooft,
{\em Nucl.\ Phys.\ B} {\bf 72}, 461 (1974).

\bibitem{Sivers:1975dg}
D.~W.~Sivers, S.~J.~Brodsky and R.~Blankenbecler,
{\em Phys.\ Rept.}\  {\bf 23}, 1 (1976).

\bibitem{deTeramond:2004qd}
G.~F.~de Teramond and S.~J.~Brodsky,
arXiv:hep-th/0409074.

\bibitem{Ji:bw}
X.~D.~Ji, F.~Yuan and J.~P.~Ma,
{\em Phys.\ Rev.\ Lett.}\  {\bf 90}, 241601 (2003).

\bibitem{Dutta:2003mk}
D.~Dutta {\it et al.}  [Jefferson Lab E940104 Collaboration],
Phys.\ Rev.\ C {\bf 68}, 021001 (2003) [arXiv:nucl-ex/0305005].

\bibitem{Aclander:2004zm}
J.~Aclander {\it et al.},
Phys.\ Rev.\ C {\bf 70}, 015208 (2004) [arXiv:nucl-ex/0405025].

\bibitem{Court:1986dh}
R.~Court {\it et al.},
Phys.\ Rev.\ Lett.\  {\bf 57}, 507 (1986).

\bibitem{Zhu:2004dy}
L.~Y.~Zhu  [Jefferson Lab E94-104 Collaboration],
arXiv:nucl-ex/0409018.

\bibitem{Frankfurt:1988nt}
L.~L.~Frankfurt and M.~I.~Strikman,
Phys.\ Rept.\  {\bf 160}, 235 (1988).

\bibitem{Jain:1996dd}
P.~Jain, B.~Pire and J.~P.~Ralston,
Phys.\ Rept.\  {\bf 271}, 67 (1996) [arXiv:hep-ph/9511333].

\bibitem{Carroll:1988rp}
A.~S.~Carroll {\em et al.},
Phys.\ Rev.\ Lett.\ {\bf 61}, 1698 (1988).

\bibitem{Mardor:1998fz}
Y.~Mardor {\em et al.},
Phys.\ Lett.\  {\bf B437}, 257 (1998) [arXiv:nucl-ex/9710002].

\bibitem{Leksanov:2001ui}
A.~Leksanov {\it et al.},
Phys.\ Rev.\ Lett.\  {\bf 87}, 212301 (2001) [arXiv:hep-ex/0104039].

\bibitem{deTeramond:1998ny}
G.~F.~de Teramond, R.~Espinoza and M.~Ortega-Rodriguez,
Phys.\ Rev.\  {\bf D58}, 034012 (1998) [arXiv:hep-ph/9708202].

\bibitem{Ralston:1986zn}
J.~P.~Ralston and B.~Pire,
Phys.\ Rev.\ Lett.\  {\bf 57}, 2330 (1986).

\bibitem{Frankfurt:1999tq}
L.~Frankfurt, G.~A.~Miller and M.~Strikman,
Found.\ Phys.\  {\bf 30}, 533 (2000) [arXiv:hep-ph/9907214].

\bibitem{Nikolaev:2000sh}
N.~N.~Nikolaev, W.~Schafer and G.~Schwiete,
Phys.\ Rev.\ D {\bf 63}, 014020 (2001) [arXiv:hep-ph/0009038].

\bibitem{Aitala:2000hc}
E.~M.~Aitala {\it et al.}  [E791 Collaboration],
Phys.\ Rev.\ Lett.\  {\bf 86}, 4773 (2001) [arXiv:hep-ex/0010044].

\bibitem{Ashery:2002jx}
D.~Ashery,
Comments Nucl.\ Part.\ Phys.\  {\bf 2}, A235 (2002).

\bibitem{Aitala:2000hb}
E.~M.~Aitala {\it et al.}  [E791 Collaboration],
Phys.\ Rev.\ Lett.\  {\bf 86}, 4768 (2001) [arXiv:hep-ex/0010043].

\bibitem{Gronberg:1998fj}
J.~Gronberg {\it et al.}  [CLEO Collaboration],
Phys.\ Rev.\ D {\bf 57}, 33 (1998) [arXiv:hep-ex/9707031].

\bibitem{Efremov:1978rn}
A.~V.~Efremov and A.~V.~Radyushkin,
Theor.\ Math.\ Phys.\  {\bf 42}, 97 (1980) [Teor.\ Mat.\ Fiz.\ {\bf 42},
147 (1980)].

\bibitem{BHDP}
S.~Brodsky, ~M.~Diehl, ~P.~Hoyer, and ~S.~Peigne, unpublished.

\bibitem{MillerFrankfurtStrikman}
L. Frankfurt, G. A. Miller, and M. Strikman, {Phys. Lett.} {\bf B304}, 1 (1993),
[arXiv:hep-ph/9305228].

\bibitem{Braun:2002wu}
V.~M.~Braun, D.~Y.~Ivanov, A.~Schafer and L.~Szymanowski,
Nucl.\ Phys.\ B {\bf 638}, 111 (2002) [arXiv:hep-ph/0204191].

\bibitem{Chernyak:2002jc}
V.~L.~Chernyak,
[arXiv:hep-ph/0206144].

\bibitem{Ivanov:2002hp}
I.~P.~Ivanov, N.~N.~Nikolaev, W.~Schafer, B.~G.~Zakharov and V.~R.~Zoller,
arXiv:hep-ph/0207045.


\bibitem{Brodsky:1995ht}
S.~J.~Brodsky and H.~J.~Lu,
arXiv:hep-ph/9506322.

\bibitem{Crewther:1972kn}
R.~J.~Crewther,
Phys.\ Rev.\ Lett.\  {\bf 28}, 1421 (1972).

\bibitem{CCFRL1}
CCFR Collaboration, W.C. Leung, {\it et al.},  Phys. Lett. {\bf B317}, 655
(1993).

\bibitem{CCFRL2}
CCFR and NuTeV Collaboration, presented by D. Harris at XXX Recontre de
Moriond, 1995,
presented by J. H. Kim at the  European Conference on High Energy Physics,
Brussels, July
1995.

\bibitem{KS}
A. L. Kataev, A.V. Sidorov,  Phys. Lett. {\bf  B331}, 179 (1994).

\bibitem{CCFRQ}
CCFR Collaboration, P.Z. Quinta, {\it et al.},  Phys. Rev. Lett. {\bf 71},
1307 (1993).

\bibitem{tHooft}
G. 't Hooft, in the Proceedings of the International School, Erice, Italy,
1977, edited
by A. Zichichi, Subnuclear Series Vol. 15 (Plenum, New York, 1979).

\bibitem{Mueller}
A.  H.  Mueller,  Phys. Lett.  {\bf B308}, 355 (1993).

\bibitem{LuOneDim}
H. J. Lu, Phys. Rev. {\bf D45}, 1217 (1992).

\bibitem{BenekeBraun}
M. Beneke  and  V. M. Braun, Phys. Lett. {\bf B348}, 513 (1995).

\bibitem{Degrassi:1992ue}
G.~Degrassi and A.~Sirlin,
%
Phys.\ Rev.\ D {\bf 46}, 3104 (1992).

\bibitem{watson}
N.J. Watson. Nucl.\ Phys.\ B {\bf 494}, 388-432 (1997).

\bibitem{prw}
J. Papavassiliou, E. de Rafael, N.J. Watson.
Nucl.\ Phys.\ B {\bf 503}, 79 (1997). hep-ph/9612237.

\bibitem{Watson:1998vw}
N.~J.~Watson,
Nucl.\ Phys.\ Proc.\ Suppl.\  {\bf 74}, 341 (1999) [arXiv:hep-ph/9812203].

\bibitem{watson2b}
N.J. Watson.
Nucl.\ Phys.\ B {\bf 552}, 461 (1999). hep-ph/9812202.

\bibitem{rafael2}
J. Papavassiliou.
Phys.\ Rev.\ Lett.\  {\bf 84}, 2782 (2000). hep-ph/9912336.

\bibitem{Binosi:2002ft}
D.~Binosi and J.~Papavassiliou,
Phys.\ Rev.\ D {\bf 66}, 111901 (2002). hep-ph/0208189.

\bibitem{Binosiqcdall}
D.~Binosi and J.~Papavassiliou,
Nucl.\ Phys.\ Proc.\ Suppl.\  {\bf 121}, 281 (2003). hep-ph/0209016.

\bibitem{Binosi:2004qe}
D.~Binosi,
arXiv:hep-ph/0401182.

\bibitem{Binger:2003by}
M.~Binger and S.~J.~Brodsky,
Phys.\ Rev.\ D {\bf 69}, 095007 (2004) [arXiv:hep-ph/0310322].


\bibitem{Sivers:1990fh}
D.~W.~Sivers,
Phys.\ Rev.\ D {\bf 43}, 261 (1991).

\bibitem{hermes0001}
HERMES Collaboration, A. Airapetian et al., Phys. Rev. Lett. {\bf 84}, 4047
(2000); Phys.
Rev. D {\bf 64}, 097101 (2001).

\bibitem{smc99}
A. Bravar, for the SMC Collaboration, Nucl.\ Phys.\ B (Proc. Suppl.) {\bf
79}, 520
(1999).

\bibitem{Weinberg:1965nx}
S.~Weinberg,
Phys.\ Rev.\  {\bf 140}, B516 (1965).

\bibitem{Collins}
J.~C.~Collins, Phys.\ Lett.\ B {\bf 536}, 43 (2002). [arXiv:hep-ph/0204004].

\bibitem{Ji:2002aa}
X.~d.~Ji and F.~Yuan,
interactions?,''
arXiv:hep-ph/0206057.



\bibitem{Drell:1970wh}
S.~D.~Drell and T.~M.~Yan,
Phys.\ Rev.\ Lett.\  {\bf 25}, 316 (1970) [Erratum-ibid.\  {\bf 25}, 902
(1970)].

\bibitem{Bodwin:1981fv}
G.~T.~Bodwin, S.~J.~Brodsky and G.~P.~Lepage,
Phys.\ Rev.\ Lett.\  {\bf 47}, 1799 (1981).

\bibitem{Bodwin:1988fs}
G.~T.~Bodwin, S.~J.~Brodsky and G.~P.~Lepage,
Phys.\ Rev.\ D {\bf 39}, 3287 (1989).


\bibitem{Brodsky:1996nj}
S.~J.~Brodsky, A.~Hebecker and E.~Quack,
Phys.\ Rev.\ D {\bf 55}, 2584 (1997).

\bibitem{Brandenburg:1994mm}
A.~Brandenburg, V.~V.~Khoze and D.~Muller,
Phys.\ Lett.\ B {\bf 347}, 413 (1995).

\bibitem{Muller:1998fv}
D.~Muller, D.~Robaschik, B.~Geyer, F.~M.~Dittes and J.~Horejsi,
Fortsch.\ Phys.\  {\bf 42}, 101 (1994).

\bibitem{Ji:1996ek}
X.~D.~Ji,
Phys.\ Rev.\ Lett.\  {\bf 78}, 610 (1997).

\bibitem{E70496}
E704 Collaboration, A. Bravar et al., Phys.\ Rev.\ Lett.\ {\bf  77}, 2626
(1996).

\bibitem{lambda}
K. Heller, in Proceedings of Spin 96, C. W. de Jager, T. J. Ketel and P.
Mulders, Eds.,
World Scientific (1997).

\bibitem{Glauber:1955qq}
R.~J.~Glauber, Phys.\ Rev.\  {\bf 100}, 242 (1955).

\bibitem{Gribov:1968gs}
V.~N.~Gribov, Sov.\ Phys.\ JETP {\bf 30}, 709 (1970) [Zh.\ Eksp.\ Teor.\
Fiz.\  {\bf 57},
1306 (1969)].

\bibitem{Stodolsky:1966am}
L.~Stodolsky,
Phys.\ Rev.\ Lett.\  {\bf 18}, 135 (1967).

\bibitem{Brodsky:1969iz}
S.~J.~Brodsky and J.~Pumplin,
Phys.\ Rev.\  {\bf 182} (1969) 1794.

\bibitem{Ioffe:1969kf}
B.~L.~Ioffe,
%
Phys.\ Lett.\ B {\bf 30}, 123 (1969).

\bibitem{Frankfurt:1988zg}
L.~L.~Frankfurt and M.~I.~Strikman, Nucl.\ Phys.\ B {\bf 316}, 340 (1989).

\bibitem{Kopeliovich:1998gv}
B.~Z.~Kopeliovich, J.~Raufeisen and A.~V.~Tarasov, Phys.\ Lett.\ B {\bf
440}, 151 (1998)
[arXiv: hep-ph/9807211].

\bibitem{Kharzeev:2002fm}
D.~E.~Kharzeev and J.~Raufeisen, nucl-th/0206073, and references therein.


\bibitem{Mueller:2004se}
A.~H.~Mueller and A.~I.~Shoshi,
Nucl.\ Phys.\ B {\bf 692}, 175 (2004) [arXiv:hep-ph/0402193].

\bibitem{Qiu:2004da}
J.~w.~Qiu and I.~Vitev,
arXiv:hep-ph/0405068.

\bibitem{Martin:2004xw}
A.~D.~Martin, M.~G.~Ryskin and G.~Watt,
arXiv:hep-ph/0406224.

\bibitem{Adloff:1997sc}
C.~Adloff {\it et al.}  [H1 Collaboration],
Z.\ Phys.\ C {\bf 76}, 613 (1997) [arXiv:hep-ex/9708016].


\bibitem{Ruspa:2004jb}
M.~Ruspa,
Acta Phys.\ Polon.\ B {\bf 35}, 473 (2004).

\bibitem{Kuti:1971ph}
J.~Kuti and V.~F.~Weisskopf,
Phys.\ Rev.\ D {\bf 4}, 3418 (1971).

\cite{Arneodo:1992wf}
\bibitem{Arneodo:1992wf}
M.~Arneodo,
Phys.\ Rept.\  {\bf 240}, 301 (1994).

\bibitem{Brodsky:2004bg}
S.~J.~Brodsky, I.~Schmidt and J.~J.~Yang,
arXiv:hep-ph/0409279.

\bibitem{McFarland:2003gx}
K.~S.~McFarland {\it et al.},
Int.\ J.\ Mod.\ Phys.\ A {\bf 18}, 3841 (2003).

\bibitem{Petti}
R. Petti, (NOMAD collaboration) presented at ICHEP(2004).

\bibitem{Geer}
Steve Geer, hep-ph/0210113, Talk given at 4th NuFact '02 Workshop (Neutrino
Factories
based on Muon Storage Rings), London, England, 1-6 Jul 2002.

\bibitem{Berger:1986dd}
E.~L.~Berger,
ANL-HEP-CP-86-78
{\it Invited paper given at Int. Symp. on Weak and Electromagnetic
Interactions in
Nuclei, Heidelberg, West Germany, Jul 1-5, 1986}

\bibitem{Accardi:2003be}
A.~Accardi {\it et al.},
arXiv:hep-ph/0308248.


\end{thebibliography}
\end{document}